\documentclass[a4paper,11pt]{article}
\usepackage{jheppub}
\usepackage{amsmath,amssymb}
\usepackage{amsthm}
\usepackage{ascmac}
\usepackage{subcaption}
\usepackage{float}
\usepackage{physics}
\usepackage{comment}
\numberwithin{equation}{section}
\newcommand{\ee}{\mathrm{e}}
\newcommand{\ii}{\mathrm{i}}
\newcommand{\cut}{\mathrm{cut}}

\newcommand {\nn} {\nonumber}

\title{Quantum decoherence
  in the Caldeira--Leggett model by the real-time path integral
on a computer}

\author[a,b]{Jun Nishimura,}
\emailAdd{jnishi@post.kek.jp}

\author[a,c,d]{Hiromasa Watanabe}
\emailAdd{hiromasa.watanabe@yukawa.kyoto-u.ac.jp}

\affiliation[a]{KEK Theory Center,
Institute of Particle and Nuclear Studies,\\
High Energy Accelerator Research Organization,
1-1 Oho, Tsukuba, Ibaraki 305-0801, Japan}
\affiliation[b]{Graduate Institute for Advanced Studies, SOKENDAI,
1-1 Oho, Tsukuba, Ibaraki 305-0801, Japan}
\affiliation[c]{Yukawa Institute for Theoretical Physics, Kyoto University,\\
Kitashirakawa Oiwakecho, Sakyo-ku, Kyoto 606-8502, Japan}
\affiliation[d]{Department of Physics, and Research and Education Center for Natural Sciences, Keio University,\\
4-1-1 Hiyoshi, Yokohama, Kanagawa 223-8521, Japan}

\preprint{KEK-TH-2694, YITP-23-139}

\abstract{We propose first-principle calculations of an open system
  based on the real-time path integral formalism
  treating the environment as well as the system of our interest together on a computer.
  The sign problem that occurs in applying Monte Carlo methods
  can be overcome in general
  by using the so-called Lefschetz thimble method, which has been
  developed over the past decade.
  Here we focus on the Caldeira--Leggett model, which
  is well known, in particular, as a model of quantum decoherence.
  In this case, the calculation simplifies drastically
  since the path integral becomes Gaussian for typical initial conditions.
  The relevant saddle point, which is unique and complex, can be determined
  by solving a linear equation with a huge but sparse
  coefficient matrix,
  and the integration over the Lefschetz thimble can be performed analytically.
  Thus we obtain, without assumptions or approximations,
  the reduced density matrix
  after a long-time evolution,
  tracing out a large number of harmonic oscillators in the environment.
  In particular, we confirm the dependence of the decoherence time on
  the coupling constant and the temperature
  that has been predicted from the master equation in a certain parameter regime.
}

\begin{document}
\maketitle
\flushbottom

\section{Introduction}

Quantum mechanics is one of the most successful theories in physics,
which enables us to understand various nontrivial phenomena
that occur in our Universe.
In particular,
since the dawn of quantum mechanics in 1920s,
experiments on microscopic or even mesoscopic systems have
become gradually possible,
and the theory of measurements in quantum mechanics
has developed significantly.
Nowadays it is widely
recognized that
the presence of the measuring device
and the
environment
cannot be ignored
since they disturb the system through
interactions and the subsequent entanglement.
Tracing out the environment after the measurement
disables
the interference that the system can potentially exhibit.
This effect,
which is called
{\it quantum decoherence} (See \emph{e.g.},
Refs.~\cite{Schlosshauer,Zurek:1991vd}),
plays a crucial role
in accounting for the outcome of measurements
consistently
by calculations
in quantum mechanics.
Since decoherence is a source of quantum noise in quantum computation
and experiments
such as the gravitational wave detection which use quantum technologies,
realistic modeling of the influence
by the environment
is important
in reducing the uncontrollable uncertainties
due to the noise.

Theoretically,
the decoherence can be expressed as the disappearance of the
off-diagonal elements of the density matrix
in a basis associated with
the measurement during the time evolution,
which causes
the quantum-to-classical transition.
Thus the decoherence
is important
in interpreting
the ``collapse of wave functions'' properly
as well as
in answering fundamental questions
such as ``How does the microscopic quantum theory turn into
the macroscopic classical theory, which describes
our real world?'' \cite{Schlosshauer,Zurek:1991vd}.

A common strategy for studying a system coupled to some environment
is to use the master equation \cite{Lindblad:1975ef,Gorini:1975nb}
that describes the non-unitary time evolution
of the reduced density matrix of the system
after tracing out the environment.
(See also Section 4 of Ref.~\cite{Schlosshauer} and references therein.)
However,
master equations are obtained in general
only under some assumptions such as high temperature
with some approximations
such as the Born and Markov approximations.
It is therefore desirable to develop alternative methods that do not rely
on such assumptions and approximations.

As a possible approach, one can think of
investigating
the unitary time evolution of the whole system
including the environment
either by
solving the Schr\"{o}dinger equation or
by evaluating
the Feynman path integral.
The first option has been pursued
in the context of
decoherence,
for instance, in
Refs.~\cite{Nagele:2020kef,adami_negulescu_2012}.
However, the required computational cost is
similar to that for
diagonalizing the Hamiltonian,
and it grows exponentially with the number of degrees of freedom
in the whole system.
Here we consider the second option,
namely evaluating the path integral based on
Monte Carlo (MC) simulation,\footnote{Apart from evaluating the real-time
path integral numerically by Monte Carlo methods, there is also
an analytic approach based on perturbation theory, which enables us to
calculate the time evolution of the reduced density matrix in
quantum field theories \cite{Burrage:2018pyg,Kading:2022jjl,Kading:2025cwg}.}
which has a potential to reduce
the computational cost
to a power-law growth.
The obstacle in that case is the notorious sign problem due to
the oscillating phase in the integrand.
See Refs.~\cite{Woodward:2022pet,Nishimura:2023dky,Blum:2023wnb,Alvestad:2023jgl,Steingasser:2023gde,Ai:2023yce,Steingasser:2024ikl,Mou:2024oca,Garbrecht:2024end,Feldbrugge:2025alg,Kavaki:2025hcu}
for recent development in the path integral approach to the real-time quantum evolution.

MC simulation is a well-established practical
method
to evaluate a multi-variable integral.
The idea is to
generate variables, let's say $\Phi$,
with the probability density proportional to
the Boltzmann weight $\ee^{-S[\Phi]}$ with the action $S[\Phi]$
and to compute the expectation values.
This method is based on the fact that the action $S[\Phi]$ is
real so that the Boltzmann weight can be regarded as the probability density.
When the action becomes complex, a naive implementation
of the MC method becomes
problematic.
A straightforward approach would be to use
the reweighting method in which
one
generates the ensemble with the probability density
$\ee^{-\Re S[\Phi]}$
and includes the effects of the complex phase $\ee^{-\ii \Im S[\Phi]}$ in
evaluating the expectation values.
Unfortunately,
this complex phase becomes highly oscillatory as the system size increases,
and forces us to generate a prohibitively
large number of sample configurations by MC to estimate physical quantities with
sufficient precision.
This is called the sign problem.

The situation with the sign problem
has drastically changed
over the last decades, however.
It has been widely recognized that 
a promising way to overcome the problem is to complexify the integration variables
and to deform the integration contour based on Cauchy's theorem
in such a way that the sign problem is ameliorated.
The Lefschetz thimble
method~\cite{Witten:2010cx,Cristoforetti:2012su,Cristoforetti:2013wha,Fujii:2013sra} is a typical method in that direction.
In particular, 
after the proposal of 
the
generalized Lefschetz thimble method (GTM)~\cite{Alexandru:2015sua},
various important techniques have been
developed \cite{Fukuma:2017fjq,Fukuma:2019uot,Fukuma:2020fez,Fukuma:2021aoo,Fujisawa:2021hxh,Fukuma:2023eru,Nishimura:2024bou},
which enabled, for instance,
the investigation of quantum tunneling \cite{Nishimura:2023dky},
quantum cosmology \cite{Chou:2024sgk}
and string theory \cite{Chou:2025moy} based on the real-time path integral.
It is expected that this method is useful also in investigating
a system coupled to its environment
since the computational cost grows only by a power law with the system size $V$.
More precisely, the generation of sample configurations requires the cost of O($V$),
whereas the calculation of the Jacobian for reweighting requires the cost of O($V^3$).
%


As a first step towards such calculations,
we focus on
the Caldeira--Leggett (CL) model  \cite{Caldeira:1982iu,Caldeira:1982uj},
which has been studied intensively
in investigating decoherence and dissipation in quantum
systems \cite{Unruh:1989dd,Paz:1992pn,Zurek:1992mv}
although it was originally proposed to describe the quantum Brownian motion.
See also a textbook~\cite{Schlosshauer},
reviews~\cite{Zurek:2003zz,Schlosshauer:2019ewh},
and references therein.
In fact, the calculations simplify in this case drastically since
the path integral
to be evaluated
for typical initial conditions
is nothing but a multi-variable Gaussian integral.
  The relevant saddle point, which is unique and complex, can be determined
  by solving a linear equation with a huge but sparse
  coefficient matrix,
  and the integration over the Lefschetz thimble can be performed analytically.
  Thus we obtain, without assumptions or approximations,
  the reduced density matrix
  after a long-time evolution,
  tracing out a large number of harmonic oscillators in the environment.


The purpose of this paper is
to confirm the
usefulness
of our formalism
by
reproducing
the nature of decoherence in the CL model
predicted
from the master equation.
%
For that,
we discuss in detail
the correspondence between
the parameters in the master equation
and those in our formalism,
and confirm the scaling with respect to the number of harmonic oscillators
in the environment.
%
In particular, we focus on the high temperature and weak coupling region,
where the description by the master equation is expected to be valid qualitatively.
%
Our results indeed confirm
the dependence of the decoherence time
on the parameters of the model which is predicted by the master equation.
Let us stress, however, that
our method
is applicable to 
a general parameter region in which the master equation is not valid.
We also observe the thermalization of the system of our interest
into a canonical distribution due to the interaction with the environment.
Thus
this work
paves the way to
a new possibility of investigating
the time evolution of an open quantum system
by performing the real-time path integral of the whole system
including the environment numerically.
Part of our results were reported briefly in our previous
publication \cite{Nishimura:2024one} emphasizing our finding
that quantum decoherence can be captured by complex saddle points
just like quantum tunneling can be captured by instantons,
which are real saddle points in the imaginary-time path integral formalism.
(See also Ref.~\cite{Nishimura:2023dky} for a new picture of quantum tunneling
based on complex saddle points in the real-time path integral formalism.)

The rest of this paper is organized as follows. 
In Section~\ref{sec:CL_review}, we first review
the CL model and discuss its properties
concerning quantum
decoherence
based on the master equation.
In Section~\ref{sec:numerical_setup}, we discuss
how to perform the real-time path integral for the model
with discretized time and a finite number of harmonic oscillators.
In Section~\ref{sec:main_results_case1},
we show our numerical results for a single wave packet,
and discuss how the decoherence takes
place in that case.
In Section \ref{sec:numerical_analysis},
we discuss how to extend our calculations
to the case of two wave packets
analogous to the double-slit experiment
and show our numerical results, which clearly indicate
the effects of decoherence
as the fading of the interference pattern.
%
Section~\ref{sec:conclusion} is devoted to a summary
and discussions.
In Appendix~\ref{sec:limits_suppl}, we discuss
the effect of finite lattice spacing introduced in discretizing time
and that of using a finite number of harmonic oscillators in the environment,
which discretizes their frequency spectrum.
In Appendix \ref{sec:more_double-slit}, we
discuss the decoherence for the two wave packets
from the viewpoint of the off-diagonal elements of the reduced density matrix.
%
%


\section{Brief review of decoherence in the Caldeira--Leggett model}
\label{sec:CL_review}

The Caldeira--Leggett (CL) model~\cite{Caldeira:1982iu,Caldeira:1982uj}
was originally introduced to describe
the quantum Brownian motion, which is
caused by the effect of friction in a quantum system.
After the original work, it was pointed out that this model also exhibits
significant
quantum decoherence~\cite{Unruh:1989dd}.
In this section, we briefly review the CL model and
its properties concerning quantum
decoherence
based on Ref.~\cite{Caldeira:1982iu}.
Although our discussion shall be
given in the path integral formalism,
which is used in our numerical method later on,
similar discussions are also possible in the operator formalism
as one can see in Ref.~\cite{PhysRevLett.113.200403}, for instance.
For more comprehensive reviews,
see
Refs.~\cite{Schlosshauer:2003zy,Schlosshauer:2019ewh,Schlosshauer,Hartle:1992as,Grabert:1988yt}.

The CL model is defined
by the Hamiltonian
\begin{align}
    H    &=    H_0 + H_\mathcal{E} + H_\mathrm{int} \ ,
    \label{eq:Hamiltonian_CL_1} \\
    H_0  &= 
    \frac{p^2}{2M} + V(x) \ ,
    \qquad
    H_\mathcal{E}
    =
    \sum_{k=1}^{N_\mathcal{E}} 
    \qty(
        \frac{p_k^2}{2m} + \frac{1}{2} m\, \omega_k^2 \, q_k^2
    ) \ ,
    \qquad
    H_\mathrm{int}
    = -  x\sum_{k=1}^{N_\mathcal{E}}c_k \, q_k \ ,
    \label{eq:Hamiltonian_CL_2}
\end{align}
where $N_\mathcal{E}$ represents the number of
harmonic oscillators
in the environment $\mathcal{E}$, and
$c_k$ are the coupling constants 
between the system $\mathcal{S}$ and the environment $\mathcal{E}$.
In what follows,
we assume that the system $\mathcal{S}$ is also a harmonic oscillator defined by
\begin{equation}
    V(x) = \frac{1}{2} \, M \, \omega_\mathrm{b}^2 \, x^2 \ ,
    \label{eq:CL_model_potential_HO}
\end{equation}
where $\omega_\mathrm{b}$ represents the ``bare'' frequency
as opposed to the ``renormalized'' one defined later in
\eqref{eq:omega_ren_large_N}.
We also assume that 
the initial condition
for the density matrix is given by
\begin{equation}
    \hat{\rho}(t=0) 
    =
    \hat{\rho}_\mathcal{S}(0)\otimes\hat{\rho}_\mathcal{E}(0) \ ,
\end{equation}
meaning that $\mathcal{S}$ and $\mathcal{E}$ are separable.
The initial density matrix $\hat{\rho}_\mathcal{S}(0)$ for the system
$\mathcal{S}$ shall be specified later \eqref{init-rho-system-gen},
whereas
the initial density matrix $\hat{\rho}_\mathcal{E}(0)$ for the environment
is assumed to be
the canonical ensemble with the temperature $T = \beta^{-1}$.

In view of the form of the interaction term $H_\mathrm{int}$ 
in \eqref{eq:Hamiltonian_CL_2},
we will work in the position basis in what follows.
In particular, the initial density matrix $\hat{\rho}_\mathcal{E}(0)$
for the environment can be written explicitly as\footnote{Throughout this paper, we set $\hbar = 1$.}
\begin{align}
    \rho_\mathcal{E}(q,\tilde{q};0)
    &=
    \mel{\tilde{q}}{\hat{\rho}_\mathcal{E}(0)}{q}
    =
    \prod_{k=1}^{N_\mathcal{E}}
        \rho_\mathcal{E}^{(k)}(q_k,\tilde{q}_k) \ ,
    \label{eq:Equilib_env_1}
    \\
        \rho_\mathcal{E}^{(k)}(q_k,\tilde{q}_k) 
    &=
    \sqrt{\frac{m\omega_k}{2\pi\sinh\beta\omega_k}}
    \exp\left[
    - \frac{m\omega_k}{2\sinh\beta\omega_k}
    \left\{
    \left(q_k^2+\tilde{q}_k^2 \right)
    \cosh\beta\omega_k - 2q_k\tilde{q}_k\right\}
    \right] \ .
    \label{eq:Equilib_env_2}
\end{align}
Note that Eq.~\eqref{eq:Equilib_env_2} is nothing but
the propagator for harmonic oscillators along the imaginary time and
can be derived, e.g., from Mehler's formula.
On the other hand,
the total density matrix for the whole system
obeys the time evolution given by
\begin{equation}
    \rho(x,q;\tilde{x},\tilde{q};t)
    =
    \mel{x,q}{\hat{\rho}(t)}{\tilde{x},\tilde{q}}
    =
    \mel{x,q}{\ee^{-\ii \hat{H} t}\hat{\rho}(0)\ee^{\ii \hat{H} t}}{\tilde{x},\tilde{q}} \ .
\end{equation}
Inserting the complete set of basis, we obtain
\begin{equation}
    \rho(x,q;\tilde{x},\tilde{q};t)
    =
    \int \dd x' \dd\tilde{x}' \dd q' \dd \tilde{q}'
    \mel{x,q}{\ee^{-\ii \hat{H} t}}{x',q'}
    \mel{x',q'}{\hat{\rho}(0)}{\tilde{x}',\tilde{q}'}
    \mel{\tilde{x}',\tilde{q}'}{\ee^{\ii \hat{H} t}}{\tilde{x},\tilde{q}} \ ,
    \label{time-evolve-rho-decompose}
\end{equation}
which can be rewritten further by using the propagator
\begin{align}
    \mel{x,q}{\ee^{-\ii \hat{H} t}}{x',q'}
    &=
    \int \mathcal{D}x\mathcal{D}q\; \ee^{\ii S[x(t),q(t)]} \ ,
\end{align}
where the action is given by
\begin{align}
    S[x,q]  &= S_0[x] + S_\mathcal{E}[q] + S_\mathrm{int}[x,q] \ ,
    \label{eq:action_CL_original}
\\
    S_0[x] 
&=    \int_0^t\dd\tau 
    \left\{ \frac{1}{2}  M  \dot{x}(\tau)^2 -
  \frac{1}{2} \, M \, \omega_\mathrm{b}^2 \, x^2 
  \right\} \ , \\
    S_\mathcal{E}[q]
    &=   \int_0^t\dd\tau 
    \sum_{k=1}^{N_\mathcal{E}}
    \left\{
        \frac{1}{2} \, m  \, \dot{q}_k(\tau)^2
        -  \frac{1}{2}\, m \, \omega_k^2 \, q_k(\tau)^2  \right \} \ ,
    \\
    S_\mathrm{int}[x,q]
    &=  \int_0^t\dd\tau \,  x(\tau)\,
     \sum_{k=1}^{N_\mathcal{E}} c_k \, q_k(\tau) \ .
\label{eq:action_CL_original_components}
\end{align}
We have defined $\dot{x}(\tau) \equiv \dd{x}\!(\tau)/\dd \tau$,
$\dot{q}_k (\tau) \equiv \dd{q}_k\!(\tau)/\dd\tau$
and
the boundary conditions are given by
$x(0) = x'$, $x(t) = x$, $q(0) = q'$ and $q(t) = q$.
The other factor $\mel{\tilde{x}',\tilde{q}'}{\ee^{\ii \hat{H} t}}{\tilde{x},\tilde{q}}$
in \eqref{time-evolve-rho-decompose} can be rewritten in a similar way.

    In order to
    investigate the decoherence,
    we trace out the environment
    by integrating out all the degrees of freedom in $\mathcal{E}$,
    and obtain the reduced density matrix of $\mathcal{S}$ as
\begin{align}
    \rho_\mathcal{S}(x_\mathrm{F},\tilde{x}_\mathrm{F};t)
    =
    \int
     \dd{q}
    \mel{x_\mathrm{F},q}{\hat{\rho}(t)}{\tilde{x}_\mathrm{F},q}
    =
    \int\dd x_\mathrm{I} \dd \tilde{x}_\mathrm{I}\;
    J(x_\mathrm{F},\tilde{x}_\mathrm{F};t;x_\mathrm{I},\tilde{x}_\mathrm{I};0)\,
    \rho_\mathcal{S}(x_\mathrm{I},\tilde{x}_\mathrm{I};0) \ ,
    \label{eq:reduced_denisty_matrix_primary}
\end{align}
where the propagator for the reduced density matrix is given by
\begin{equation}
    J(x_\mathrm{F},\tilde{x}_\mathrm{F};t;x_\mathrm{I},\tilde{x}_\mathrm{I};0)
    =
    \int\mathcal{D}x\mathcal{D}\tilde{x}\;
    \mathrm{e}^{\mathrm{i}(S_0[x]-S_0[\tilde{x}])}
    \,\mathrm{e}^{-  W(x,\tilde{x})} 
    \label{eq:J_propagator_primary}
\end{equation}
with the boundary conditions $x(0)= x_\mathrm{I}$,
$x(t)= x_\mathrm{F}$ and
$\tilde{x}(0) = \tilde{x}_\mathrm{I}$,
$\tilde{x}(t) = \tilde{x}_\mathrm{F}$.
We have defined the Feynman--Vernon influence functional~\cite{Feynman:1963fq}
\begin{equation}
    e^{-W(x,\tilde{x})}
    =
    \int \mathcal{D}q\mathcal{D}\tilde{q}\, 
    \exp\qty[\ii\qty(S_\mathcal{E}[q] - S_\mathcal{E}[\tilde{q}]
      + S_\mathrm{int}[x,q] - S_\mathrm{int}[\tilde{x},\tilde{q}])] \, 
        \rho_\mathcal{E}(q(0),\tilde{q}(0);0) \ ,
    \label{Feynman-Vernon}
\end{equation}
which contains all the effect of the environment $\mathcal{E}$ that
has been traced out.
Here
the path integral is performed over
the variables in the environment $q$ and $\tilde{q}$
with the boundary condition $q(t)=\tilde{q}(t)$.

In the case at hand
with the initial condition
\eqref{eq:Equilib_env_1},
the path integral \eqref{Feynman-Vernon}
can be evaluated
explicitly as \cite{Caldeira:1982iu,Feynman:1963fq}
%
\begin{equation}
    W(x,\tilde{x})
    =
    \int_0^t\dd\tau\int_0^\tau\dd s
    \left\{ x(\tau)-\tilde{x}(\tau)\right\}
    \left\{
        \alpha(\tau-s)x(s)
        -
        \alpha^*(\tau-s)\tilde{x}(s)
    \right\}
      \ ,
\end{equation}
where
the kernel $\alpha(\tau)$
is given by
\begin{gather}
    \Re \alpha(\tau) 
    =  \sum_{k=1}^{N_\mathcal{E}}
    \frac{c_k^2}{2m\omega_k}\coth(\frac{\beta\omega_k}{2})
    \cos(\omega_k\tau) \ ,
    \qquad
    \Im \alpha(\tau)
    =  -    \sum_{k=1}^{N_\mathcal{E}}
    \frac{c_k^2}{2m\omega_k}\sin(\omega_k\tau) \ .
    \label{eq:kernel_Re_Im}
\end{gather}
Plugging this in \eqref{eq:J_propagator_primary},
we can rewrite the propagator
$J(x_\mathrm{F},\tilde{x}_\mathrm{F};t;x_\mathrm{I},\tilde{x}_\mathrm{I};0)$
in the form
\begin{align}
    J(x_\mathrm{F},\tilde{x}_\mathrm{F};t;x_\mathrm{I},\tilde{x}_\mathrm{I};0)
    &=
    \int\mathcal{D}x\mathcal{D}\tilde{x}\; 
    \ee^{\ii (S_0[x] - S_0[\tilde{x}])}
    \ee^{
    -\ii W_\mathrm{I}(x,\tilde{x};t)
    -W_\mathrm{R}(x,\tilde{x};t)} \ ,
    \label{eq:J_propagator_kernel}
\\
    W_\mathrm{R}(x,\tilde{x};t)
&=
    \int_0^t\dd\tau\int_0^\tau\dd{s}\;
    \{ x(\tau)-\tilde{x}(\tau) \} \Re\alpha(\tau-s) \{ x(s)-\tilde{x}(s) \} \ ,
    \label{eq:W_real}
    \\
    W_\mathrm{I}(x,\tilde{x};t)
  &=
    \int_0^t\dd\tau\int_0^\tau\dd{s}\;
    \{ x(\tau)-\tilde{x}(\tau) \} \Im\alpha(\tau-s) \{ x(s)+\tilde{x}(s) \}  \ .
    \label{eq:W_imag}
\end{align}

Note that the influence functional given by 
\eqref{eq:W_real} and \eqref{eq:W_imag}
involves integration over $\tau$ and $s$,
which implies that it is non-local with respect to time.
We can
simplify it
by taking the $N_\mathcal{E} \to \infty$ limit as follows.
First
we introduce the spectral density as
\begin{equation}
  \rho(\omega)  =
    \sum_{k=1}^{N_\mathcal{E}} \delta(\omega-\omega_k) \ ,
    \qquad
  \rho(\omega) \, C(\omega)^2
  =
    \sum_{k=1}^{N_\mathcal{E}}
    c_k^2 \, 
    \delta(\omega-\omega_k) \ ,
    \label{def-spectral-density}
\end{equation}
and rewrite the sums in \eqref{eq:kernel_Re_Im} as
\begin{align}
    \Re \alpha(\tau)
    & =
    \int_0^\infty
        \dd{\omega}
    \frac{\rho(\omega)\, C(\omega)^2}{2m\omega}
    \coth\left(\frac{\beta\omega}{2}\right)\cos(\omega \tau)\ ,
\label{re-alpha-def}
    \\
    \Im \alpha(\tau)
    & =
    -\int_0^\infty
        \dd{\omega}
        \frac{\rho(\omega)\, C(\omega)^2}{2m\omega}\sin(\omega \tau) \ .
\label{im-alpha-def}
\end{align}
Let us assume
a specific form of the density called
the Ohmic spectrum\footnote{
The reason for being Ohmic can be seen from the fact that 
the spectral density
$ J(\omega) = \sum_k \frac{c_k^2}{2m_k \omega_k} \delta(\omega-\omega_k) $,
which is used in standard textbooks instead of \eqref{def-spectral-density},
becomes linear
$J(\omega) \propto \omega $ for $\omega \le \omega_\cut$.}
as
\begin{equation}
    \rho(\omega) \, C^2(\omega)
    =
    \left\{
    \begin{array}{cc}
        \displaystyle
                8mM\gamma \, \omega^2/(2\pi) &\qquad \omega \le \omega_\cut \ ,
        \\
        \displaystyle
        0 &\qquad \omega > \omega_\cut \ ,
    \end{array}
    \right. 
    \label{eq:Ohmic_spectrum}
\end{equation}
which simplifies the kernel $\alpha(\tau)$
and validates
the Markov approximation~\cite{Schlosshauer:2019ewh,Schlosshauer}.
Within this assumption,
there are still
two parameters
$\gamma$, and $\omega_\cut$. 
Note that $\gamma$
%
governs
the strength of the coupling between the system and the environment.
As we will see shortly in \eqref{eq:omega_ren_large_N},
the cutoff frequency $\omega_\cut$, as well as $\gamma$,
appears in the physical parameters of the system as the ``renormalization''
of the frequency.


Under the assumption~\eqref{eq:Ohmic_spectrum},
the imaginary part of the kernel
\eqref{im-alpha-def}
becomes
\begin{align}
    \Im \alpha(\tau)
    &=
    - 4 M \gamma
    \int_0^{\omega_\mathrm{cut}} \frac{\dd\omega}{2\pi}\;\omega\sin(\omega\tau)
    =
    2 M \gamma
   \dv{\tau} f(\tau) \ , 
\end{align}
where we have defined
\begin{align}
  f(\tau) =
  \int_{-\omega_\mathrm{cut}}^{\omega_\mathrm{cut}} \frac{\dd\omega}{2\pi}\;
 \cos(\omega\tau)
 \label{def-f-function}
  \ ,
\end{align}
which can be approximated by the delta function $\delta(\tau)$
for $\omega_\mathrm{cut}\gg 1/\tau_0$ with $\tau_0$ being the typical
time resolution.
Then the integral
over
$s$ in \eqref{eq:W_imag} can be performed by parts as 
\begin{align}
    W_\mathrm{I}(x,\tilde{x};t)
  & \approx - 2 M \gamma
    \int_0^t\dd\tau
    \;
    \left[ \{ x^2(\tau)-\tilde{x}^2(\tau) \}  f(0)
    -  \frac{1}{2} \{ x(\tau)-\tilde{x}(\tau) \}
    \{ \dot{x}(\tau)+\dot{\tilde{x}}(\tau) \} \right]
    \nn
    \\
  &  \approx
    -\frac{2M\gamma \, \omega_\mathrm{cut}}{\pi}
    \int_0^t\dd{\tau} \{ x^2(\tau)-\tilde{x}^2(\tau) \} 
    +
    M\gamma\int_0^t\dd{\tau}
    \{ x(\tau)-\tilde{x}(\tau)\} \{ \dot{x}(\tau)+\dot{\tilde{x}}(\tau) \}  \ ,
\label{W-imag-result}
\end{align}
where we have used $f(0)= \omega_\mathrm{cut}/\pi$.
Note that the first term can be absorbed
by shifting the
parameter $\omega_\mathrm{b}^2$
of the potential \eqref{eq:CL_model_potential_HO}
in $S_0[x]$ and $S_0[\tilde{x}]$
that appear in \eqref{eq:J_propagator_kernel} as
\begin{align}
  \label{eq:omega_ren_large_N}
  \omega_\mathrm{r}^2
  =  \omega_\mathrm{b}^2
  - (\Delta\omega)^2 \ ,
\qquad
   (\Delta\omega)^2
    =
    \frac{4\gamma \, \omega_\cut}{\pi} \ .
\end{align}
%
Thus, this term corresponds to ``renormalization''
of the parameter in the original Hamiltonian
due to the presence of the environment.
This issue shall be discussed
in the next section
from a more general point of view.

Similarly the real part of the kernel
\eqref{re-alpha-def}
simplifies under the assumption~\eqref{eq:Ohmic_spectrum} as
\begin{align}
    \Re \alpha(\tau)
    &= 4 M \gamma
    \int_0^{\omega_\mathrm{cut}} \frac{\dd\omega}{2\pi}\;
       \omega \coth\left( \frac{\beta\omega}{2}\right) \, \cos(\omega\tau)
    \approx
    \frac{4 M \gamma}{\beta}
    f(\tau) \ ,
    \label{eq:Re-alpha}
\end{align}
where we have used
$\coth\qty(\frac{\beta\omega}{2})\approx\frac{2}{\beta\omega}$
assuming
$\beta \omega_\mathrm{cut} \ll 1 $ \cite{Caldeira:1982iu}.
Then the integral over $s$ in \eqref{eq:W_real}
can be performed as
\begin{align}
    W_\mathrm{R}(x,\tilde{x};t)
    &=
    \frac{4M\gamma}{\beta}
    \int\dd{\tau}\dd{s} \{ x(\tau)-\tilde{x}(\tau) \} 
    f(\tau - s)
    \{ x(s)-\tilde{x}(s) \} 
    \notag\\
    &\approx
    \frac{2M\gamma}{\beta}\int_0^t\dd{\tau} \{ x(\tau)-\tilde{x}(\tau) \} ^2  \ ,
    \label{W-real-result}
\end{align}
where we have approximated \eqref{def-f-function} by the delta function again.

Using \eqref{W-imag-result} and \eqref{W-real-result},
the propagator~\eqref{eq:J_propagator_kernel}
for the reduced density matrix
becomes~\cite{Caldeira:1982iu,Feynman:1963fq},
\begin{align}
    & J(x_\mathrm{F},\tilde{x}_\mathrm{F};t;x_\mathrm{I},\tilde{x}_\mathrm{I};0) \nn \\
    = &      \int \mathcal{D}x\mathcal{D}\tilde{x}\,
    \exp\left[
        \ii S_\mathrm{r}[x] - \ii S_\mathrm{r}[\tilde{x}]
        -
        \ii M\gamma \int_0^t \dd\tau
         \{ x(\tau)-\tilde{x}(\tau) \} \{ \dot{x}(\tau)+\dot{\tilde{x}}(\tau) \} 
    \right]
    \notag\\
    & \quad \times 
    \exp\left[
        -\frac{2M\gamma}{\beta}
        \int_0^t \dd\tau\,\left\{  x(\tau) - \tilde{x}(\tau)
        \right \} ^2
    \right] , 
    \label{eq:J_propagator_CL}
\end{align}
where the effect of the renormalization appears in the action as
\begin{equation}
    S_\mathrm{r} [x]
    = \frac{1}{2}  M  \dot{x}^2 - 
         \frac{1}{2} \, M\omega_\mathrm{r}^2 \, x^2 \ ,
\end{equation}
which is obtained by replacing $\omega_\mathrm{b}^2$
in $S_0[x]$ by \eqref{eq:omega_ren_large_N}.


Plugging the obtained propagator \eqref{eq:J_propagator_CL}
in \eqref{eq:reduced_denisty_matrix_primary},
one can deduce the
master equation
for the time evolution of the reduced density matrix
as\footnote{This master equation can be expressed in a 
  simpler form
  in the operator
  formalism~\cite{PhysRevLett.113.200403,Schlosshauer:2019ewh,Schlosshauer} as
\begin{align}
    \dv{\hat{\rho}_\mathcal{S}(t)}{t}
    &=
    -\ii \comm{\hat{H}_\mathrm{r}}{\hat{\rho}_\mathcal{S}(t)}
    -\ii\gamma \comm{\hat{x}}{\left\{\hat{p}, \hat{\rho}_\mathcal{S}(t)\right\}}
    -\frac{2M\gamma}{\beta}\comm{\hat{x}}
      {\comm{\hat{x}}{\hat{\rho}_\mathcal{S}(t)}} \ ,
\end{align}
where the Hamiltonian $\hat{H}_\mathrm{r}$ is defined as
$ \hat{H}_\mathrm{r} = \hat{H}_0 - \frac{1}{2}M(\Delta\omega)^2 x^2 $
with the shifted frequency \eqref{eq:omega_ren_large_N}.
We should also mention
that this master equation
does not maintain
the positivity of
the reduced density matrix
due to the high temperature approximation.
The positivity can be recovered \cite{Diosi_1993,DIOSI1993517}, however,
by adding an $O(\beta)$ term
$\propto \gamma\beta\comm{\hat{p}}{\comm{\hat{p}}{\hat{\rho}_\mathcal{S}(t)}}$,
which makes the master equation of the Lindblad type.
}
%
%
%
%
%
%
\begin{align}
&     \dv{t}\rho_\mathcal{S}(x,\tilde{x};t)
    =
    K(x,\tilde{x})
    \rho_\mathcal{S}(x,\tilde{x},t) \ , \\
&    K(x,\tilde{x}) 
    = 
    \frac{\ii}{2M}\qty(\pdv[2]{x}-\pdv[2]{\tilde{x}})
    -
    \frac{\ii}{2}M\omega_\mathrm{r}^2(x^2-\tilde{x}^2)
    - 
    \gamma(x-\tilde{x})\qty(\pdv{x}-\pdv{\tilde{x}})
    -
    \frac{2M\gamma}{\beta}(x-\tilde{x})^2 \ .
    \label{eq:CL_master_eq}
\end{align}
The first two terms correspond to the standard Liouville-von Neumann terms
describing the unitary evolution.
The third term describes the dissipation
due to the momentum damping, where
its typical time scale is controlled
by the effective coupling $\gamma$.

At high temperature (small $\beta$), the last term dominates
in the master equation~\eqref{eq:CL_master_eq}, and one obtains
\begin{equation}
    \dv{t}\rho_\mathcal{S} (x,\tilde{x};t)
    \sim    -\frac{2M \gamma}{\beta}(x-\tilde{x})^2\rho_\mathcal{S}(x,\tilde{x};t) \ ,
    \label{rho-last-term}
\end{equation}
which implies the behavior
\begin{equation}
    \rho_\mathcal{S}(x, \tilde{x};t)
    =    \rho_\mathcal{S}(x, \tilde{x} ;0)\,
    \ee ^{- \frac{2M \gamma}{\beta} (x - \tilde{x})^2 t } \ .
    \label{eq:rho_S_time-evol_highT}
\end{equation}
Thus the off-diagonal elements of the density matrix in the position basis
decay exponentially with time.
This corresponds to the decoherence
caused by
the environment,
where its typical time scale $\tau_\mathrm{d}$ can be defined by
\begin{equation}
    \tau_\mathrm{d}^{-1}(x,\tilde{x})
    =    \frac{2M\gamma}{\beta}(x-\tilde{x})^2
    = \gamma \left( \frac{x-\tilde{x}}{\lambda_\mathrm{dB}} \right)^2 \ .
    \label{eq:decoherence_timescale}
\end{equation}
Here we have defined
the thermal de Broglie wavelength $\lambda_\mathrm{dB} = \sqrt{{\beta}/{2M}}$
of the particle with mass $M$ under temperature $T=\beta^{-1}$.
Let us emphasize that the decoherence time $\tau_\mathrm{d}$
becomes shorter
for longer distance between the positions $x$ and $\tilde{x}$.

One of our main purposes of this work is to investigate the decoherence
from first-principle calculations and to confirm, in particular,
the formula \eqref{eq:decoherence_timescale} obtained above
from the master equation.
The importance of this is clear
since the simple form of the master equation~\eqref{eq:CL_master_eq}
for the reduced density matrix $\rho_\mathcal{S}(x,\tilde{x})$
is arrived at only with a specific setup and some assumptions.
%
In particular,
in order to validate the Markov approximation,
we have taken the large $N_\mathcal{E}$ limit
assuming a specific form of the spectral density
\eqref{eq:Ohmic_spectrum}
for the environment and imposed a condition
$1/\tau_0 \ll \omega_\mathrm{cut} \ll \beta^{-1}$,
where $\tau_0$ is the typical time resolution.
%
While the simple master equation derived in this way
is indeed useful in understanding
the non-unitary and dissipative dynamics of the system,
it is important to clarify
how and to what extent such behaviors are realized 
in the unitary time-evolution of
the whole system, \emph{i.e.}, the system $\mathcal{S}$ and the
environment $\mathcal{E}$.
\section{The real-time path integral with discretized time}
\label{sec:numerical_setup}

In this section, we
first discuss how we discretize
the Caldeira-Leggett model in the real-time path integral formalism
in a way which is useful
in comparing our results
with
the decoherence
predicted by the master equation.
Then we explain how to perform explicit calculations in
the discretized model.

\subsection{Choosing the parameters of the model}
\label{sec:Lagrangian_in_continuum}

The Lagrangian corresponding to the Hamiltonian \eqref{eq:Hamiltonian_CL_1}
is given by
\begin{align}
    L  &=  L_\mathcal{S} + L_\mathcal{E}  + L_\mathrm{int} \ ,
  \\
    L_\mathcal{S}
    &=
    \frac{1}{2}\, \dot{x}^2
    - \frac{1}{2} \,  \omega_\mathrm{b}^2 \, x^2 \ ,
    \quad
    L_\mathcal{E}
    =
    \sum_{k=1}^{N_\mathcal{E}}
    \left\{ \frac{1}{2} \, (\dot{q}^k)^2
    -\frac{1}{2} \, \omega_k^{\;2} \, (q^k)^2  \right\} \ ,
    \quad
    L_\mathrm{int}
    =
    c\ x \sum_{k=1}^{N_\mathcal{E}} \,q^k \ ,
    \label{eq:Lagrangian}
\end{align}
where we denote the coordinate of the $k$-th harmonic oscillator
as $q^k$ with the upper suffix from now on reserving the lower suffix
for the discretized time.
We set the coupling constants $c_k$ between the
system $\mathcal{S}$ and the $k$-th harmonic oscillator in the
environment $\mathcal{E}$
to a constant $c$
since the results in the large $N_\mathcal{E}$ limit
with the spectral density \eqref{def-spectral-density}
depends only on the combination $\rho(\omega) C(\omega)^2$,
which implies that
one can fix $C(\omega)=c$ without loss of generality.\footnote{This is true for
any form of the spectral density, and it does not have to be
the specific form \eqref{eq:Ohmic_spectrum}.}
We have also absorbed the mass parameters $M$ and $m$ in
\eqref{eq:action_CL_original}
by rescaling $x \to x/ \sqrt{M}$, $q^k \to q^k/\sqrt{m}$ and
$c \to c \sqrt{Mm}$.
%



The frequencies $\omega_k$ of the harmonic oscillators
in the environment can be determined
by requiring that the Ohmic spectrum \eqref{eq:Ohmic_spectrum}
is reproduced in the large $N_\mathcal{E}$ limit.
For that, we introduce a function
$\omega=g(\kappa)$
of $\kappa = \frac{k}{N_\mathcal{E}}$, which gives
$\dd \omega = (\dd g / \dd \kappa) \, \dd \kappa$.
Since the distribution of the harmonic oscillators
with respect to $\kappa$ is uniform,
the Ohmic spectrum \eqref{eq:Ohmic_spectrum} is reproduced if
\begin{align}
\left( \frac{\dd g}{\dd \kappa} \right)^{-1} \propto \omega^2 = g(\kappa)^2 \ ,
\end{align}
which implies $g(\kappa) \propto \kappa^{1/3}$.
Thus we obtain
\begin{equation}
    \omega_k  = 
    \omega_\mathrm{cut}
    \left(\frac{k}{N_\mathcal{E}}\right)^{1/3} \ ,
    \label{eq:omega_k_finiteNenv}
\end{equation}
where $\omega_\mathrm{cut}$ is the cutoff parameter
introduced in \eqref{eq:Ohmic_spectrum}.
The spectral density one obtains in the large $N_\mathcal{E}$ limit
is given by
\begin{align}
  \rho(\omega) = \frac{3 N_\mathcal{E}}{\omega_\mathrm{cut}^3} \, \omega^2 \ ,
  \qquad
  \rho(\omega)\, C(\omega)^2
  = \frac{3 c^2 N_\mathcal{E}}{\omega_\mathrm{cut}^3} \, \omega^2 \ ,
  \label{rho-C2-discrete}
\end{align}
where $\omega \le \omega_\mathrm{cut}$.
Comparing \eqref{rho-C2-discrete} with \eqref{eq:Ohmic_spectrum},
one obtains the asymptotic behavior of the coupling constant $c$
at large $N_\mathcal{E}$ as
\begin{align}
  c^2 \sim \frac{4 \,  \omega_\mathrm{cut}^3}{3 \pi N_\mathcal{E}} \, \gamma  \ .
  \label{c-gamma-rel-infN}
\end{align}

In order to determine the coupling constant $c$ at finite $N_\mathcal{E}$,
we reconsider the physical origin of the shift \eqref{eq:omega_ren_large_N}
in the frequency.
For that, 
we complete the square
with respect to $q^k$ in the Lagrangian \eqref{eq:Lagrangian} as
\begin{equation}
    L  = \frac{1}{2} \, \dot{x}^2
    -
    \frac{1}{2}\, \tilde{\omega}_\mathrm{r}^2 x^2
    +
    \sum_{k=1}^{N_\mathcal{E}}
    \qty[
    \frac{1}{2} \,  (\dot{q}^k)^2
    -
    \frac{1}{2} \, \omega_k^2 \qty(q^k - \frac{c}{\omega_k^2}x)^2
    ] \ ,
    \label{eq:Lagrangian_coupled_osc}
\end{equation}
where we have defined
\begin{equation}
    \tilde{\omega}_\mathrm{r}^2 
    =  \omega_\mathrm{b}^2 -  c^2
    \sum_{k=1}^{N_\mathcal{E}} 
    \frac{1}{\omega_k^2} \ .
    \label{eq:omega_ren_discrete}
\end{equation}
Since the harmonic oscillators $q^k$ in the environment
are expected to oscillate around the potential minimum
$cx/\omega_k^2$ when $x$ varies slowly with time,
the frequency $\omega_\mathrm{b}$
of the system $\mathcal{S}$ is shifted to
\eqref{eq:omega_ren_discrete}
due to the environment $\mathcal{E}$ even at finite $N_\mathcal{E}$.
Identifying 
$\omega_\mathrm{r}$ in \eqref{eq:omega_ren_large_N} with
$\tilde{\omega}_\mathrm{r}$ in \eqref{eq:omega_ren_discrete}
and using the frequency spectrum \eqref{eq:omega_k_finiteNenv},
we obtain the relationship between the coupling constant $c$
at finite $N_\mathcal{E}$ 
and the effective coupling $\gamma$ as
\begin{equation}
    c^2     = 
    \frac{4\gamma}{\pi} \omega_\cut^3 
    \left\{
    \sum_{k=1}^{N_\mathcal{E}} \qty(\frac{N_\mathcal{E}}{k})^{2/3}
    \right\}^{-1} \ .
    \label{eq:coupling_scaling}
\end{equation}
This is indeed consistent with the asymptotic behavior
\eqref{c-gamma-rel-infN} at large $N_\mathcal{E}$.
\subsection{The effective action for the Caldeira-Leggett model}

In order to put the system on a computer,
we discretize the time $\tau$ as
$\tau_n = n \, \epsilon$ $(n=0,\cdots, N_t)$
and
$t \equiv \tau_{N_t}$.
Accordingly the variables
$x(\tau)$ and $q_k(\tau)$ are also discretized
as $x_n = x(\tau_n)$ and $q^k_n =q_k(\tau_n)$.
The action with the discretized time can be written as
\begin{align}
S(x,q)  &= S_\mathcal{S}(x) + S_\mathcal{E}(q) + S_\mathrm{int}(x,q) \ , \\
    S_\mathcal{S}(x)
    &= 
    \frac{1}{2} \, \epsilon
    \sum_{n=0}^{N_t-1}
    \qty[
    \qty(\frac{x_n-x_{n+1}}{\epsilon})^2
    -
    \omega_\mathrm{b}^2  \frac{x_n^2+x_{n+1}^2}{2}
    ] \ ,
    \\
    S_\mathcal{E}(q)
    &= 
    \frac{1}{2} \, \epsilon
    \sum_{k=1}^{N_\mathcal{E}}
    \sum_{n=0}^{N_t-1}
    \qty[
    \qty(\frac{q_n^k-q_{n+1}^k}{\epsilon})^2
    -
    \omega_k^2
    \frac{(q_n^k)^2+(q_{n+1}^k)^2}{2}
    ] \ ,
    \\
    S_\mathrm{int}(x,q)
    &=
        c \, \epsilon   \sum_{k=1}^{N_\mathcal{E}}
    \sum_{n=0}^{N_t-1}  \frac{x_n q_n^k+x_{n+1} q_{n+1}^k}{2}  \ .
    \end{align}
\begin{figure}[t]
	\centering
	\includegraphics[width=0.65\textwidth]{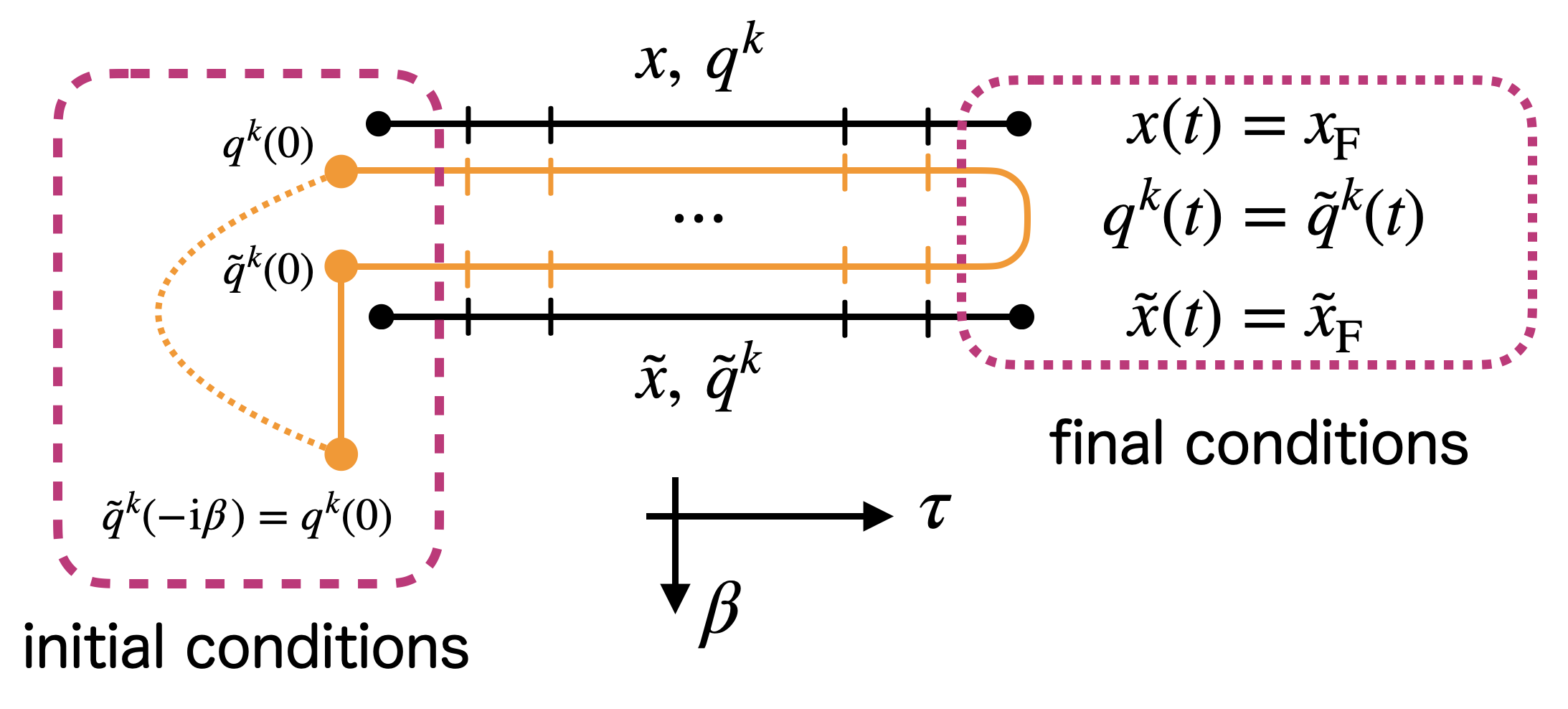}
	\caption{
            Schematic picture of the reduced density matrix of the system in the position space~\eqref{eq:rho_S_position_basis}.
            The boundary conditions are imposed at the initial time and the final time.
	}
        \label{fig:lattice-reduced-density-matrix}
\end{figure}

The reduced density matrix of the system $\mathcal{S}$ in the position basis
can be given by
\begin{equation}
    \rho_\mathcal{S}(x_\mathrm{F},\tilde{x}_\mathrm{F};t)
    = \int\mathcal{D}x\mathcal{D}\tilde{x}
    \left( \prod_{k=1}^{N_\mathcal{E}} \mathcal{D}q^k\mathcal{D}\tilde{q}^k \right)
    \rho_\mathcal{S}(x_0,\tilde{x}_0;0)
    \left( \prod_{k=1}^{N_\mathcal{E}}  \rho_\mathcal{E}^{(k)}(q^k_0,\tilde{q}^k_0,\beta) \right)\, 
    \mathrm{e}^{\mathrm{i} (S(x,q)-S(\tilde{x},\tilde{q}))} \ ,
    \label{eq:rho_S_position_basis}
\end{equation}
where the path integral measure is understood as the ordinary
integration measure for discretized variables.
At the final time, we impose
$x_{N_t} = x_\mathrm{F}$, $\tilde{x}_{N_t} = \tilde{x}_\mathrm{F}$
corresponding to the element of the reduced density matrix to be calculated,
and $q_{N_t}^k =\tilde{q}_{N_t}^k$
corresponding to taking the trace with respect to the environment $\mathcal{E}$ as
in \eqref{eq:reduced_denisty_matrix_primary}.

As the initial condition
for the system $\mathcal{S}$, we consider a Gaussian wave
packet\footnote{The following discussions can be
easily generalized to the case with the initial wave function
$ \psi _\mathrm{I} (x)  =  \exp(-\frac{1}{4 \sigma^2} (x-x_0)^2 + \ii p x)$
and to the case with two wave packets
as we discuss in Section \ref{sec:numerical_analysis}.}
\begin{align}
    \rho_\mathcal{S}(x,\tilde{x};0) 
    &=
    \psi_\mathrm{I}
    (x)\psi^* _\mathrm{I}
    (\tilde{x}) \ ,
    \label{init-rho-system-gen}
    \\
    \psi
    _\mathrm{I}
    (x)  &= 
    \exp(-\frac{1}{4 \sigma^2} x^2) \ .
    \label{init-rho-system-gaussian}
\end{align}
The initial condition of the environment $\mathcal{E}$
is given by the canonical ensemble \eqref{eq:Equilib_env_1}
with temperature $T = \beta^{-1}$.
Here we introduce an additional path for the variables $\tilde{q}^k$ in
the imaginary time direction as depicted
in Fig.~\ref{fig:lattice-reduced-density-matrix}
with
the free Euclidean action\footnote{
This is actually not needed in the present case,
where the initial density matrix is given explicitly by \eqref{eq:Equilib_env_1}.
Here we choose to do this in order to demonstrate that our formalism is useful also
in more general cases, where the initial density matrix is not given explicitly.}
\begin{equation}
    S_0(\tilde{q}) = 
\frac{1}{2} \, \tilde{\epsilon}
    \sum_{k=1}^{N_\mathcal{E}}
    \sum_{j=0}^{N_\beta-1}
    \qty[
        \qty(\frac{\tilde{q}^k_0(j+1)-\tilde{q}^k_0(j)}{\tilde{\epsilon}})^2
        +
       \omega_k^2
          \frac{   \tilde{q}^k_0(j+1)^2 + \tilde{q}^k_0(j)^2 }{2} ] \ ,
\end{equation}
where we define $\tilde{q}^k_0(j) \equiv \tilde{q}^k(-\ii(j\tilde{\epsilon}))$
and impose $\tilde{q}_0^k =\tilde{q}_0^k(0)$ and
$q_0^k =\tilde{q}_0^k(N_\beta)$
with the inverse temperature represented as
$\beta = N_\beta \, \tilde{\epsilon}$.





To summarize, the reduced density matrix \eqref{eq:rho_S_position_basis}
of the system $\mathcal{S}$
can be written as
\begin{equation}
    \rho_\mathcal{S}(x_\mathrm{F},\tilde{x}_\mathrm{F};t)
    = \int\mathcal{D}x\mathcal{D}\tilde{x}
    \left( \prod_{k=1}^{N_\mathcal{E}} \mathcal{D}q^k\mathcal{D}\tilde{q}^k \mathcal{D}\tilde{q}_0^k \right)
    \, \mathrm{e}^{-S_{\rm eff}(x,\tilde{x},q,\tilde{q},\tilde{q}_0)}
    \ ,
    \label{eq:rho_S_position_basis-summary}
\end{equation}
where the effective action is given by
\begin{align}
  S_{\rm eff} (x,\tilde{x},q,\tilde{q},\tilde{q}_0)
  &= - \mathrm{i} \left\{ S(x, q) - S(\tilde{x}, \tilde{q}) \right\}
   + S_0(\tilde{q}_0) + \frac{1}{4 \sigma^2} (x_0 ^2 + \tilde{x}_0 ^2)  \ .
  \label{def-eff-action}
    \end{align}
Note that the first two terms are purely imaginary,
whereas the last two terms are real.
Since the integrand of \eqref{eq:rho_S_position_basis-summary}
is complex, the sign problem occurs
when one applies Monte Carlo methods naively.

\subsection{Performing the path integral by the Gaussian integral}
\label{sec:path-integral_on_thimble}

In the present case, the effective action \eqref{def-eff-action} is
quadratic with respect to
the integration variables, and it can be written as
\begin{align}
  S_{\rm eff} (x,\tilde{x},q,\tilde{q},\tilde{q}_0)
    = 
    \frac{1}{2}&X_\mu \mathcal{M}_{\mu\nu}X_\nu - C_\mu X_\mu  + B \ ,
    \label{eq:primitive_lattice_action}
\end{align}
where $X_\mu$ ($\mu = 1 , \cdots , D$) represents the integration variables collectively
and the number of integration variables is
$D = 2N_t (1 + N_\mathcal{E})  + N_\beta N_\mathcal{E}$.
Note that $\mathcal{M}$ is a $D\times D$  complex symmetric matrix,
which is independent of $x_\mathrm{F}$ and $\tilde{x}_\mathrm{F}$, whereas
$C_\mu$ and $B$ are defined by
\begin{align}
C_\mu X_\mu  &= 
   -  \frac{\mathrm{i}}{\epsilon}
    \qty(x_\mathrm{F}x_{N_t-1} - \tilde{x}_\mathrm{F}\tilde{x}_{N_t-1})
    +   \frac{\mathrm{i}}{2} c \, \epsilon \sum_k
    \qty(x_\mathrm{F} - \tilde{x}_\mathrm{F}) q^k_{N_t} \ , 
    \notag \\
    B &=   -  \frac{\mathrm{i}}{2} \, b  \, (x_\mathrm{F}^2 - \tilde{x}_\mathrm{F}^2 )  \ ,
    \quad \quad
    \mbox{~where~~~}
    b =    \frac{1}{\epsilon}-\frac{\omega_\mathrm{b}^2\epsilon}{2}   \ .
    \label{eq:def-C-B}
\end{align}
Since $C_\mu$ is linear in $x_\mathrm{F}$ and $\tilde{x}_\mathrm{F}$,
let us write them as
\begin{align}
  C_\mu  &=  \mathrm{i} (c_\mu x_\mathrm{F} - \tilde{c}_\mu \tilde{x}_\mathrm{F} ) \ .
    \label{eq:def-c-mu}
\end{align}


The saddle point of this effective action
is given by
\begin{equation}
    \bar{X}_\mu = \mathcal{M}^{-1}_{\mu\nu} \, C_\nu \ ,
    \label{eq:saddle-pt_eq}
\end{equation}
and redefining the integration variables as $Y_\mu = X_\mu - \bar{X}_\mu$,
the effective action becomes
\begin{equation}
  S_{\rm eff} (x,\tilde{x},q,\tilde{q},\tilde{q}_0)
  =   \frac{1}{2}Y_\mu \mathcal{M}_{\mu\nu}Y_\nu
 +  \left(  B - \frac{1}{2} \bar{X}_\mu \mathcal{M}_{\mu\nu} \bar{X}_\nu   \right)  \ .
\end{equation}
Integrating out $Y_\mu$, we obtain
\begin{align}
  \rho_\mathcal{S}(x_\mathrm{F}, \tilde{x}_\mathrm{F};t)
  &= \frac{1}{\sqrt{\det \mathcal{M}}} \ee^{-\mathcal{A}} \ ,
\label{eq:rho-detM}
  \\
  \mathcal{A}
  &=  B - \frac{1}{2} \, \bar{X}_\mu \mathcal{M}_{\mu\nu} \bar{X}_\nu
  \nn  \\
  &=  B - \frac{1}{2} \,  C_\mu  \left( \mathcal{M}^{-1} \right)_{\mu\nu}
  C_\nu  \nn \\
  &=  \frac{1}{2}
  \begin{pmatrix}
x_\mathrm{F}  & \tilde{x}_\mathrm{F} 
\end{pmatrix}
\begin{pmatrix}
-\mathrm{i}  b + c_\mu (\mathcal{M}^{-1})_{\mu\nu} c_{\nu}
  &  - c_\mu (\mathcal{M}^{-1})_{\mu\nu} \tilde{c}_{\nu} \\
 - \tilde{c}_\mu (\mathcal{M}^{-1})_{\mu\nu} c_{\nu}
  & \mathrm{i} b + \tilde{c}_\mu (\mathcal{M}^{-1})_{\mu\nu} \tilde{c}_{\nu}  \\
\end{pmatrix}
\begin{pmatrix}
x_\mathrm{F}  \\ \tilde{x}_\mathrm{F} 
\end{pmatrix}
 \ .
\end{align}
%
%
Let us consider the magnitude
$| \rho_\mathcal{S}(x_\mathrm{F}, \tilde{x}_\mathrm{F};t) |$,
which is determined by
\begin{align}
 {\rm Re } \mathcal{A}
  &=  \frac{1}{2}
  \begin{pmatrix}
x_\mathrm{F}  & \tilde{x}_\mathrm{F} 
\end{pmatrix}
  \begin{pmatrix}
    J   &   -K  \\
  -K   &  J   \\
\end{pmatrix}
\begin{pmatrix}
x_\mathrm{F}  \\ \tilde{x}_\mathrm{F} 
\end{pmatrix}
\nn \\
&=  \frac{1}{4} \,
\{ (J-K)( x_\mathrm{F}  +  \tilde{x}_\mathrm{F}  )^2
+ (J+K)( x_\mathrm{F}  -  \tilde{x}_\mathrm{F}  )^2  \} \ ,
\end{align}
where we have defined
\begin{align}
  J =  {\rm Re} \{   c_\mu (\mathcal{M}^{-1})_{\mu\nu} c_{\nu} \} 
= {\rm Re} \{ \tilde{c}_\mu (\mathcal{M}^{-1})_{\mu\nu} \tilde{c}_{\nu}  \}  \  , \\
K = {\rm Re} \{ c_\mu (\mathcal{M}^{-1})_{\mu\nu} \tilde{c}_{\nu}\} =
  {\rm Re} \{ \tilde{c}_\mu (\mathcal{M}^{-1})_{\mu\nu} c_{\nu} \} \ .
    \label{eq:def-J-K}
\end{align}

In order to investigate the decoherence, we define the quantities
\begin{align}
  \Gamma_\mathrm{diag}(t) &= 2 (J-K) \ , \\
  \Gamma_\mathrm{off\mathchar`-diag}(t) &= 2 (J+K) \ ,
    \label{eq:def-Gamma}
\end{align}
which characterize the fall-off of the magnitude of the matrix element
in the diagonal and off-diagonal directions, respectively, as 
\begin{align}
  | \rho_\mathcal{S}(x_\mathrm{F}, \tilde{x}_\mathrm{F};t) |
  &\simeq
  \exp \left\{ - \frac{1}{2}
  \Gamma_\mathrm{diag}(t)
 \left( \frac{x_\mathrm{F}  +  \tilde{x}_\mathrm{F}}{2}  \right)^2
 - \frac{1}{2}  \Gamma_\mathrm{off\mathchar`-diag}(t)
 \left( \frac{x_\mathrm{F}  -  \tilde{x}_\mathrm{F}}{2}  \right)^2
 \right\} \ ,
    \label{eq:rho_S_gaussian}
\end{align}
omitting the prefactor
independent of $x_\mathrm{F}$ and $\tilde{x}_\mathrm{F}$.
From \eqref{eq:rho_S_time-evol_highT}, we expect the behavior\footnote{This equation
corresponds to Eq.~(29) in the letter version \cite{Nishimura:2024one}
of this work, where we had $8$ instead of $16$ by mistake. This affects the
prediction from the master equation by a factor of $2$
in subsequent discussions in Ref.~\cite{Nishimura:2024one}
although the main conclusion of that paper remains the same.
This correction is crucial in the agreement with the prediction
observed in Section \ref{sec:cutoff_dep}.}
\begin{equation}
  \Gamma_\mathrm{off\mathchar`-diag}(t) \sim
  \frac{16\gamma}{\beta}t \ ,
    \label{eq:decoherence_from_Gamma_offdiag}
\end{equation}
which enables us to probe the decoherence at high temperature.

\section{Numerical results for a single wave packet}
\label{sec:main_results_case1}

In this section we consider the case in which the initial wave function of
the system $\mathcal{S}$ is assumed
to be
the ground state of the harmonic oscillator with the renormalized frequency
$\omega_\mathrm{r}$, which corresponds
to \eqref{init-rho-system-gaussian} with $\sigma^2 = \frac{1}{2 \omega_\mathrm{r}^2}$.
Throughout this section,
we set\footnote{The bare frequency $\omega_\mathrm{b}$
is determined
by \eqref{eq:omega_ren_large_N},
whereas the coupling constant $c$ is determined by \eqref{eq:coupling_scaling}.}
$\omega_\mathrm{r} =0.08$.
The lattice spacing in the time direction
is chosen to be $\epsilon=0.05$, whereas 
the lattice spacing in the temperature direction
is chosen to be
$\tilde{\epsilon}=0.05$ for $\beta\ge 0.2$,
and $\tilde{\epsilon}=\beta/4$ for $\beta\le 0.2$.
In Appendix \ref{sec:limits_suppl},
we present some evidence that changing the lattice spacing does not alter our main conclusion.

\begin{figure}[t]
	\centering
	\scalebox{0.23}{\includegraphics{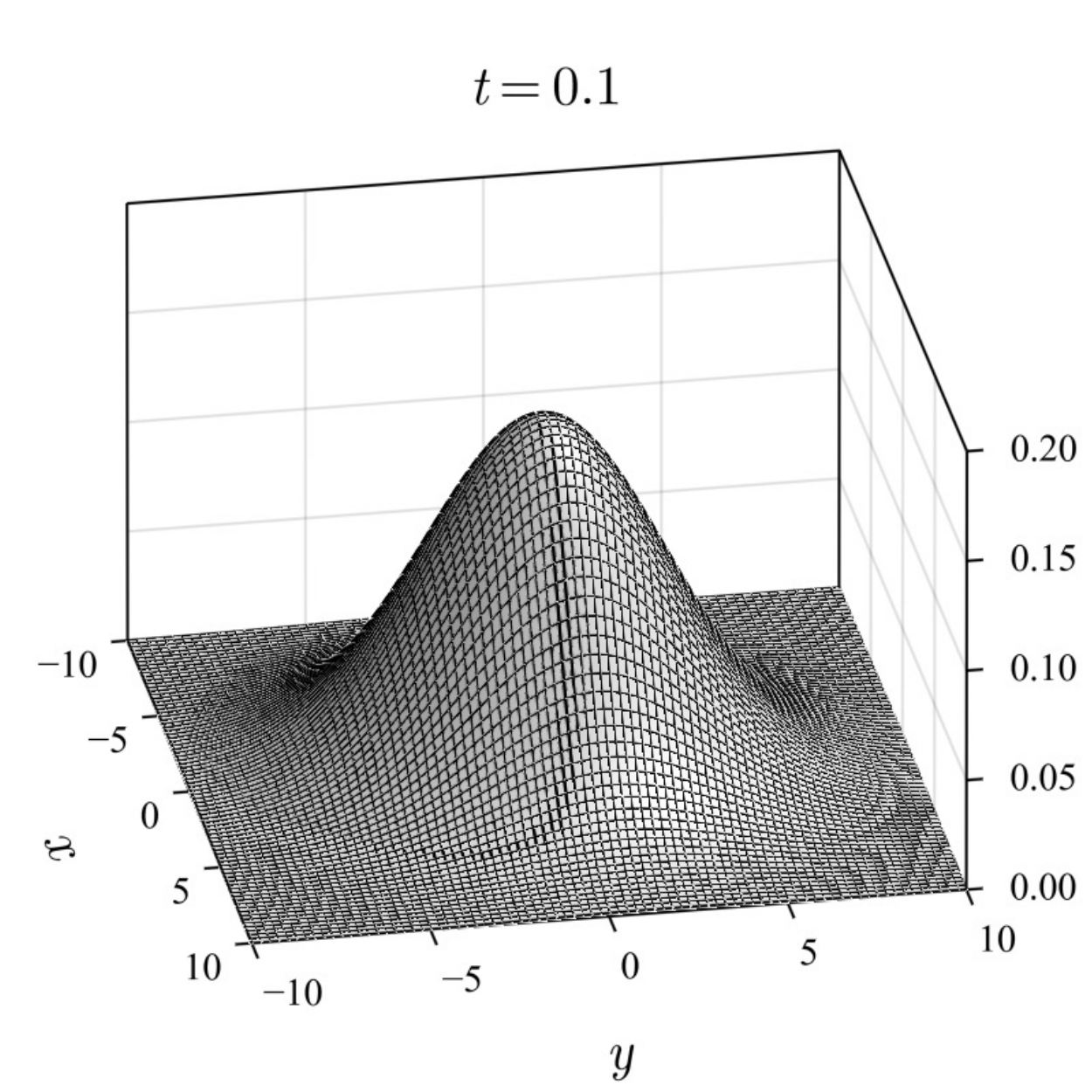}}
	\scalebox{0.23}{\includegraphics{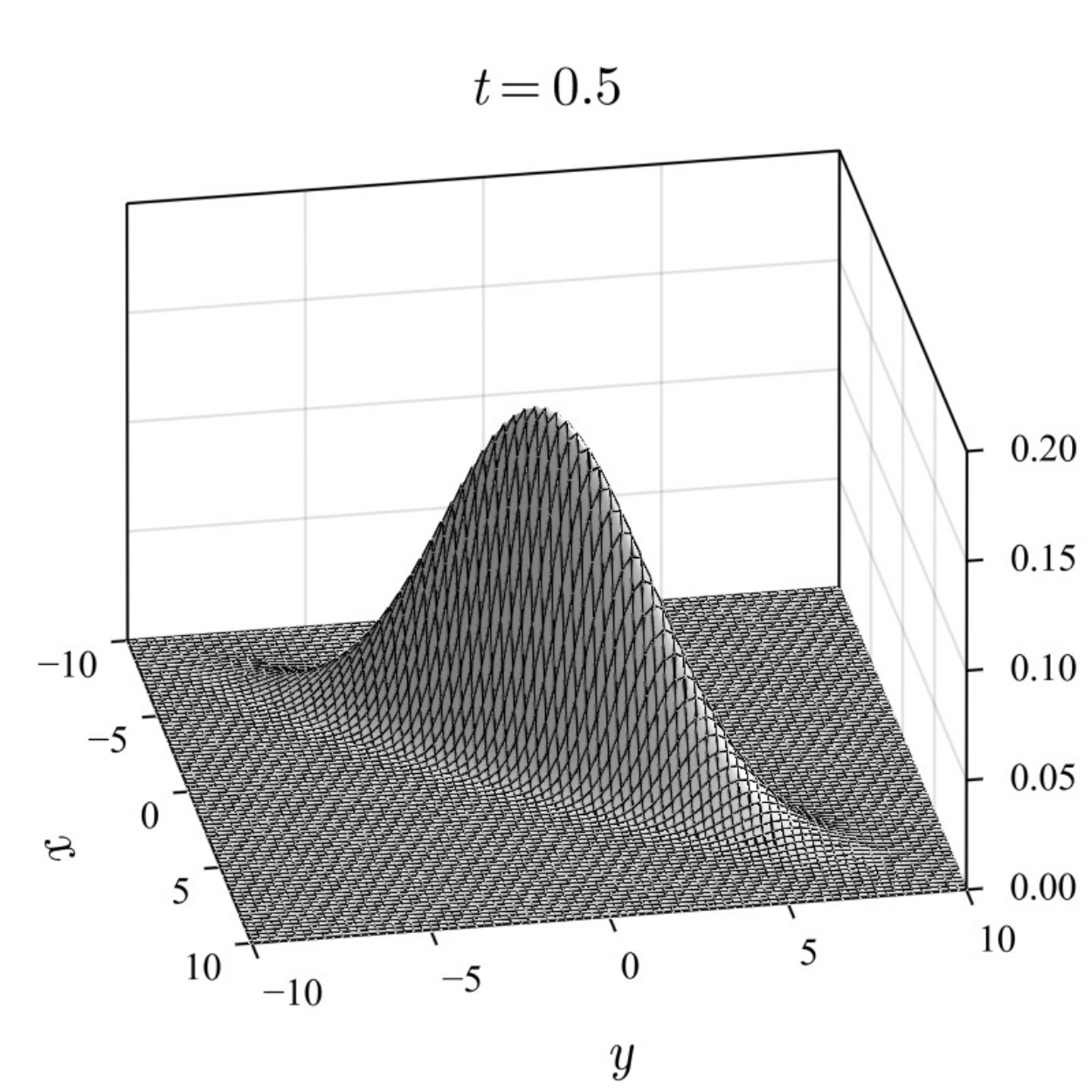}}
	\scalebox{0.23}{\includegraphics{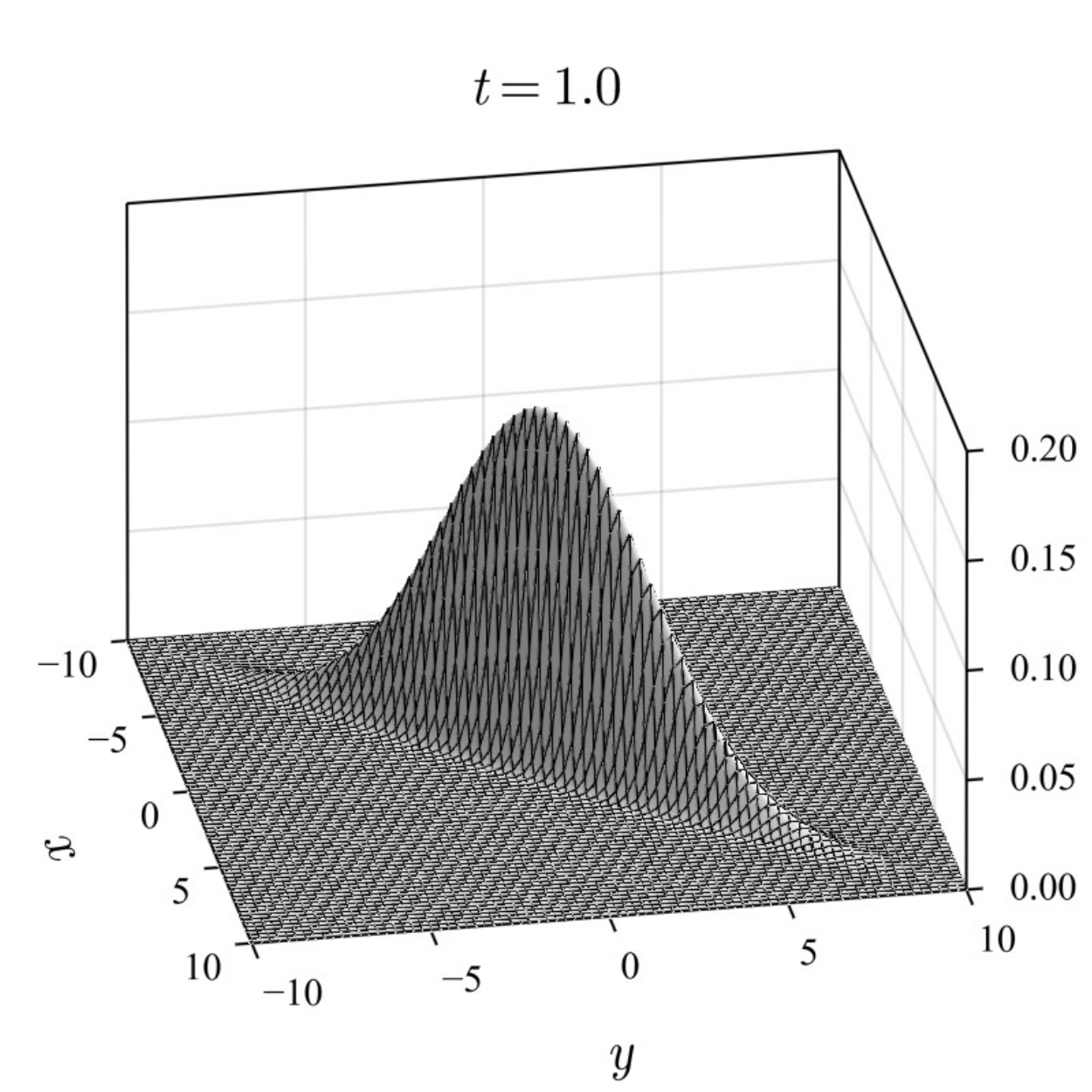}}
	\scalebox{0.23}{\includegraphics{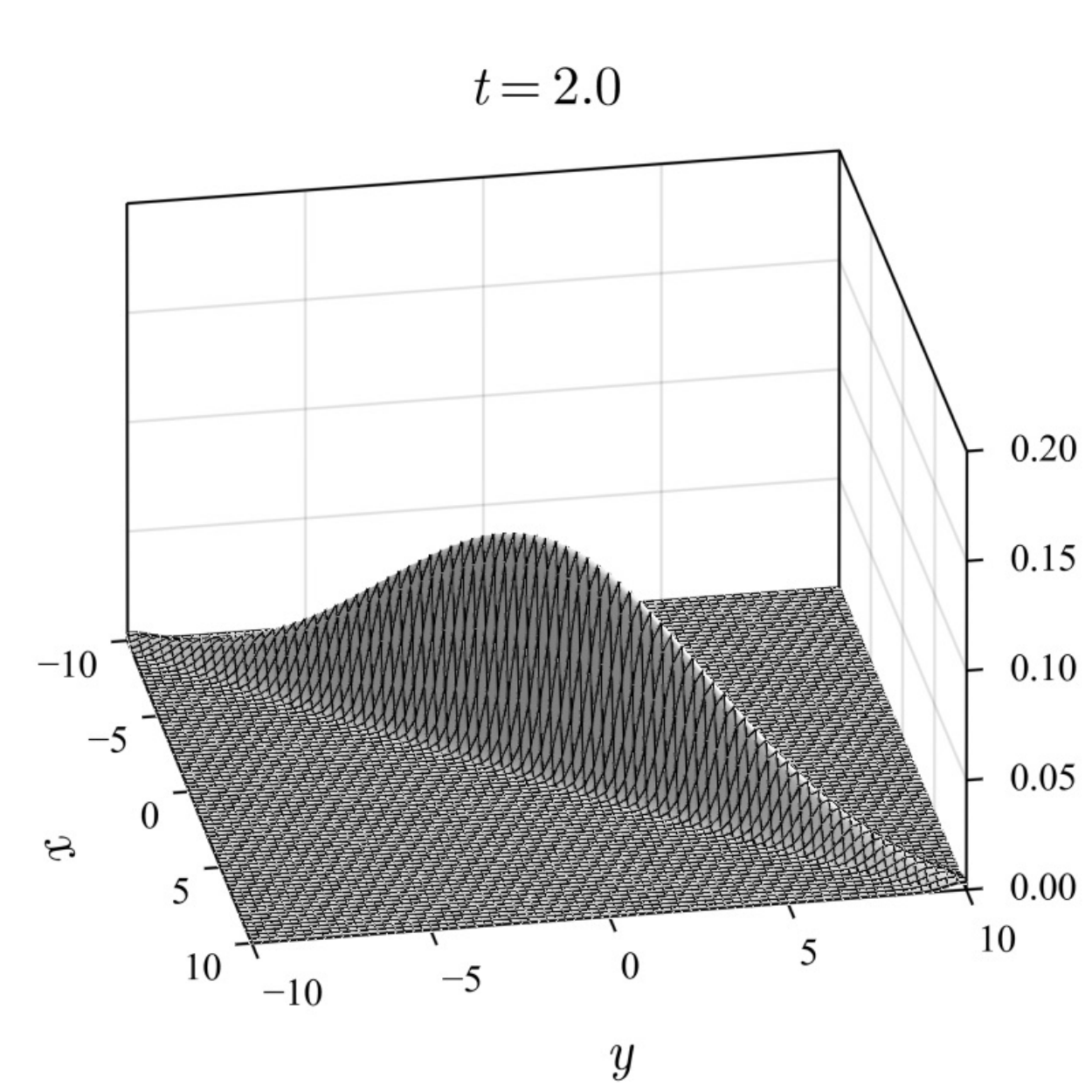}}
	\scalebox{0.23}{\includegraphics{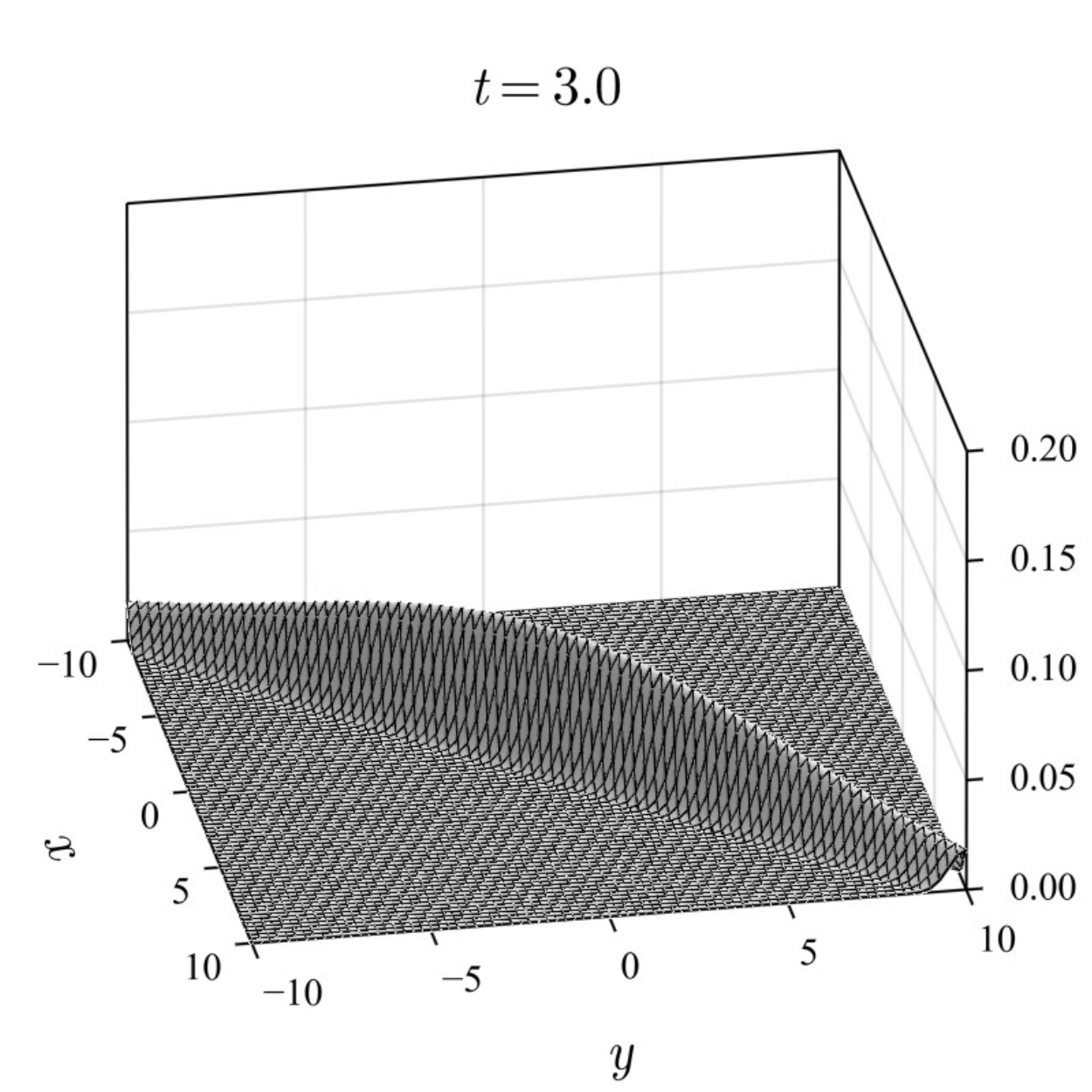}}
	\caption{The reduced density matrix $  | \rho_\mathcal{S}(x  , y ;t) |$  
          is plotted in the $xy$ plane for $t=0.1, 0.5, 1.0, 2.0, 3.0$
          with $N_\mathcal{E} = 64$, $\beta = 0.05$, $\gamma = 0.1$ and $\omega_\cut= 2$.}
        \label{fig:Gaussian_width_Nenv=64}
\end{figure}

In Fig.~\ref{fig:Gaussian_width_Nenv=64}, we plot
the behavior of the reduced density matrix~\eqref{eq:rho_S_gaussian}
for $N_\mathcal{E}=64$, $\beta = 0.05$, $\gamma = 0.1$ and $\omega_\cut= 2$,
which satisfy
the hierarchy
$\omega_\mathrm{r} \ll  \omega_\cut  \ll  \beta^{-1} = T$
required in
deriving the master equation.
Here the reduced density matrix $\rho_\mathcal{S}(x,y;t)$ is correctly
normalized by the factor $\sqrt{ 2\pi / \Gamma_\mathrm{diag}(t)}$.
At $t=0$,
the Gaussian distribution is symmetric since
$\Gamma_\mathrm{diag}(0) =
\Gamma_\mathrm{off\mathchar`-diag}(0)=2\, \omega_\mathrm{r}$.
Up to $t \simeq 1$, the diagonal elements do not change much
while the off-diagonal elements
decrease with $t$,
which clearly indicates the effect of decoherence.
In order to investigate it
more quantitatively,
we discuss the time evolution
of $\Gamma_{\mathrm{diag}}(t)$
and $\Gamma_{\mathrm{off\mathchar`-diag}}(t)$
for various $N_\mathcal{E}$, $\beta$, $\gamma$ and $\omega_\cut$ in what follows.



\subsection{Increasing the number of harmonic oscillators}
\label{sec:N_env-dependence}

\begin{figure}[p]
	\centering
	\includegraphics[width=\textwidth]{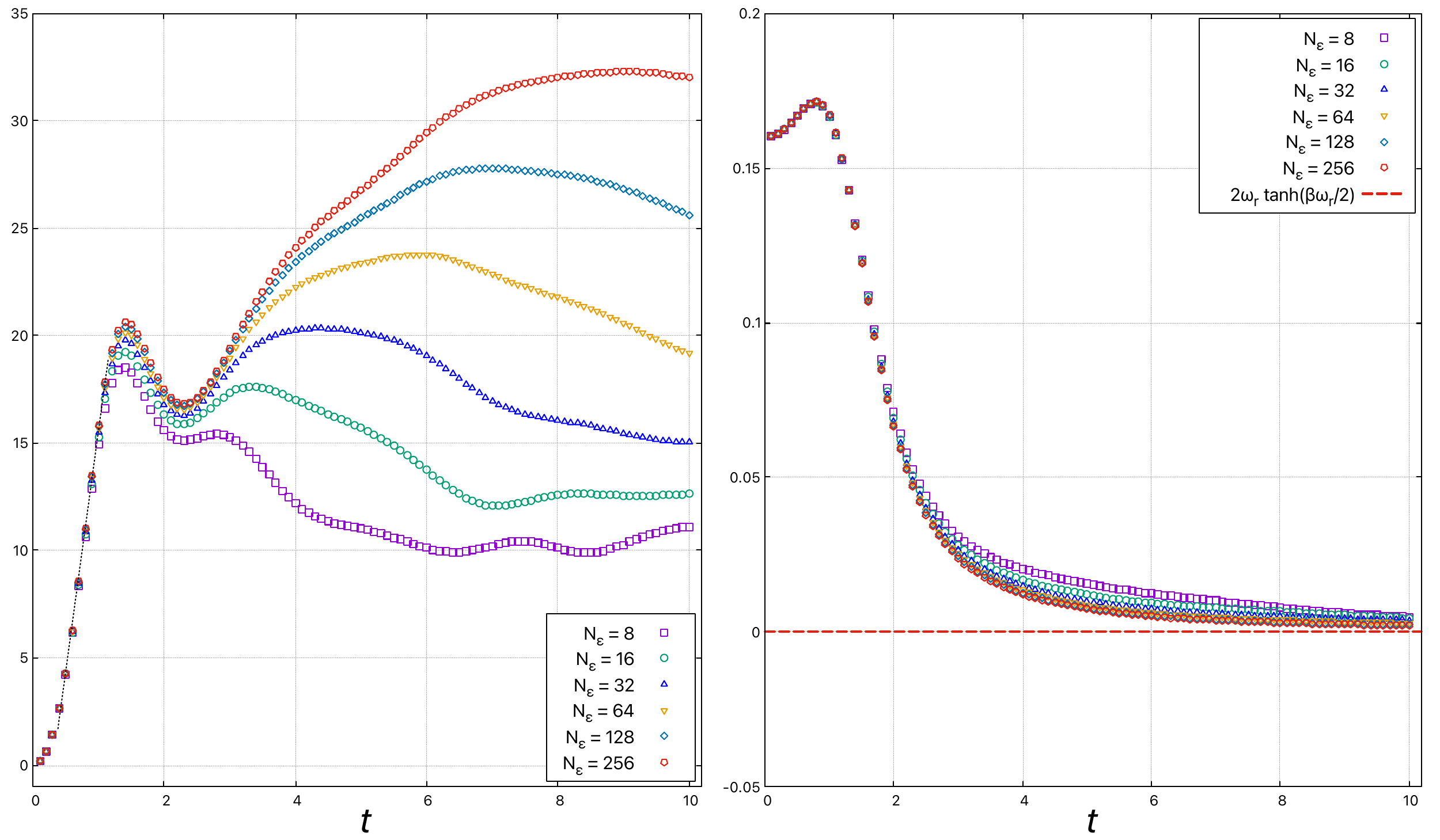}
	\includegraphics[width=0.6\textwidth]{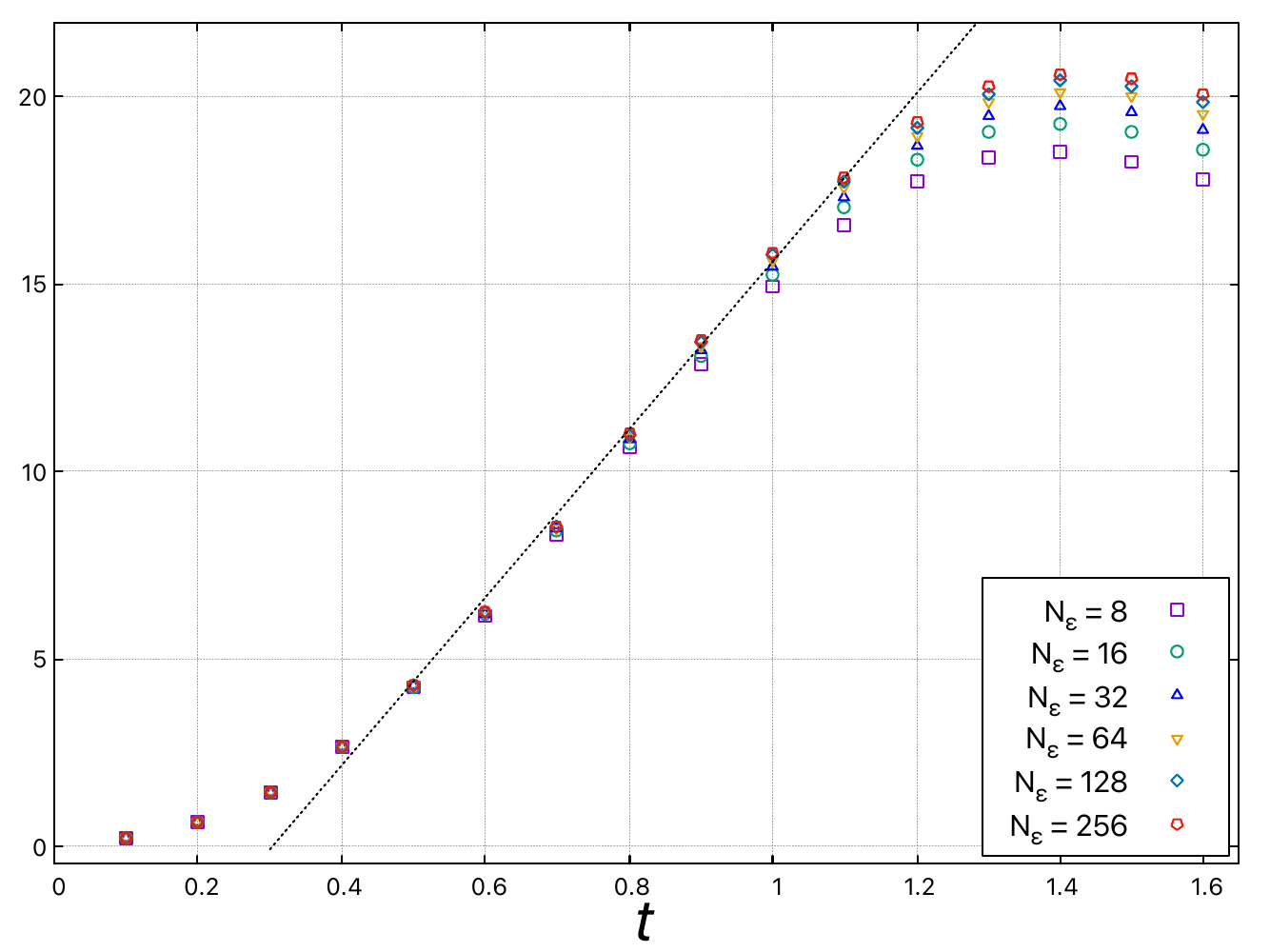}
	\caption{The quantities
          $\Gamma_\mathrm{off\mathchar`-diag}(t)$ (Top-Left) and
          $\Gamma_\mathrm{diag}(t)$ (Top-Right)
          are plotted against $t$
for $N_\mathcal{E} = 8, 16, \cdots , 256$ 
with $\beta = 0.05$, $\gamma = 0.1$ and $\omega_\cut= 2$. 
In the Top-Left panel, the dotted line represents a fit
of the $N_\mathcal{E} =256$ data within the region $0.4\le t \le 1.1$
to the behavior $A \frac{16\gamma}{\beta} t + B$.
In the Top-Right panel, the dashed red line represents
$C_{\rm th} = 2\omega_\mathrm{r}\tanh\frac{\beta\omega_\mathrm{r}}{2}=0.00032$
predicted by thermalization at the inverse temperature $\beta=0.05$,
which is the value expected to be approached at $t\rightarrow \infty$.
(Bottom) A zoom-in
of the Top-Left panel, where
we present $\Gamma_{\mathrm{off\mathchar`-diag}}(t)$ for $t \le 1.6$.
	}
        \label{fig:Nenv_dependency}
\end{figure}

Let us first present our results
for increasing
$N_\mathcal{E}$,
the number of harmonic oscillators
in the environment $\mathcal{E}$.
In Fig.~\ref{fig:Nenv_dependency}
we plot
$\Gamma_{\mathrm{diag}}(t)$
and $\Gamma_{\mathrm{off\mathchar`-diag}}(t)$
for various $N_\mathcal{E}$
with $\beta = 0.05$, $\gamma = 0.1$ and $\omega_\cut= 2$.
We see a clear converging behavior to $N_\mathcal{E}=\infty$
for $t\lesssim 3$.

\begin{figure}[H]
	\centering
	\scalebox{0.23}{\includegraphics{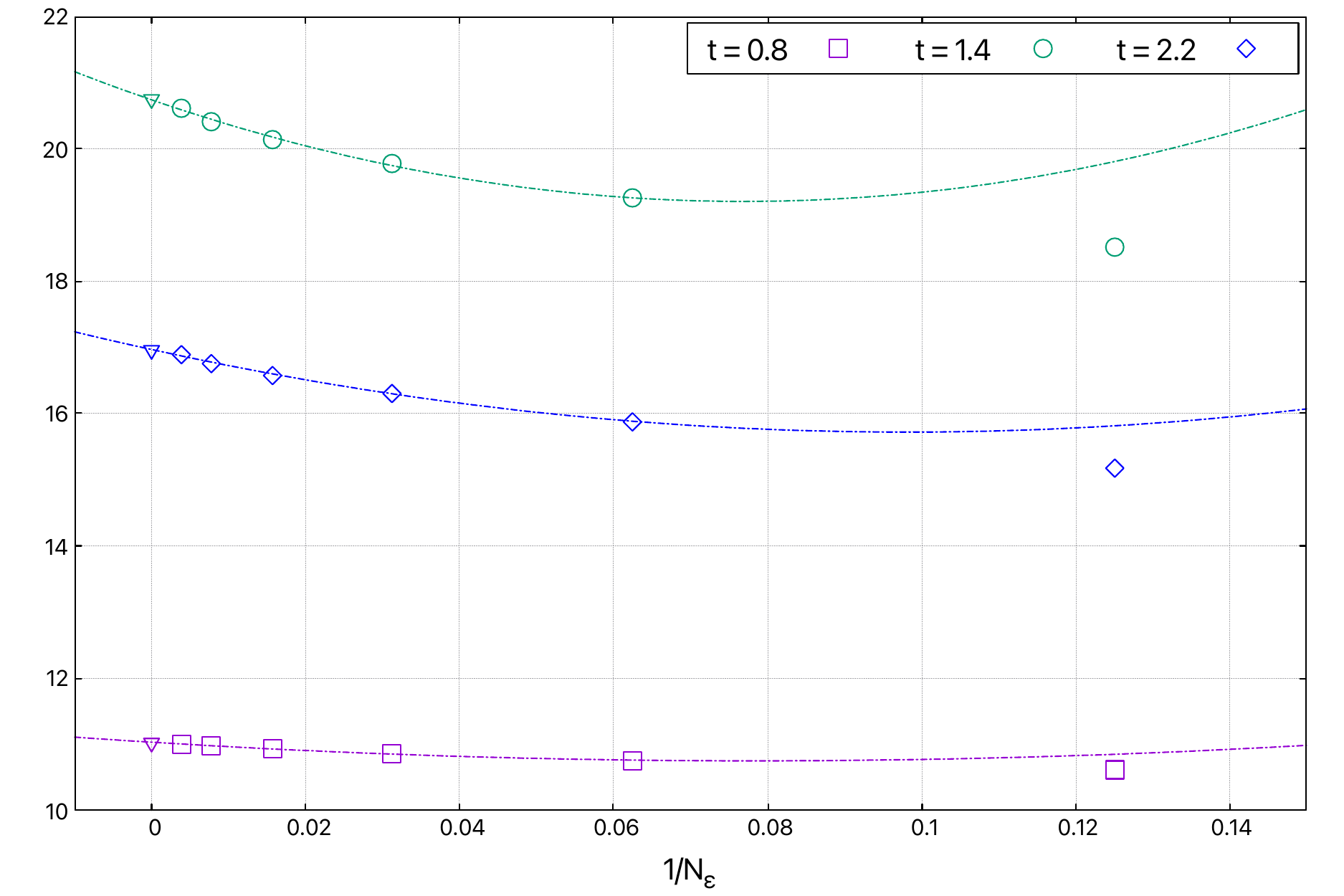}}
 	\scalebox{0.23}{\includegraphics{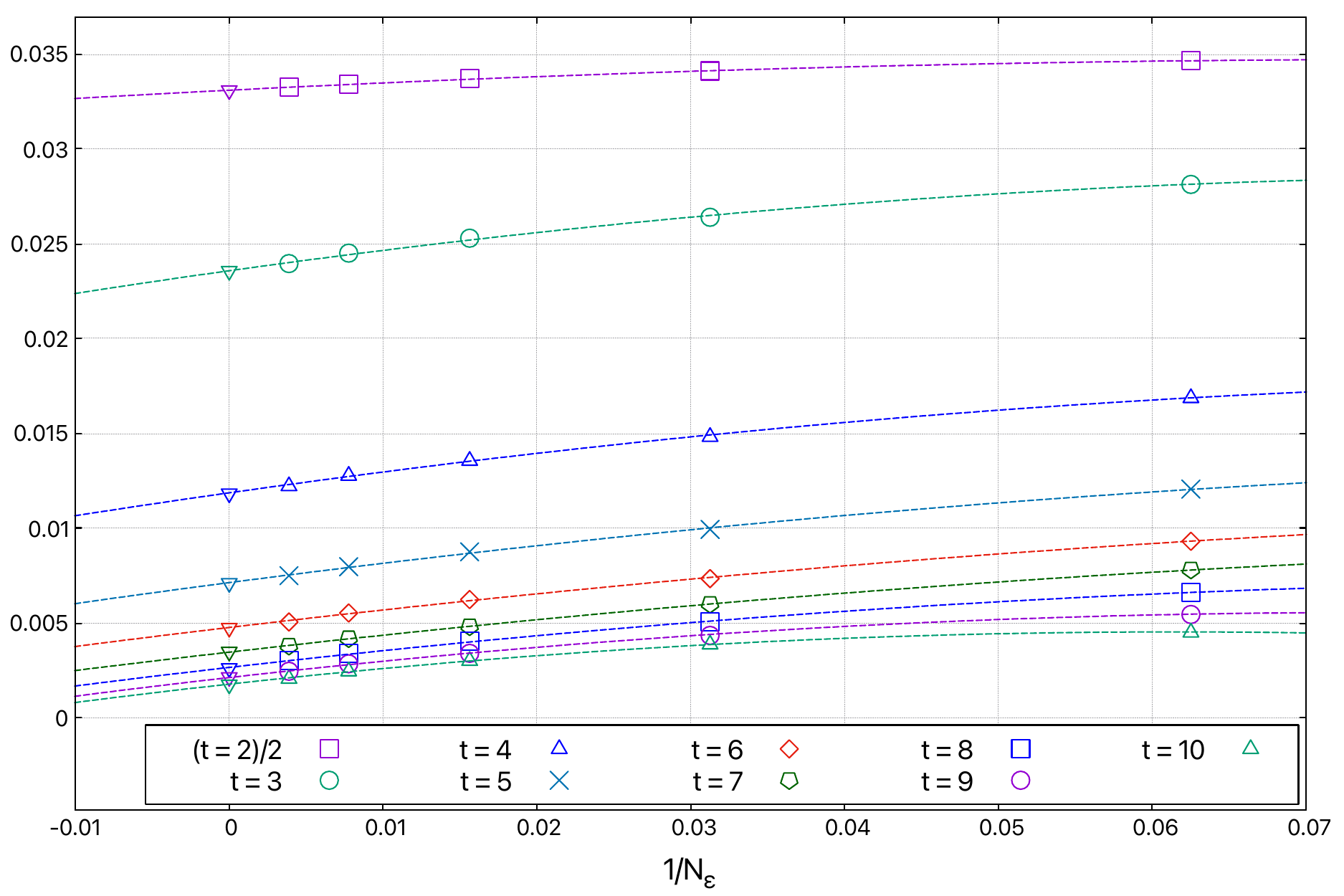}}
	\caption{(Left) The quantity $\Gamma_{\mathrm{off\mathchar`-diag}}(t)$ is plotted
          against $1/N_\mathcal{E}$ at $t = 0.8$, $1.4$, $2.2$.
          (Right) The quantity $\Gamma_{\mathrm{diag}}(t)$ is plotted against $1/N_\mathcal{E}$
          at $t = 2$, $3,  \cdots, 10$.
          The results for $t=2$ are divided by 2 for the sake of visualization.
          In both panels, the lines represent quadratic fits to our data
          at large $N_\mathcal{E}$ and
          the inverted triangles represent the extrapolated values
          at $N_\mathcal{E}=\infty$.}
        \label{fig:Nenv_dependency_extrap}
\end{figure}

In Fig.~\ref{fig:Nenv_dependency_extrap}
we plot $\Gamma_\mathrm{off\mathchar`-diag}(t)$ (Left)
and $\Gamma_\mathrm{diag}(t)$ (Right)
against $1/N_\mathcal{E}$ for various $t$.
We find that our data for the chosen values of $t$
can be nicely fitted to quadratic functions
at sufficiently large $N_\mathcal{E}$, which enables us to
extrapolate our data to $N_\mathcal{E}=\infty$
as represented by the inverted triangles.
This confirms the validity of our choice \eqref{eq:coupling_scaling}
of the coupling constant $c$ for finite $N_\mathcal{E}$ .


The linear growth of $\Gamma_{\mathrm{off\mathchar`-diag}}(t)$
appears
after $t = 0.2$, which is consistent with
the typical time scale of the decoherence
$\tau_\mathrm{d} = \frac{\beta}{2\gamma} = 0.25$,
and continues until $t \sim 1$, where 
$\Gamma_{\mathrm{diag}}(t)$ start to decrease rapidly,
indicating that the effects of the environment other than decoherence
are coming into play.
The dotted line in Fig.~\ref{fig:Nenv_dependency} (Top-Left) and (Bottom)
represents a fit of our data in the region $0.4\le t \le 1.1$ 
to the behavior $A \frac{16 \gamma}{\beta} t + B$
yielding $A\sim 0.69$, which is smaller than
the predicted value $1$.
In fact, the fitted value of $A$ increases toward the predicted value
with increasing $\omega_\mathrm{cut}$ as we will see in Section \ref{sec:cutoff_dep}.

\begin{figure}[H]
	\centering
	\scalebox{0.33}{\includegraphics{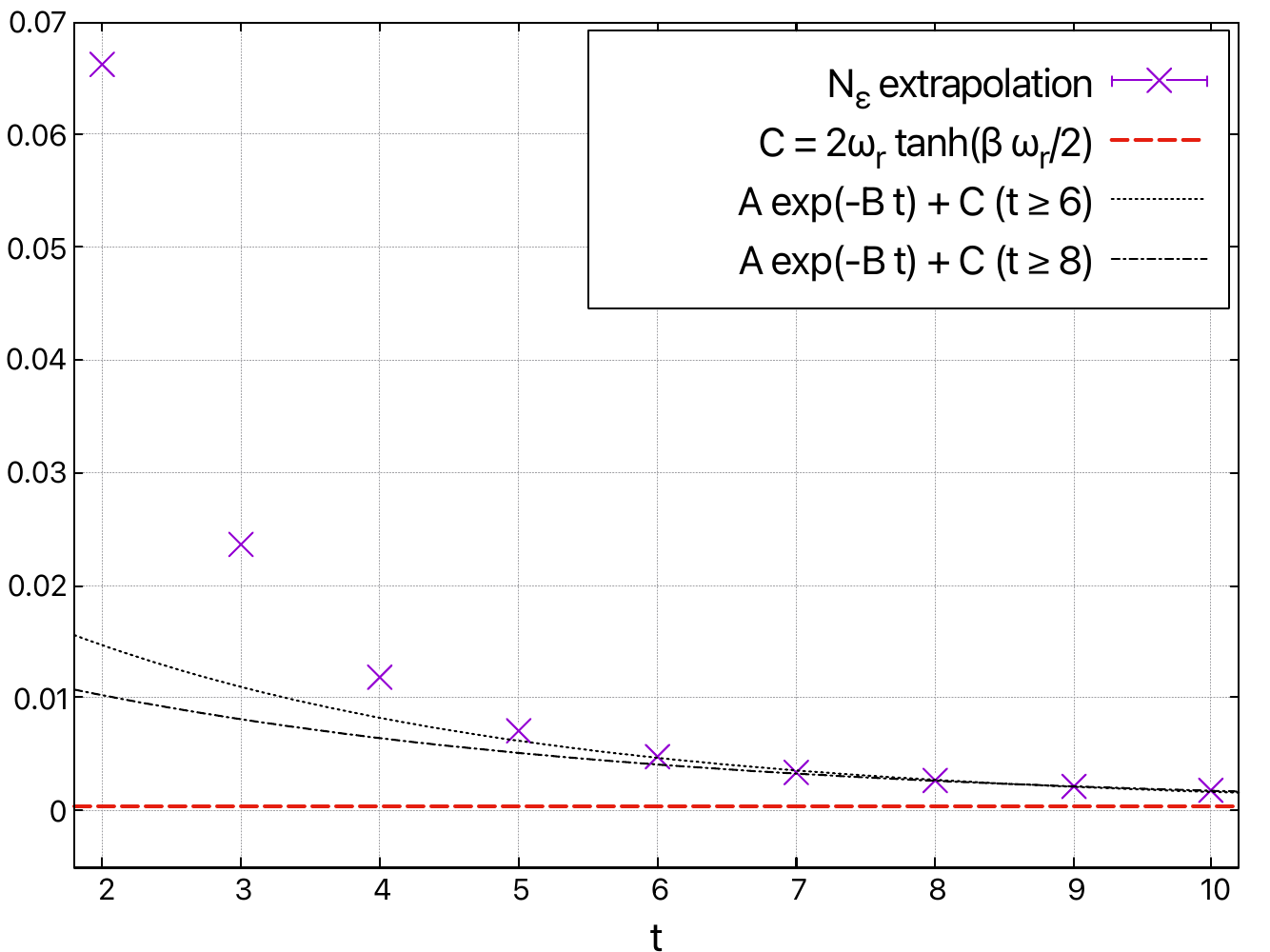}}
 	\scalebox{0.33}{\includegraphics{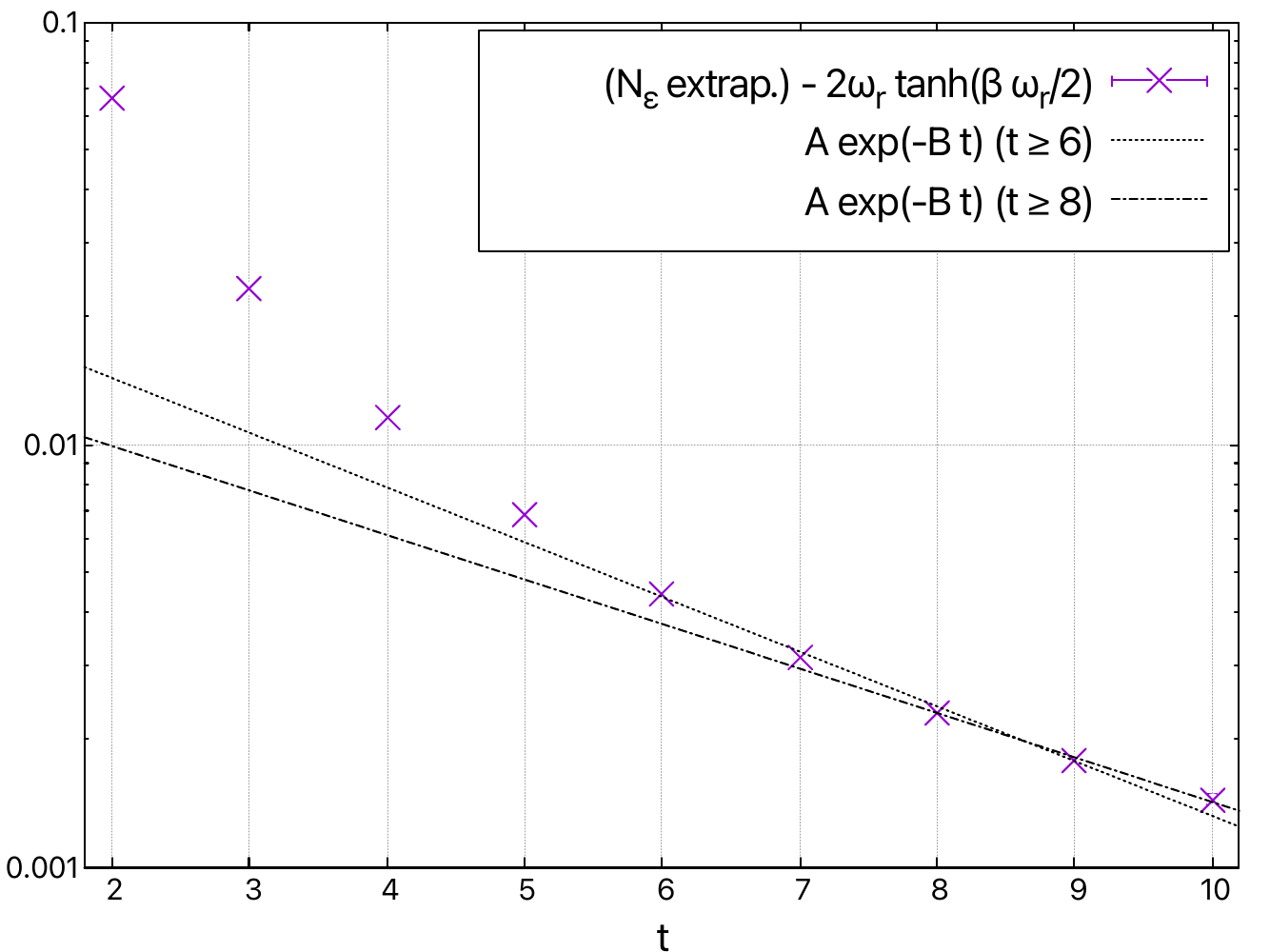}}
	\caption{
          (Left) The large-$N_\mathcal{E}$ extrapolated results for
          $\Gamma_{\mathrm{diag}}(t)$ at $t = 2,$ $3, \cdots, 10$
          obtained from Fig.~\ref{fig:Nenv_dependency_extrap} are plotted
          as a function of $t$.
          The dashed red line represents
          the value
          $C_{\rm th}
          = 2\omega_\mathrm{r}\tanh\frac{\beta\omega_\mathrm{r}}{2}=0.00032$
          predicted by thermalization at the inverse temperature $\beta=0.05$,
          whereas the dotted line and the dash-dotted line represent
          fits to $C_{\rm th} + A \exp(-B t)$ with the fitting range
          $t\ge 6$ and $t\ge 8$, respectively.
          (Right) The deviation of $\Gamma_{\mathrm{diag}}(t)$
          from the predicted value $C_{\rm th}$
          is plotted against $t$ in the log scale.
          The dotted line and the dash-dotted line represent linear fits
          to the data for $t \ge 6$ and $t\ge 8$, respectively.}
        \label{fig:therm_t-extrap_diag}
\end{figure}

From Fig.~\ref{fig:Nenv_dependency} (Top-Right), we find that
$\Gamma_\mathrm{diag}(t)$ becomes constant at late times,
which may be interpreted as thermalization
since the environment can be regarded as a thermal bath
with the inverse temperature $\beta$
in the $N_\mathcal{E} \rightarrow \infty$ limit.
If so, it is expected that the reduced density matrix
$\rho_\mathcal{S}(x,\tilde{x},t)$ approaches
\begin{equation}
  \lim_{t \rightarrow \infty}
  \rho_\mathcal{S}(x,\tilde{x},t)
    =
    \sqrt{\frac{\omega_\mathrm{r}}{2\pi\sinh\beta\omega_\mathrm{r}}}
    \exp\qty[-\frac{\omega_\mathrm{r}}{2\sinh\beta\omega_\mathrm{r}}
      \qty{\qty(x^2+\tilde{x}^2)\cosh\beta\omega_\mathrm{r} - 2x \tilde{x}}] 
\end{equation}
similarly to \eqref{eq:Equilib_env_2}.
In particular, it is expected that
\begin{equation}
  \lim_{t \rightarrow \infty}
  \Gamma_{\mathrm{diag}}(t)
  = 2 \omega_\mathrm{r} \tanh\frac{\beta\omega_\mathrm{r}}{2} \ , \quad \quad
  \lim_{t \rightarrow \infty}
  \Gamma_\mathrm{off\mathchar`-diag}(t) 
  = 2 \omega_\mathrm{r} \coth\frac{\beta\omega_\mathrm{r}}{2} \ .
    \label{eq:rho_thermal_prediction}
\end{equation}
In Fig.~\ref{fig:therm_t-extrap_diag}, we plot
the large-$N_\mathcal{E}$ extrapolated
results for $\Gamma_{\mathrm{diag}}(t)$ 
against time,
which are consistent with the behavior
$ 2\omega_\mathrm{r}\tanh\frac{\beta\omega_\mathrm{r}}{2}+A \exp(-B t)$.

On the other hand,
the off-diagonal part
shown in Fig.~\ref{fig:Nenv_dependency} (Top-Left)
has significant $N_\mathcal{E}$ dependence
at late times.
This may be due to the
recurrence for a finite environment.
(For similar discussions, see Section 2.10 of
Ref.~\cite{Schlosshauer} and references therein.) 
According to \eqref{eq:rho_thermal_prediction},
$\Gamma_\mathrm{off\mathchar`-diag}(t)$ is expected to approach
$2\omega_\mathrm{r}\coth \frac{\beta\omega_\mathrm{r}}{2} \approx 80$
at late times for sufficiently large $N_\mathcal{E}$.
In order to confirm this,
we need to increase $N_\mathcal{E}$ and the time $t$ further,
which we leave for future investigations.

Here let us comment on the computational efforts required for such an analysis.
To compute the reduced density matrix at each $t$, we solve the equation~\eqref{eq:saddle-pt_eq}
numerically by the LU decomposition\footnote{If we use an iterative method for
solving the linear equation \eqref{eq:saddle-pt_eq}, the cost may be reduced to $O(D)$
due to the sparseness of the matrix $\mathcal{M}_{\mu\nu}$.}, whose cost scales as $O(D^3)$ for the matrix size
$D = 2N_t (1 + N_\mathcal{E})  + N_\beta N_\mathcal{E}$.
In order to add another data point with increased time $t= N_t \, \epsilon$ for fixed $\epsilon$,
we need the computational time of $O(N_t ^3)$ since $D = O(N_t)$.
In order to obtain results for larger $N_\mathcal{E}$,
we need the computational time of $O(N_\mathcal{E}^3)$ since $D = O(N_\mathcal{E})$.
Therefore, taking both limits
$N_\mathcal{E} \rightarrow \infty$ and $t \rightarrow \infty$ seems to be a big challenge.

\begin{figure}[t]
  \centering
  \includegraphics[width=\textwidth]{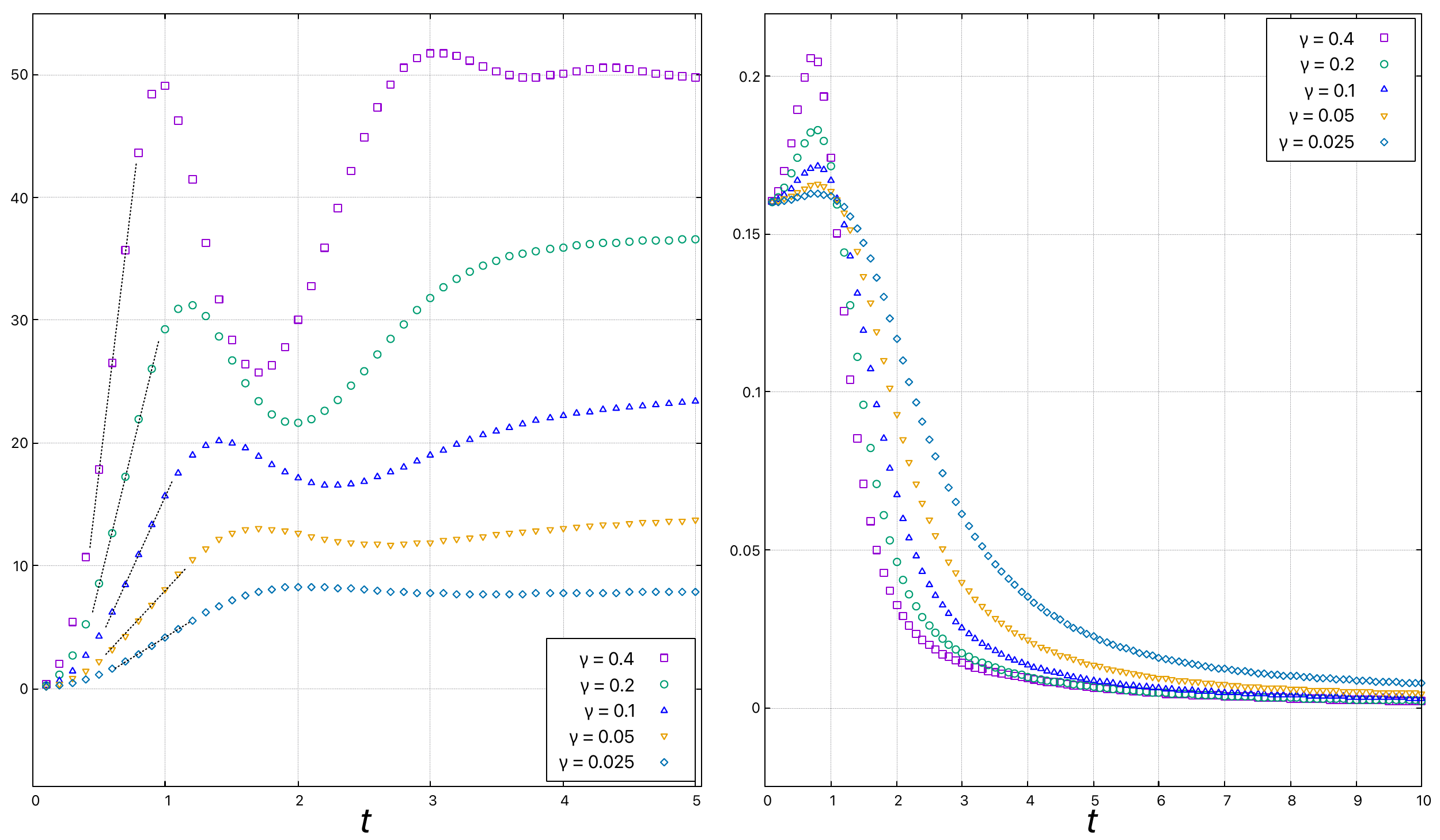}
  \includegraphics[width=0.6\textwidth]{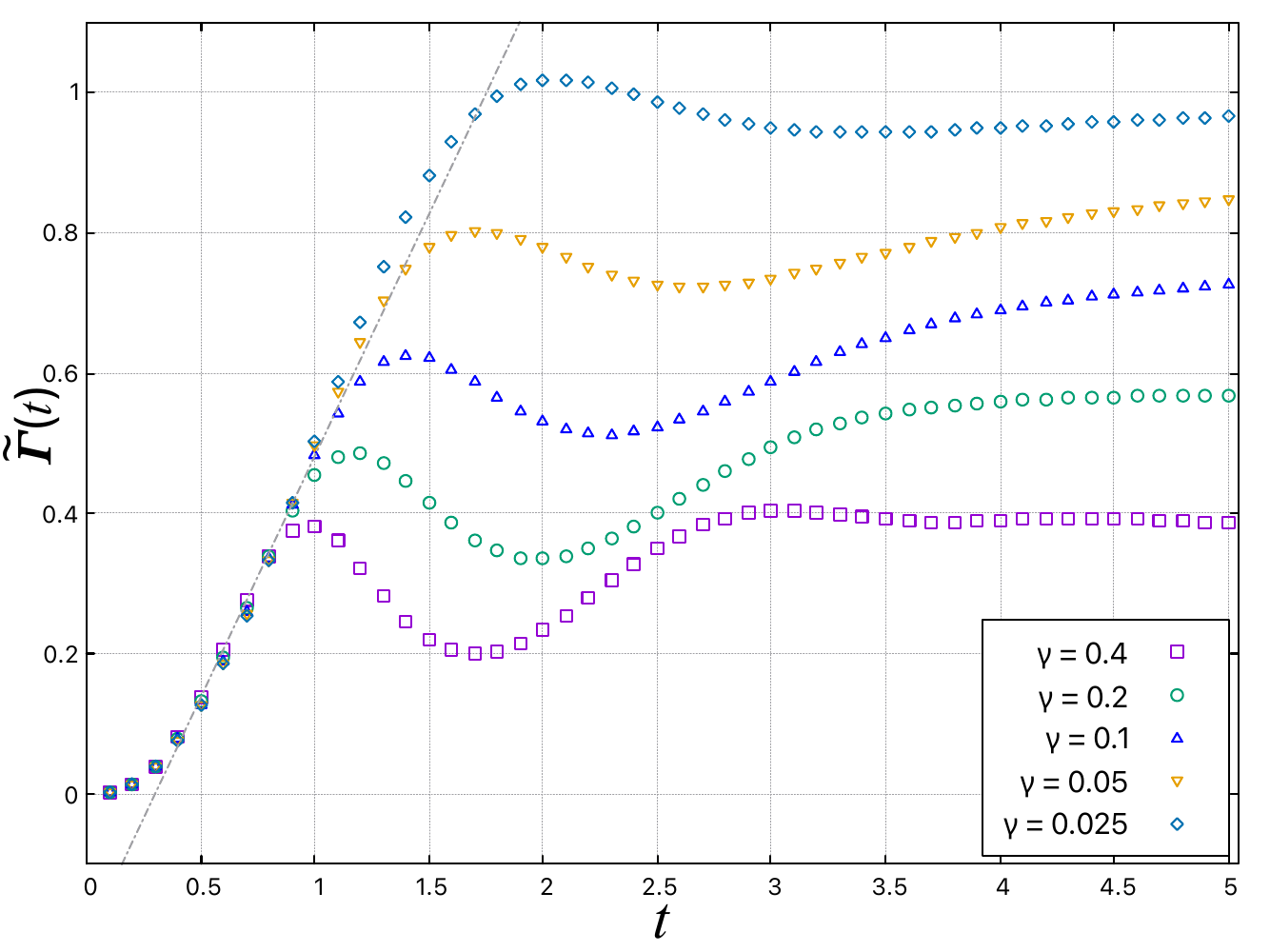}
  \caption{The quantities
        $\Gamma_\mathrm{off\mathchar`-diag}(t)$ (Top-Left)
    and
    $\Gamma_\mathrm{diag}(t)$ (Top-Right)
          are plotted against $t$
          for $\gamma=0.025$, $0.05, \cdots , 0.4$ with
          $N_\mathcal{E}=64$, $\beta=0.05$ and $\omega_\cut= 2$ fixed.
          The dotted lines in the Top-Left panel represent fits
          to the behavior
          $A  \frac{16\gamma}{\beta} t + B$
          at early times.
          (Bottom)
The rescaled quantity $ \tilde{\Gamma}(t)$ defined by \eqref{tilde-Gamma-def}
    is plotted against $t$, which reveals a clear scaling behavior.
              The dash-dotted line represents a fit of the $\gamma=0.1$ data
          within $0.6 \le t \le 1$
          to a linear behavior $At+B$, where $A\sim 0.69$ is obtained.}
        \label{fig:gamma_dependency}
\end{figure}

\subsection{Dependence on the coupling and the temperature}
\label{sec:dep-coupling-temp}

Next we discuss how our results
depend on the coupling
$\gamma$ and the temperature.
In Fig.~\ref{fig:gamma_dependency},
we plot $\Gamma_\mathrm{off\mathchar`-diag}(t)$ (Top-Left)
and $\Gamma_\mathrm{diag}(t)$ (Top-Right) against $t$
for $\gamma=0.025$, $0.05, \cdots , 0.4$ with
$N_\mathcal{E}=64$, $\beta=0.05$ and $\omega_\cut= 2$ fixed.
We observe a linear growth of
$\Gamma_\mathrm{off\mathchar`-diag}(t)$
from $t \approx 0.5$.
In the bottom panel, we plot
the rescaled quantity
\begin{align}
  \tilde{\Gamma}(t) =
  \frac{\beta}{16\gamma} \, \{\Gamma_\mathrm{off\mathchar`-diag}(t)
  -\Gamma_\mathrm{off\mathchar`-diag}(0)\} \ ,
  \label{tilde-Gamma-def}
\end{align}
which reveals a clear scaling behavior at early times.
While the scaling behavior implies
$\Gamma_\mathrm{off\mathchar`-diag}(t) \sim  A \frac{16\gamma}{\beta} t $,
which is qualitatively consistent with the prediction from the master equation,
the coefficient is $A\sim 0.69$, which is smaller than the predicted value 1. 
At $\gamma=0.05$, we start to see a deviation from a linear behavior,
which is understandable since
$\gamma/ \beta \gg 1$ is required to
justify the approximation of the master equation by \eqref{rho-last-term}.

\begin{figure}[t]
  \centering
  \includegraphics[width=\textwidth]{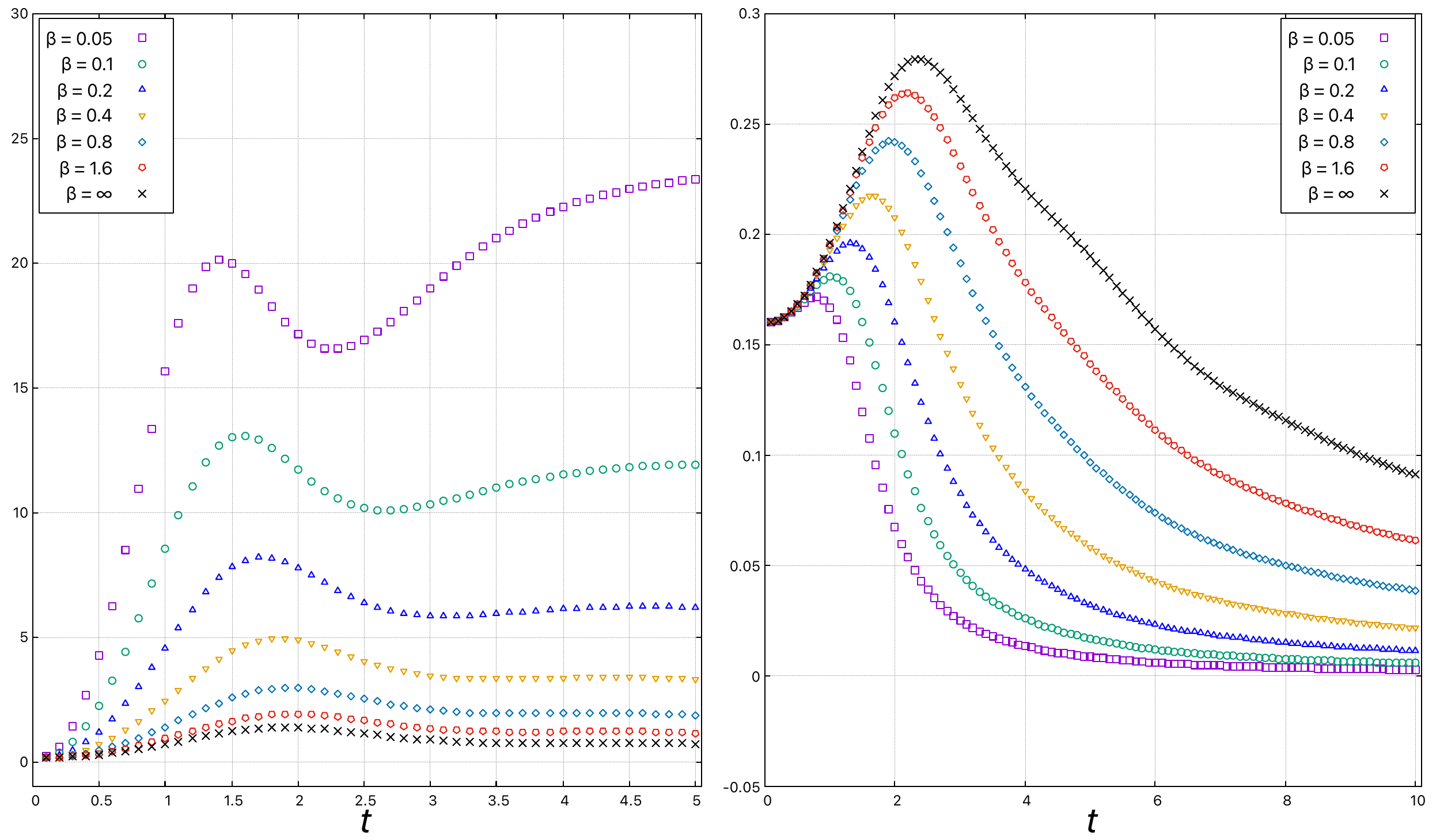}
  \includegraphics[width=0.6\textwidth]{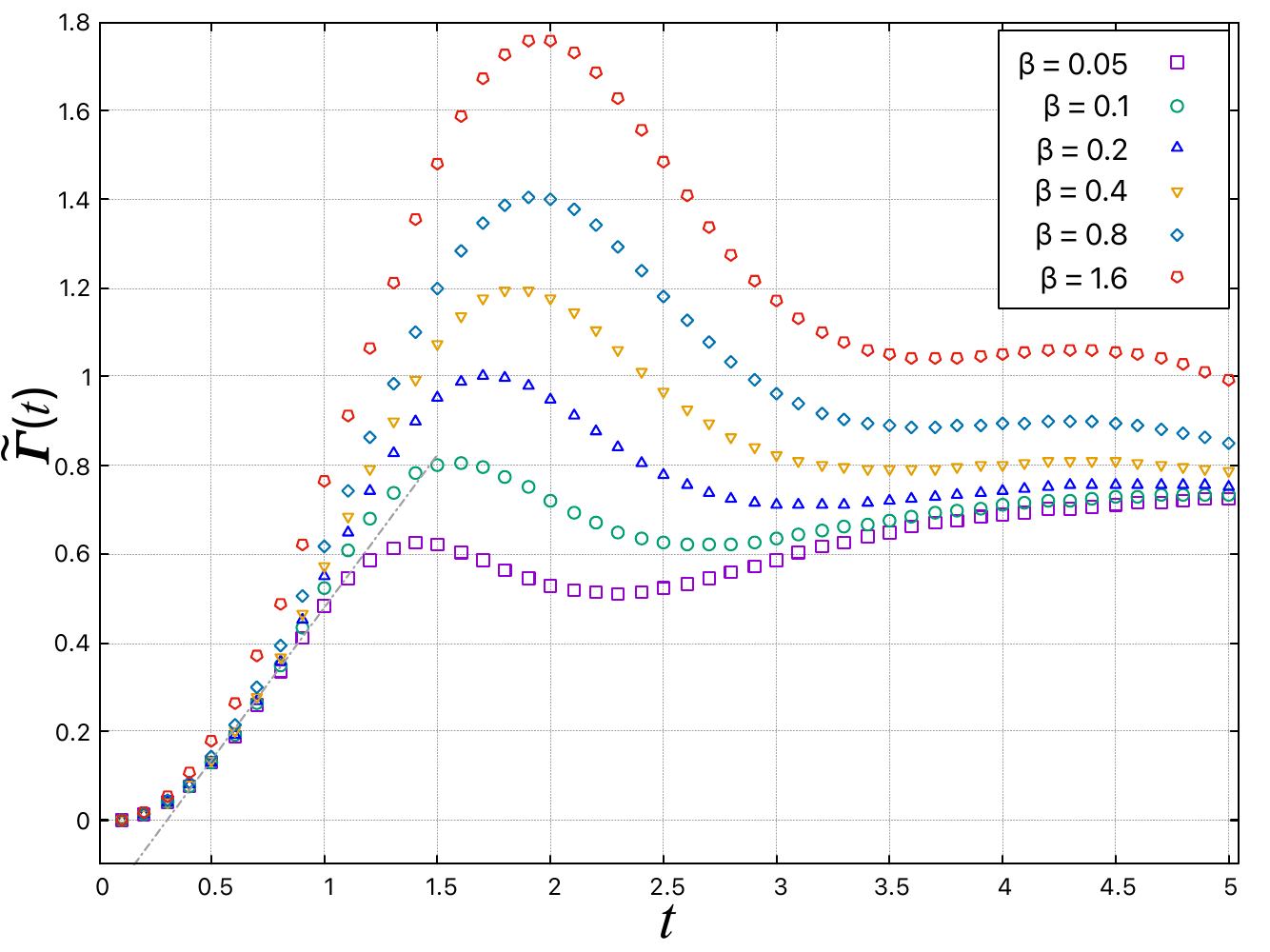}
  \caption{The quantities
    $\Gamma_\mathrm{off\mathchar`-diag}(t)$ (Top-Left) and
    $\Gamma_\mathrm{diag}(t)$ (Top-Right)
    are plotted for 
    $\beta = 0.05$, $0.1$, $0.2$, $0.4$, $0.8$, $1.6$ and $\infty$
    with $N_\mathcal{E} = 64$, $\gamma = 0.1$ and $\omega_\cut= 2$ fixed. 
    (Bottom)
The rescaled quantity $ \tilde{\Gamma}(t)$ defined by \eqref{tilde-Gamma-def}
    is plotted against $t$, which reveals a clear scaling behavior
    at small $\beta$.
    The dash-dotted line represents a fit of the $\beta=0.05$ data
              within $0.4 \le t \le 1.1$
    to a linear behavior $At+B$, where $A\sim 0.69$ is obtained.}
        \label{fig:T_dependency}
\end{figure}




In Fig.~\ref{fig:T_dependency},
we plot $\Gamma_\mathrm{off\mathchar`-diag}(t)$ (Top-Left)
and $\Gamma_\mathrm{diag}(t)$ (Top-Right)
against $t$
for $\beta = 0.05$, $0.1$, $0.2$, $0.4$, $0.8$, $1.6$
and $\infty$\footnote{In order to obtain the results at zero temperature,
which corresponds to $\beta =\infty$,
we have performed an independent calculation
by replacing the closed imaginary-time path for $\mathcal{E}$
with the ground-state wave function
$\psi_\mathrm{init}(q) = \ee^{-\sum_k \frac{1}{2} \, \omega_k (q_k)^2}.$
As $\beta$ is increased in Fig.~\ref{fig:T_dependency} (Top),
our results approach smoothly to the result at $\beta=\infty$ obtained in this way.}
with $N_\mathcal{E} = 64$, $\gamma = 0.1$ and $\omega_\cut= 2$.
In the bottom panel, we plot
the rescaled quantity \eqref{tilde-Gamma-def},
which shows the emergence of a clear scaling behavior
at early times as $\beta$ is decreased.
At lower temperature, we see
clear deviation from
the scaling behavior predicted by the master equation,
which assumes both $\beta \omega_\cut \ll 1$ (See \eqref{eq:Re-alpha}.)
and $\gamma/ \beta \gg 1$.

\subsection{Dependence on the cutoff frequency $\omega_\cut$}
\label{sec:cutoff_dep}

\begin{figure}[t]
	\centering
  \includegraphics[width=0.49\textwidth]{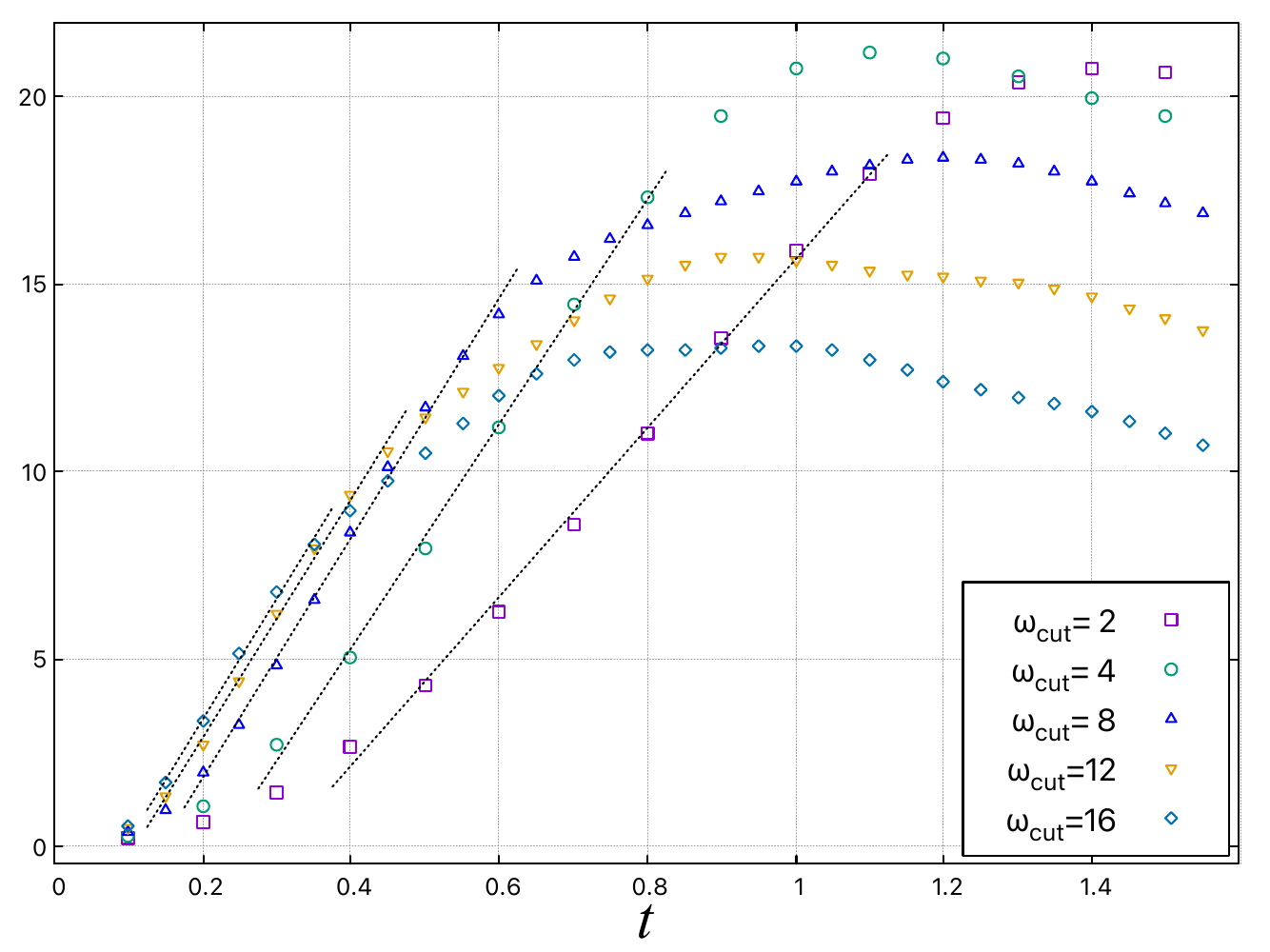}
  \includegraphics[width=0.49\textwidth]{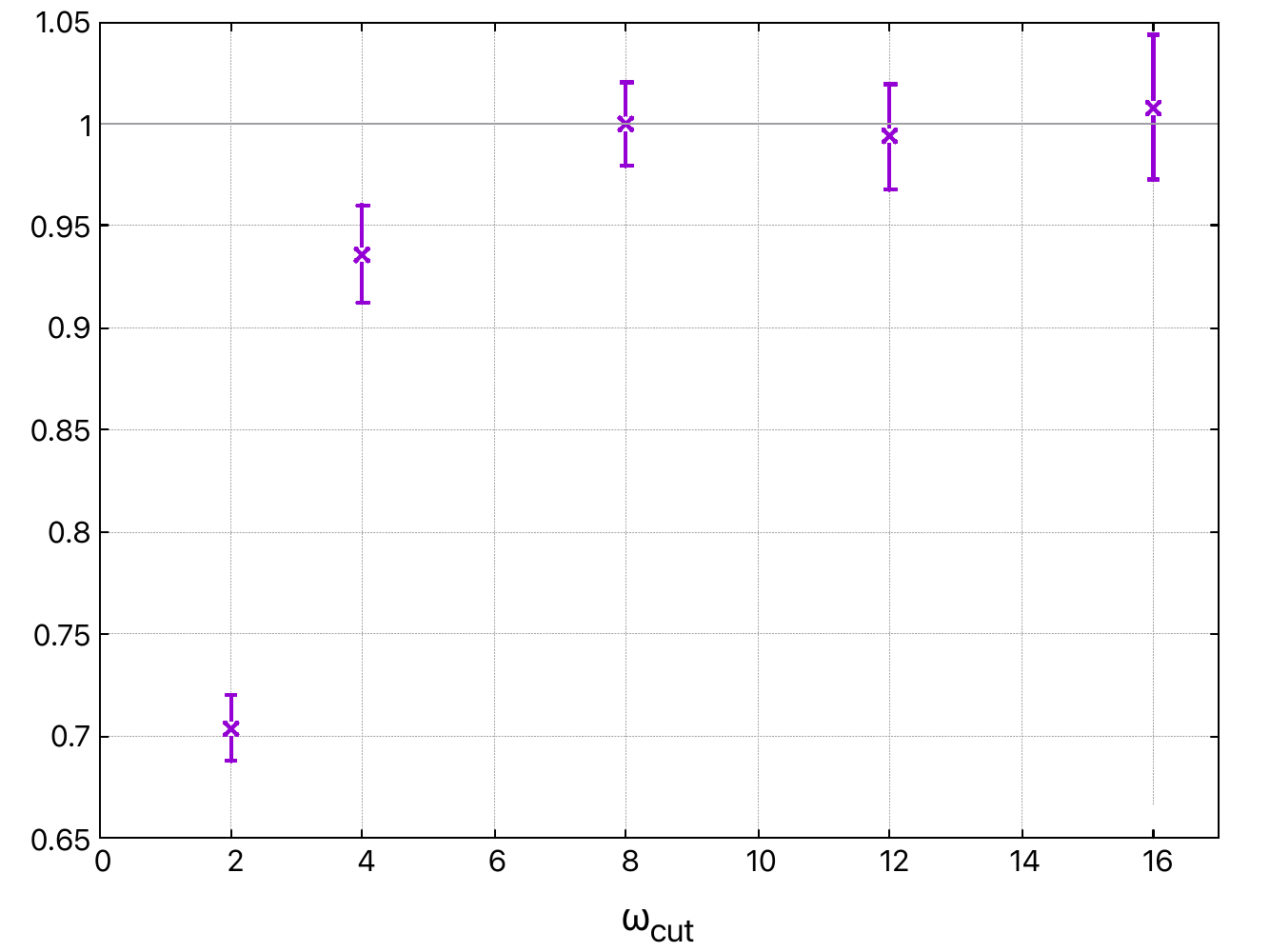}
	\caption{
    (Left) The quantity
    $\Gamma_\mathrm{off\mathchar`-diag}(t)$ 
    is plotted for 
    various $\omega_\cut$
    with $N_\mathcal{E} = 512$, $\beta = 0.05$ and $\gamma = 0.1$. 
    The dotted and dashed lines represent fits of our data at early times
    to a linear behavior $A \frac{16\gamma}{\beta}t+B$.
    (Right) The coefficient $A$
    obtained by the linear fit is plotted
    against $\omega_\cut$.
    The error bars represent the fitting errors.
    }
        \label{fig:cut_dependency}
\end{figure}

So far, we have confirmed the linear growth for $\Gamma_\mathrm{off\mathchar`-diag}(t)$
at early times with a slope proportional to $\gamma/\beta$ as predicted by the CL master equation.
However, the slope was smaller than the predicted value by 30\%.
In this section, we show that the numerical coefficient increases toward the predicted value
as we increase $\omega_\cut$.

In Fig.~\ref{fig:cut_dependency} (Left),
we plot $\Gamma_\mathrm{off\mathchar`-diag}(t)$ up to $t =1.6$ for $\omega_\cut=2,4,8,12,16$
with $N_\mathcal{E} = 512$, $\beta = 0.05$ and $\gamma = 0.1$.
We have used a larger value for $N_\mathcal{E}$ in order to keep
the frequencies \eqref{eq:omega_k_finiteNenv} in the environment dense enough.
(See appendix \ref{sec:limits_suppl} for more details.)
We fit our data to a linear behavior $A \frac{16\gamma}{\beta} t + B$ for each $\omega_\cut$.
In Fig.~\ref{fig:cut_dependency} (Right),
we plot the coefficient $A$ obtained by the fits
as a function of $\omega_\cut$,
where the error bars represent the fitting errors.
We find good agreement with the predicted value $A=1$ for $\omega_\cut = 8, 12, 16$.
This is understandable in view of the condition\footnote{Note, however, that the condition $\omega_\cut \ll \beta^{-1}=20$ is not well satisfied for $\omega_\cut = 8, 12, 16$,
which suggests that the condition $\omega_\cut \gg \omega_\mathrm{r}$
is more important for the agreement with the prediction $A=1$.}
$\omega_\cut \gg \omega_\mathrm{r} =0.08$, which is used
to approximate the function $f(\tau)$ in \eqref{def-f-function} by the delta function.
\section{Numerical results for the ``double-slit experiment''}
\label{sec:numerical_analysis}

In this section, we generalize our calculations
to the initial state with two wave packets,
which was studied also in
Refs.~\cite{Caldeira:1981rx,Caldeira:1982uj,Paz:1992pn,Zurek:1992mv}.
This enables us, in particular, to
investigate the decoherence
through the interference terms in the reduced density matrix
analogously to the situation in the well-known double-slit experiment.
We will see the effect of decoherence clearly as the fading of the interference
pattern.

\subsection{Extending the calculations to two wave packets}

Here we discuss how one can extend the calculations
in Section \ref{sec:numerical_setup} to the case of two wave packets.
Let us consider the initial wave function
\begin{align}
  \psi_\mathrm{I} (x)  &= \psi_0 (x) + \psi_1 (x) \ , \nn 
  \\
  \psi_0 (x) &=
  \exp(-\frac{1}{4 \sigma^2} (x-\xi)^2 - \ii p x) \ , \nn 
  \\
  \psi_1 (x) &=
  \exp(-\frac{1}{4 \sigma^2} (x+\xi)^2 + \ii p x) \ ,
  \label{init-rho-system-double}
\end{align}
where $\psi_0 (x)$ and $\psi_1 (x)$
represent
the wave packets that pass through the ``double slit''
separated by some distance $2 \xi>0$ and move towards $x=0$ with
some momentum $p>0$.
The corresponding reduced density matrix of the system \eqref{init-rho-system-gen}
can be decomposed as
\begin{align}
    \rho_\mathcal{S}(x,\tilde{x};0) 
    &=  \sum_{a,b=0}^1  \rho_{ab}(x,\tilde{x};0) \ ,
    \label{init-rho-system-double-2-prev}
    \\
    \rho_{ab} (x,\tilde{x};0) &=  \psi_{a} (x) \, \psi_{b} ^*(\tilde{x}) \ .
    \label{init-rho-system-double-2}
\end{align}

Plugging this in \eqref{eq:rho_S_position_basis},
we obtain the time-evolved reduced density matrix
\begin{align}
    \rho_\mathcal{S}(x,\tilde{x};t) 
    &=  \sum_{a,b=0}^1  \rho_{ab}(x,\tilde{x};t)   \ ,
    \label{evolved-rho-system-double}
\end{align}
where $\rho_{ab}(x,\tilde{x};t)$ is given by
\begin{equation}
    \rho_{ab}(x_\mathrm{F},\tilde{x}_\mathrm{F};t)
    = \int\mathcal{D}x\mathcal{D}\tilde{x}
    \left( \prod_{k=1}^{N_\mathcal{E}} \mathcal{D}q^k\mathcal{D}\tilde{q}^k \mathcal{D}\tilde{q}_0^k \right)
    \, \mathrm{e}^{-S_{ab}(x,\tilde{x},q,\tilde{q},\tilde{q}_0)}
    \ ,
    \label{eq:rho_S_position_basis-summary-double}
\end{equation}
with the effective action $S_{ab}$ given by
\begin{align}
  S_{ab} (x,\tilde{x},q,\tilde{q},\tilde{q}_0)
  &= - \mathrm{i} \left\{ S(x, q) - S(\tilde{x}, \tilde{q}) \right\}
   + S_0(\tilde{q}_0) 
  \nn \\
  & \quad  + \frac{1}{4 \sigma^2} \Big \{ x_0-(-1)^a \xi \Big\}^2
  + \ii (-1)^a p x_0
  + \frac{1}{4 \sigma^2} \Big\{ \tilde{x}_0 -(-1)^b \xi \Big\}^2
  - \ii (-1)^b p \tilde{x}_0
 \ .
  \label{def-eff-action-double}
    \end{align}
Similarly to \eqref{eq:primitive_lattice_action}, the effective action
can be written as
\begin{align}
  S_{ab} (x,\tilde{x},q,\tilde{q},\tilde{q}_0)
    & = 
  \frac{1}{2} X_\mu \mathcal{M}_{\mu\nu}X_\nu - (C_\mu+E_\mu^{ab}) X_\mu
  + \left( B + \frac{1}{2 \sigma^2} \xi^2  \right)   \ ,
    \label{eq:primitive_lattice_action-double}
    \\
  E_\mu ^{ab} 
  &=  (-1)^a \left( k  - \ii p \right)  e_\mu
  +   (-1)^b \left(  k  + \ii p \right) \tilde{e}_\mu \ ,
    \label{eq:def-E}
\end{align}
where $k = \frac{\xi}{2\sigma^2}$
and the unit vectors $e_\mu$ and $\tilde{e}_\mu$ are defined by
$x_0 = e_\mu X_\mu$ and $\tilde{x}_0 = \tilde{e}_\mu X_\mu$.

The saddle point of the action \eqref{eq:primitive_lattice_action-double}
is given by
\begin{equation}
    \bar{X}_\mu^{ab} = \mathcal{M}^{-1}_{\mu\nu} \, (C_\nu+E_\nu^{ab}) \ ,
\end{equation}
and we obtain each component of the reduced density matrix as
\begin{align}
    \rho_{ab} (x_\mathrm{F}, \tilde{x}_\mathrm{F};t)
    &= \frac{1}{\sqrt{\det \mathcal{M}}} \ee^{-\mathcal{A}_{ab}
      - \frac{1}{2\sigma^2} \xi^2 } \ ,
\label{eq:rho-detM-2}
  \\
  \mathcal{A}_{ab}  &=  B - \frac{1}{2} \, \bar{X}_\mu^{ab} \mathcal{M}_{\mu\nu} \bar{X}_\nu^{ab}
  \nn
  \\
  &=  B - \frac{1}{2} \,  (C_\mu+E_\mu^{ab})  \left( \mathcal{M}^{-1} \right)_{\mu\nu}
  (C_\nu+E_\nu^{ab})   \nn \\
  &=  \frac{1}{2}
  \begin{pmatrix}
x_\mathrm{F}  & \tilde{x}_\mathrm{F} 
\end{pmatrix}
\begin{pmatrix}
-\mathrm{i}  b + c_\mu (\mathcal{M}^{-1})_{\mu\nu} c_{\nu}
  &   - c_\mu (\mathcal{M}^{-1})_{\mu\nu} \tilde{c}_{\nu} \\
  - \tilde{c}_\mu (\mathcal{M}^{-1})_{\mu\nu} c_{\nu}
  & \mathrm{i} b + \tilde{c}_\mu (\mathcal{M}^{-1})_{\mu\nu} \tilde{c}_{\nu}  \\
\end{pmatrix}
\begin{pmatrix}
x_\mathrm{F}  \\ \tilde{x}_\mathrm{F} 
\end{pmatrix}
\nn 
\\
& \quad - \ii E_\mu^{ab} (\mathcal{M}^{-1})_{\mu\nu}
( c_{\nu} x_\mathrm{F} - \tilde{c}_{\nu} \tilde{x}_\mathrm{F} )
- \frac{1}{2} E_\mu^{ab} (\mathcal{M}^{-1})_{\mu\nu}  E_\nu^{ab}
\ .
\label{eq:rho-calA}
\end{align}
Note that there are relationships between the components as
\begin{align}
  \rho_{00} (x_\mathrm{F}, \tilde{x}_\mathrm{F};t)
  & = \rho_{11} (-x_\mathrm{F}, -\tilde{x}_\mathrm{F};t) \ , \\
  \rho_{01} (x_\mathrm{F}, \tilde{x}_\mathrm{F};t)
  & = \rho_{10} (-x_\mathrm{F}, -\tilde{x}_\mathrm{F};t) \ ,
\end{align}
due to the symmetry $x \mapsto -x $ of our chosen setup.


Here we focus on the diagonal elements of the density
matrix\footnote{See Appendix \ref{sec:more_double-slit}
  for the discussions on the off-diagonal elements of the density matrix.}
corresponding to $x_\mathrm{F} = \tilde{x}_\mathrm{F} = x$,
for which the exponent $\mathcal{A}_{ab}$ reduces to
\begin{align}
  \mathcal{A}_{ab}
  &=  \frac{1}{2}
(c_\mu-\tilde{c}_\mu) (\mathcal{M}^{-1})_{\mu\nu} (c_{\nu}-\tilde{c}_\nu) x^2
-\ii E_\mu^{ab} (\mathcal{M}^{-1})_{\mu\nu}
( c_{\nu} - \tilde{c}_{\nu}) x
 - \frac{1}{2} E_\mu^{ab} (\mathcal{M}^{-1})_{\mu\nu}  E_\nu^{ab} 
\ .
\end{align}
Thus each component of the density matrix is given by
\begin{align}
  \label{rho00-def}
\rho_{00}(x,x;t)
  &\simeq \exp \left\{ -  \frac{1}{2} K_1 \, x^2
  - \left( k K_2  - p  K_3\right)  x
+ \frac{1}{2} (
k^2  K_4
- p^2 K_5 )
\right\} \ , \\
\rho_{11}(x,x;t)
&= \rho_{00}(-x,-x;t) \ ,
\\
\rho_{01}(x,x;t)
  &\simeq \exp \left\{ -  \frac{1}{2}
K_1 \, x^2
+ \ii \left( k  K_3 + p   K_2 \right)  x
+ \frac{1}{2} (
k^2  K_5
- p^2 K_4 )
\right\} \ ,
\\
\rho_{10}(x,x;t)
&= \rho_{01}(-x,-x;t) \ , 
\end{align}
omitting the prefactors common to all components.
The real quantities $K_i$ ($i=1, \cdots , 5$) are defined as
\begin{align}
K_1 &=
(c_\mu-\tilde{c}_\mu) \,  {\rm Re}
(\mathcal{M}^{-1})_{\mu\nu} (c_{\nu}-\tilde{c}_\nu) \ , \\
K_2 &=
  (e_{\mu}+\tilde{e}_\mu) \, {\rm Im} (\mathcal{M}^{-1})_{\mu\nu}
( c_{\nu} - \tilde{c}_{\nu}) \ ,  \\
K_3 &=
 (e_{\mu}-\tilde{e}_\mu) \,  {\rm Re} (\mathcal{M}^{-1})_{\mu\nu}
  ( c_{\nu} - \tilde{c}_{\nu}) \ , \\
K_4 &=
 (e_{\mu}+\tilde{e}_\mu) \,  {\rm Re} (\mathcal{M}^{-1})_{\mu\nu}
  ( e_{\nu} + \tilde{e}_{\nu}) \ , \\
K_5 &=
 (e_{\mu}-\tilde{e}_\mu) \,  {\rm Re} (\mathcal{M}^{-1})_{\mu\nu}
( e_{\nu} - \tilde{e}_{\nu}) \ .
\end{align}

In fact, we need to normalize the reduced density matrix
as $\rho_{ab} (x, \tilde{x};t) \mapsto
\frac{1}{{\cal N}}\rho_{ab} (x, \tilde{x};t)$
with
the normalization factor defined by
\begin{align}
  {\cal N}(t)  = \int \dd x \, \rho_\mathcal{S}(x,x;t)
  &= 2 \sqrt{\frac{2\pi}{K_1}}
   \left[ \exp\left\{ \frac{1}{2} \frac{(k K_2 - p K_3)^2}{K_1}
     + \frac{1}{2} (k^2 K_4 - p^2 K_5) \right\}
  \right. \nn \\  & \qquad \qquad \left.
     + \exp\left\{ - \frac{1}{2} \frac{(k K_3 + p K_2)^2}{K_1}
     + \frac{1}{2} (k^2 K_5 - p^2 K_4) \right\} \right] \ .
   \label{normalize-rho}
\end{align}
In what follows, we assume that the reduced density matrix is
normalized in this way.




\begin{figure}[t]
	\centering
 	\includegraphics[width=0.8\textwidth]{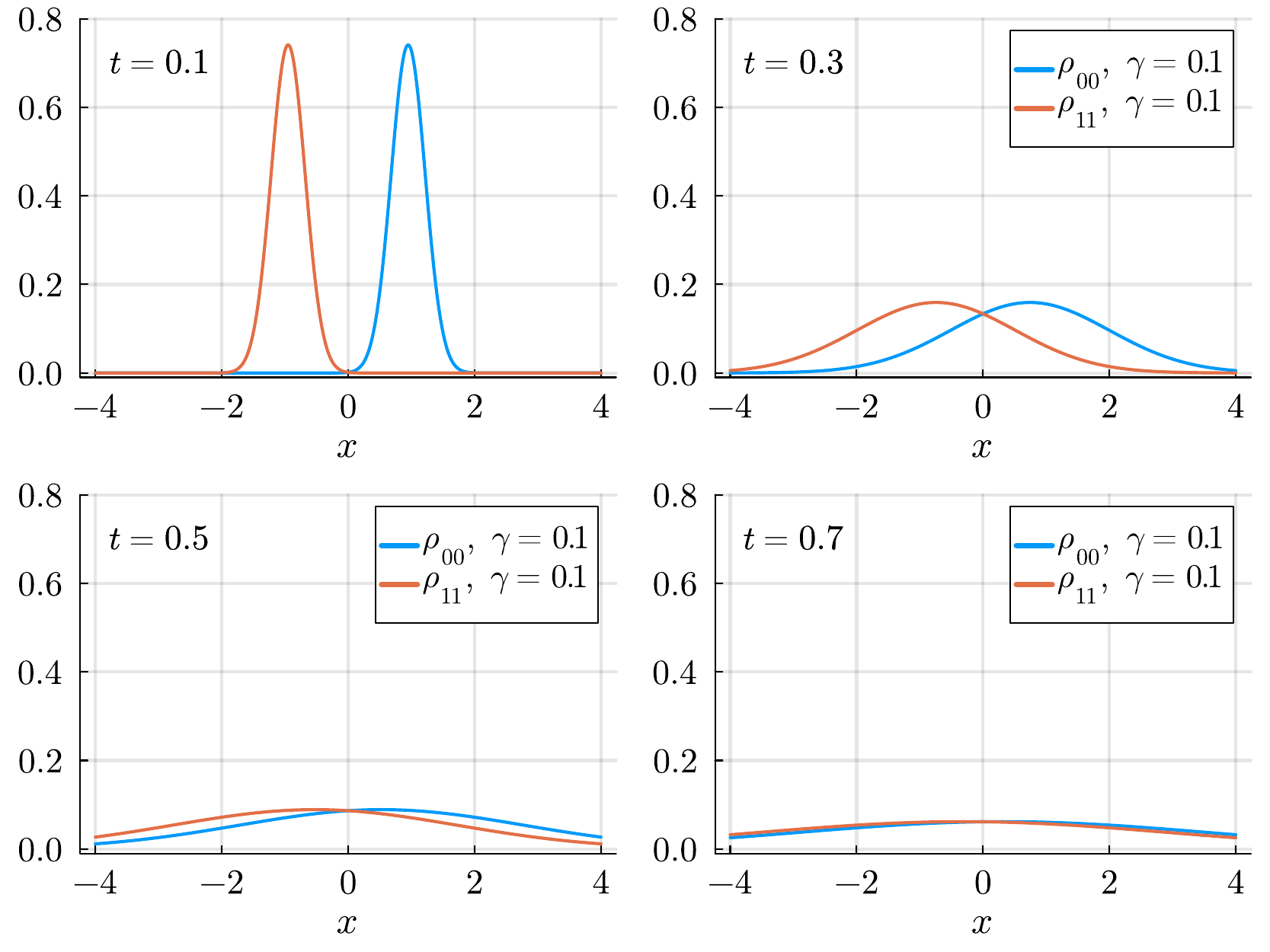}
	\caption{The direct terms $\rho_{00}(x,x;t)$ and $\rho_{11}(x,x;t)$
          are plotted against $x$ for $t=0.1$, $0.3$, $0.5$, $0.7$
          with $\gamma=0.1$.}
        \label{fig:plot_rho_00}
\end{figure}

\begin{figure}[t]
	\centering
  \scalebox{0.33}{\includegraphics{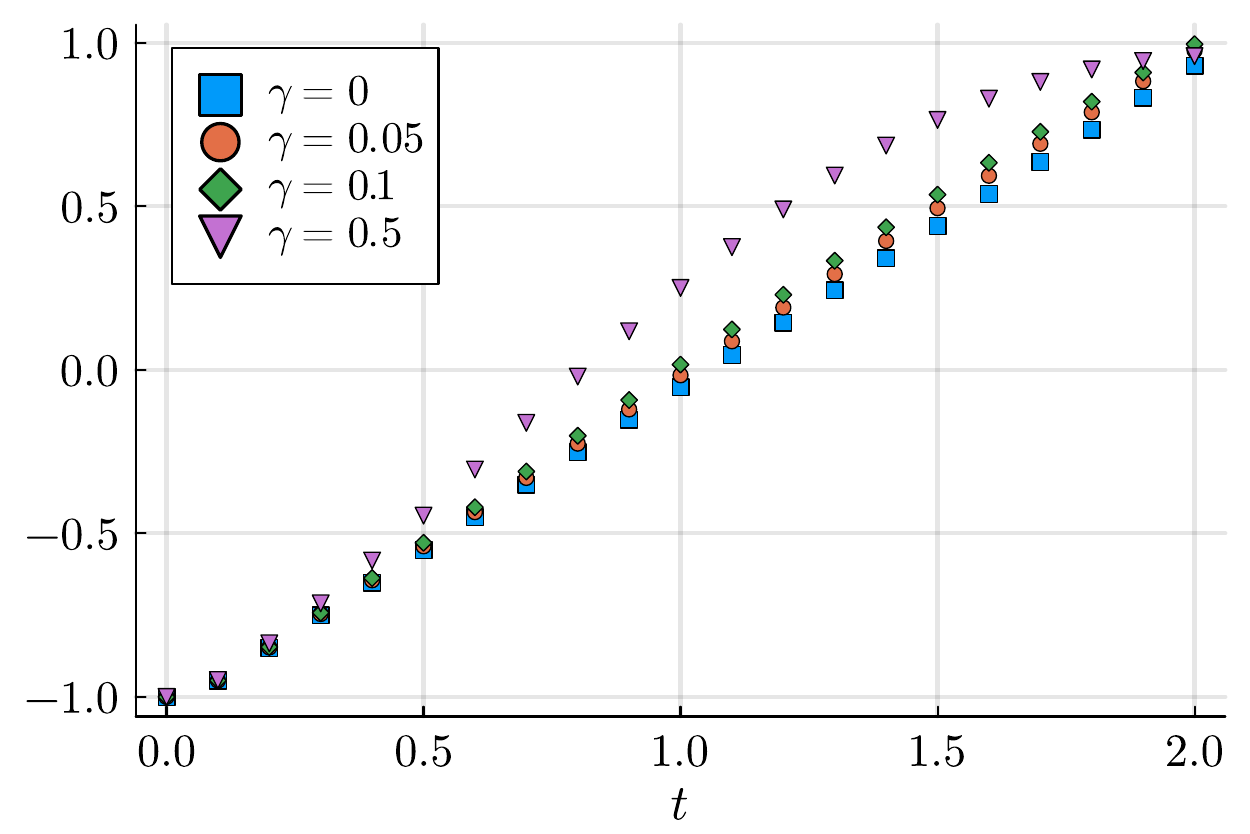}}
  \scalebox{0.33}{\includegraphics{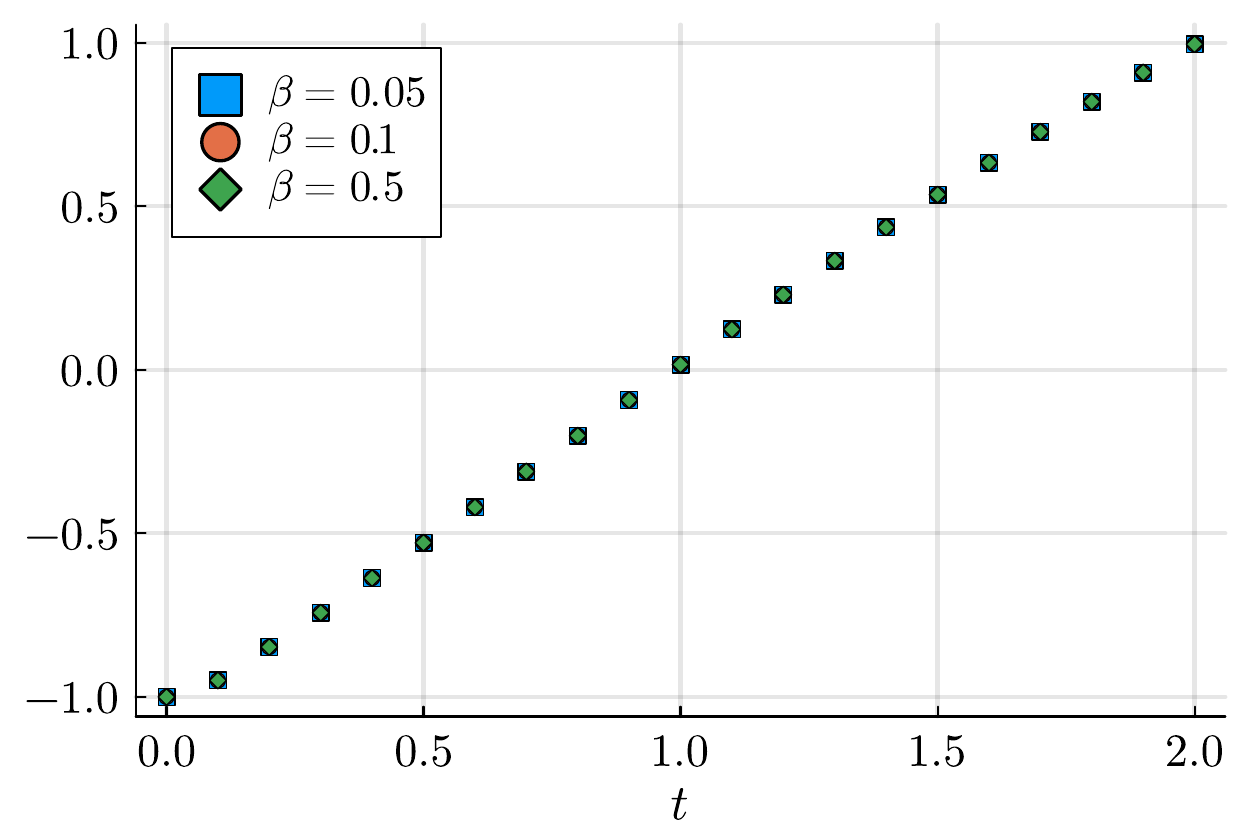}}
  \caption{The peak position of $\rho_{00}(x,x;t)$ is plotted
    for various $\gamma$ with $\beta = 0.05$ (Left)
    and for various $\beta$ with $\gamma = 0.1$ (Right).}
        \label{fig:center_wp_diagonal}
\end{figure}

\subsection{The effect of decoherence on the interference pattern}
\label{sec:main_results_case2}

Here we present our results for the time-evolved
reduced density matrix obtained in the way described in the previous section.
Throughout this section, we set the parameters as
$N_\mathcal{E}=64$, $\omega_\mathrm{cut} = 2$,
$\omega_\mathrm{r} = 0$, $\sigma=0.1$, $\xi=1$, $p=1$.
Note also that $k=\xi/(2\sigma^2)=1/(2\times 0.1^2)=50$.
The lattice spacing in the time direction
is chosen to be $\epsilon=0.05$,
whereas the lattice spacing in the temperature direction
is chosen to be $\tilde{\epsilon}=0.0125$.


Let us first discuss the direct terms $\rho_{00}(x,x;t)$ and $\rho_{11}(x,x;t)$
in the reduced density matrix,
which represent
the contribution
from $\psi_0(x)$ and $\psi_1(x)$, respectively.
In Fig.~\ref{fig:plot_rho_00}, we plot these quantities against $x$
for $t=0.1$, $0.3$, $0.5$, $0.7$.
We find that they are given by two Gaussian distributions,
which are peaked around $x=1$ and $x=-1$, respectively, at early times,
and come close to each other with time
due to the chosen initial condition \eqref{init-rho-system-double}.
The results obtained here are actually
quite close to those for $\gamma=0$.
This implies that the direct terms $\rho_{00}(x,x;t)$ and $\rho_{11}(x,x;t)$
are not sensitive to the interaction with the environment at least
for $\gamma \lesssim 0.1$.

In Fig.~\ref{fig:center_wp_diagonal}, we plot the peak position
of $\rho_{00}(x,x;t)$, which is given by $\frac{k K_2 - p K_3}{K_1}$
from \eqref{rho00-def}, against $t$.
Note that the time dependence of the peak position should
agree with the classical motion of free particles for $\gamma=0$
due to our choice $\omega_\mathrm{r} = 0$.
From our results for $\gamma > 0$, 
there seems to be some ``effective $\omega_\mathrm{r}$''
induced by the interactions with the environment on top of the expected
deceleration due to the friction-like effects.
On the other hand, we do not see clear $\beta$ dependence for a fixed $\gamma$.



\begin{figure}[H]
	\centering
  {
 	\includegraphics[width=0.8\textwidth]{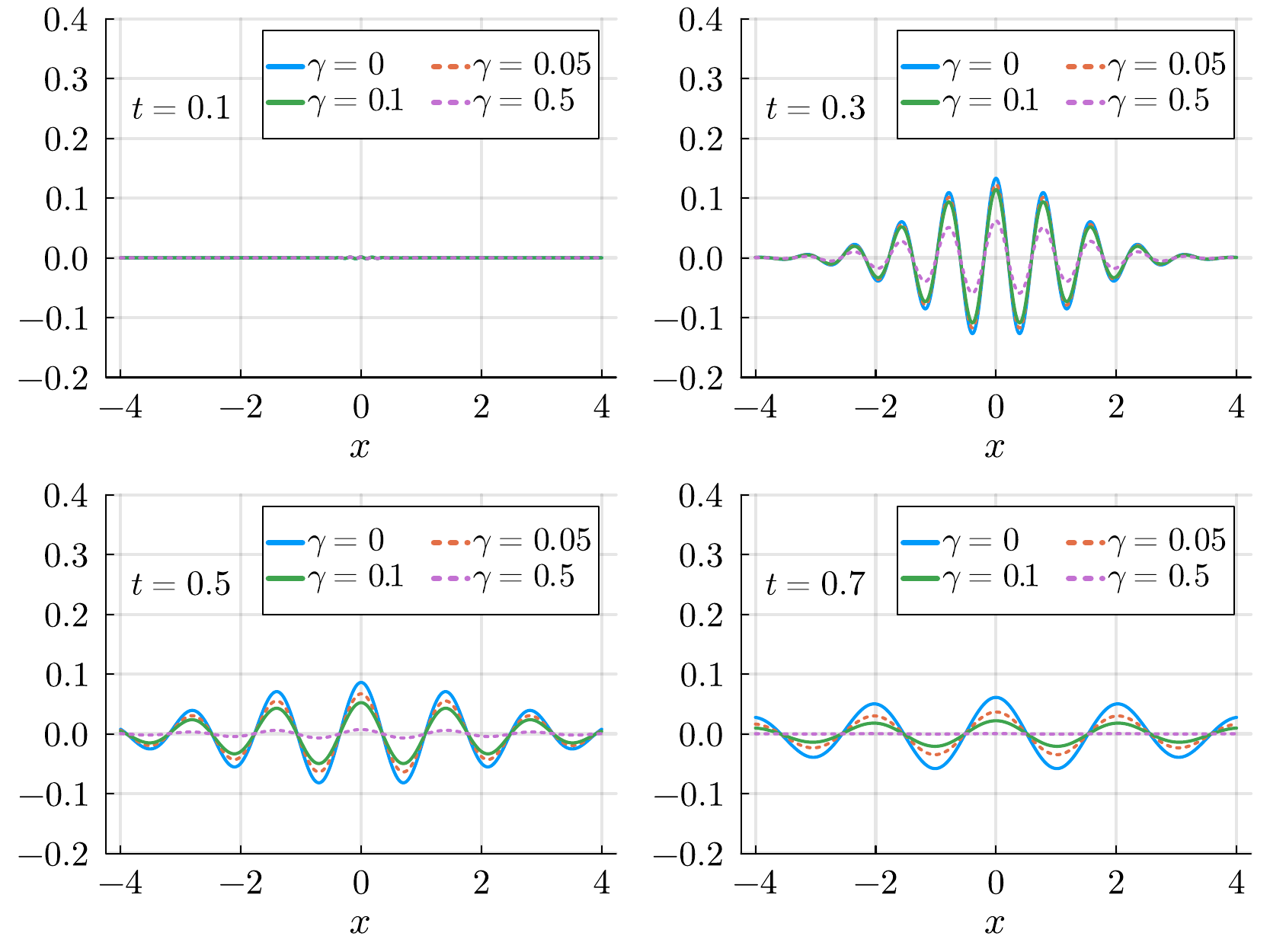}
  \subcaption{$\gamma$ dependence}
  \vspace{10pt}
  }
  {
 	\includegraphics[width=0.8\textwidth]{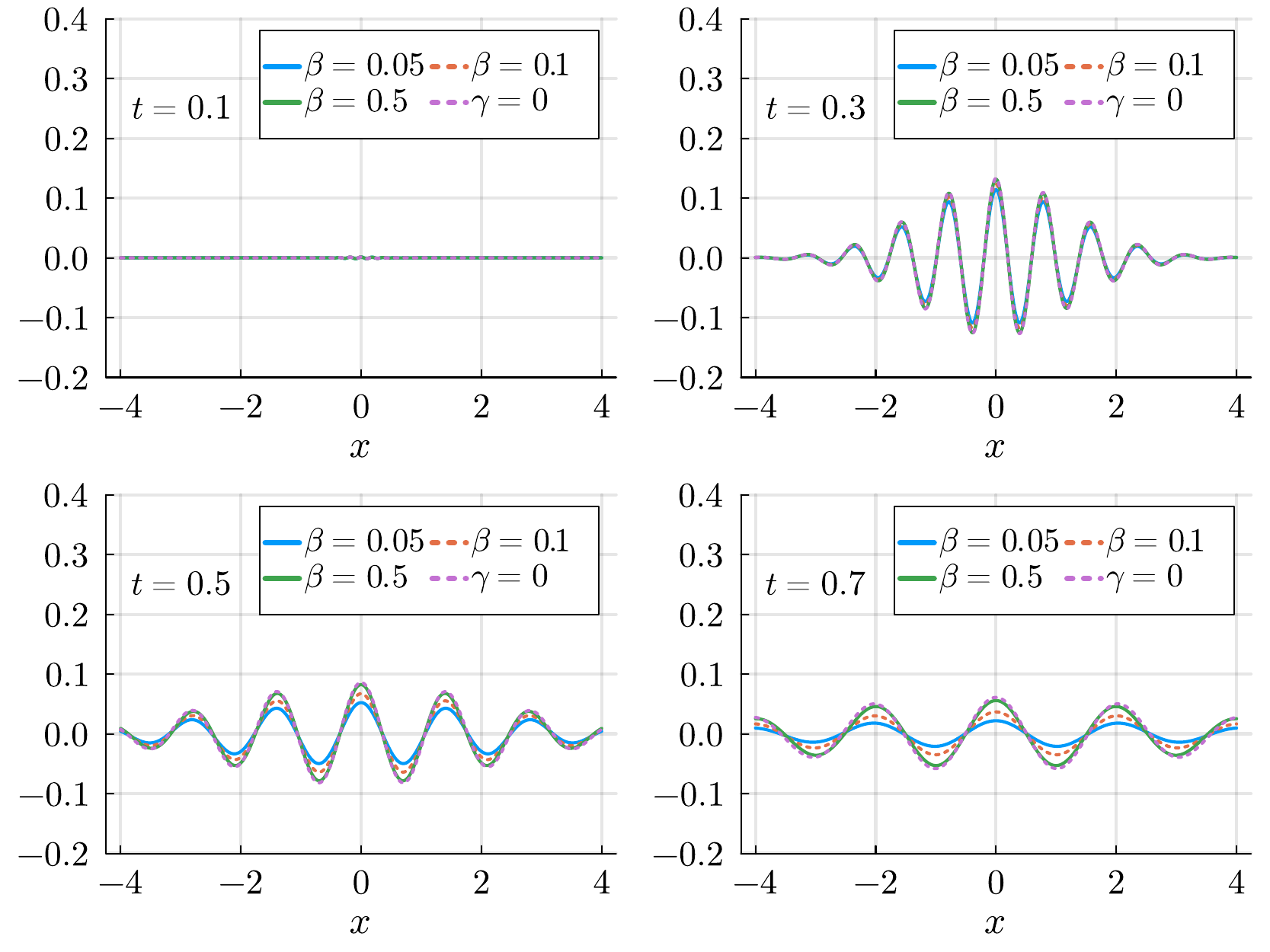}
  \subcaption{$\beta$ dependence}
  }
	\caption{The interference term ${\rm Re} \, \rho_{01}(x,x;t)$
          is plotted
          for various $\gamma$
with fixed $\beta=0.05$ (a) and 
for various $\beta$ with fixed $\gamma=0.1$ (b).
In the latter, the result for $\gamma=0$ is also shown for comparison.
}
        \label{fig:re_rho_01}
\end{figure}

Next we consider
the interference terms
\begin{align}
  \rho_{01}(x,x;t) +   \rho_{10}(x,x;t)  &=  2 \, {\rm Re} \, \rho_{01}(x,x;t) \ .
\end{align}
In Fig.~\ref{fig:re_rho_01}, we plot
${\rm Re} \,  \rho_{01}(x,x;t)$
for various $\gamma$
with fixed $\beta=0.05$ (a) and 
for various $\beta$ with fixed $\gamma=0.1$ (b).
We find that the amplitude of the oscillation gets
smaller as $\gamma/\beta$ increases.
In Fig.~\ref{fig:re_rho_01} (b),
the result for $\beta=0.5$ with $\gamma=0.1$
is almost identical to the result for $\gamma=0$,
which is shown for comparison.

\begin{figure}[H]
	\centering
 	\scalebox{0.33}{\includegraphics{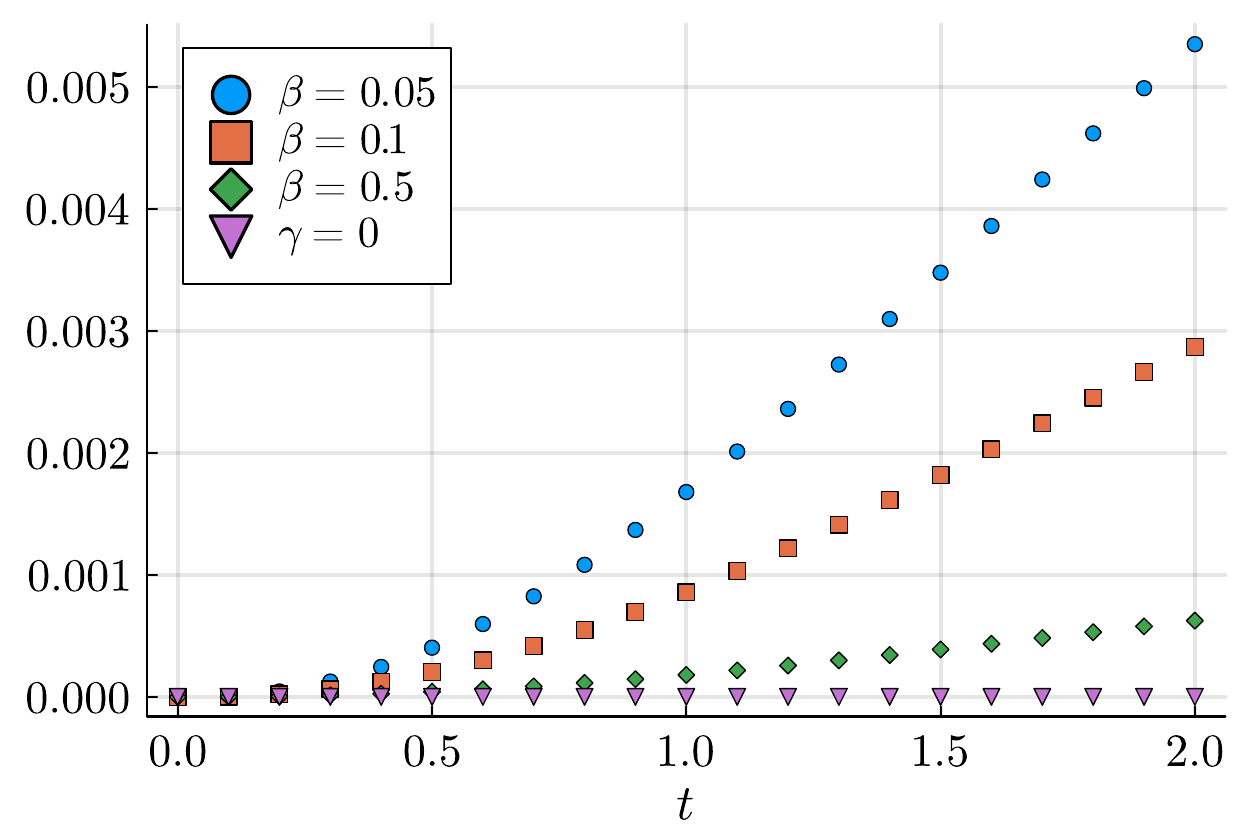}}
	\scalebox{0.33}{\includegraphics{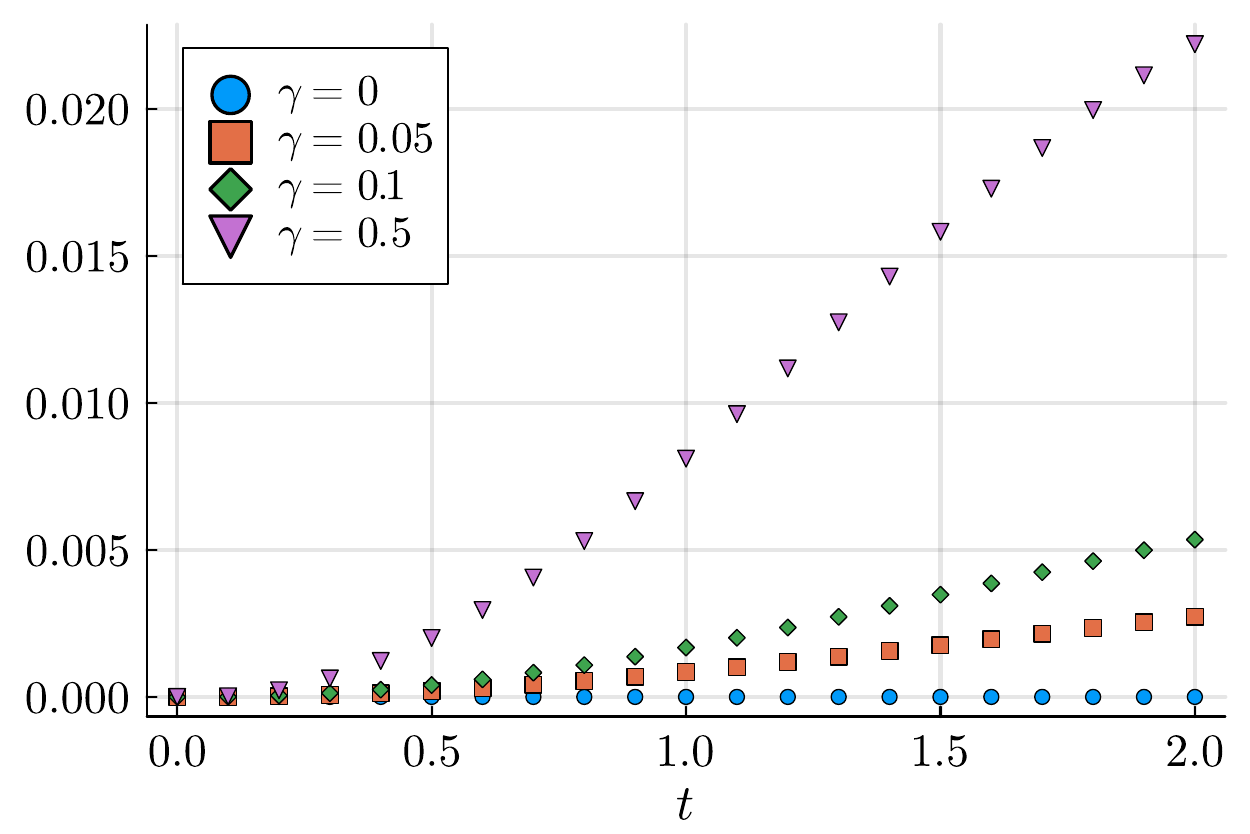}}
 	\includegraphics[width=0.5\textwidth]{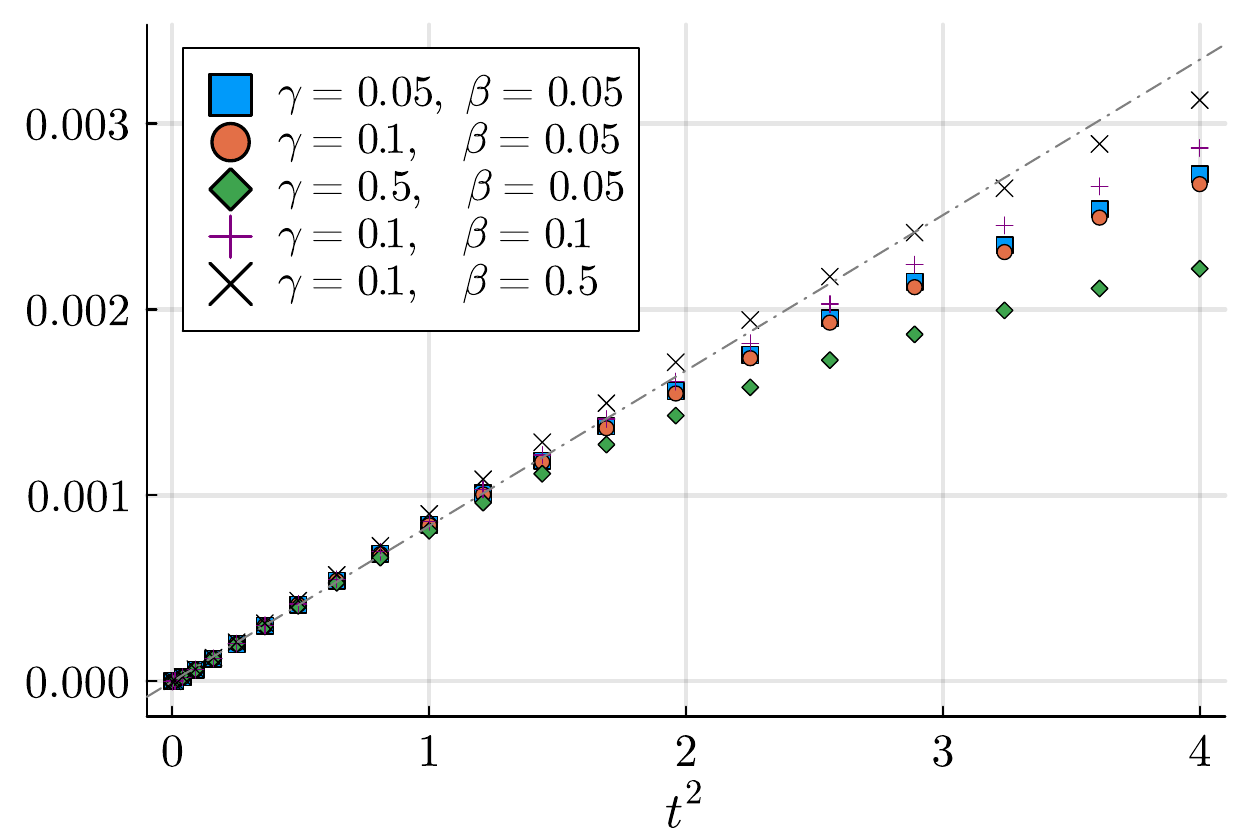}
	\caption{
          (Top) The quantity $K_4 - K_5$ is plotted against $t$
          for various $\gamma$ with $\beta=0.05$ (Left) and for various $\beta$
          with $\gamma=0.1$ (Right).
(Bottom) The rescaled quantity $\frac{\beta}{\gamma}(K_4 - K_5)$
          is plotted against $t^2$ for various $\beta$ and $\gamma$.
          }
        \label{fig:K4-K5_vs_parameters}
\end{figure}

In order to clarify
the effect of decoherence
more quantitatively,
we define the ratio
\begin{align}
\mathcal{A}(t) &=   \frac{\rho_{01}(0,0;t)}{\rho_{00}(0,0;t)}
=  \exp\left\{  - \frac{1}{2} (k^2 + p^2 ) (K_4 - K_5) \right\} \ ,
\label{def-A-ratio}
\end{align}
where $k^2 + p^2 = 50^2 + 1 = 2501$ in our setup.
In Fig.~\ref{fig:K4-K5_vs_parameters} (Top),
we plot $K_4 - K_5$ against $t$
for various values of $\gamma$ and $\beta$.
In Fig.~\ref{fig:K4-K5_vs_parameters} (Bottom), we plot
the rescaled quantity $\frac{\beta}{\gamma}(K_4 - K_5)$ against $t^2$.
We find that the data points lie on a straight line
at early times,
which implies
that the ratio $\mathcal{A}(t)$ decreases as
$\exp\{-c\, (\gamma/\beta)\, t^2\}$.
This $O(t^2)$ behavior is different from
the $O(t)$ and $O(t^3)$ behaviors observed
in different parameter regimes \cite{Caldera:1985tk}.

\begin{figure}[H]
	\centering
 	{
  \includegraphics[width=0.8\textwidth]{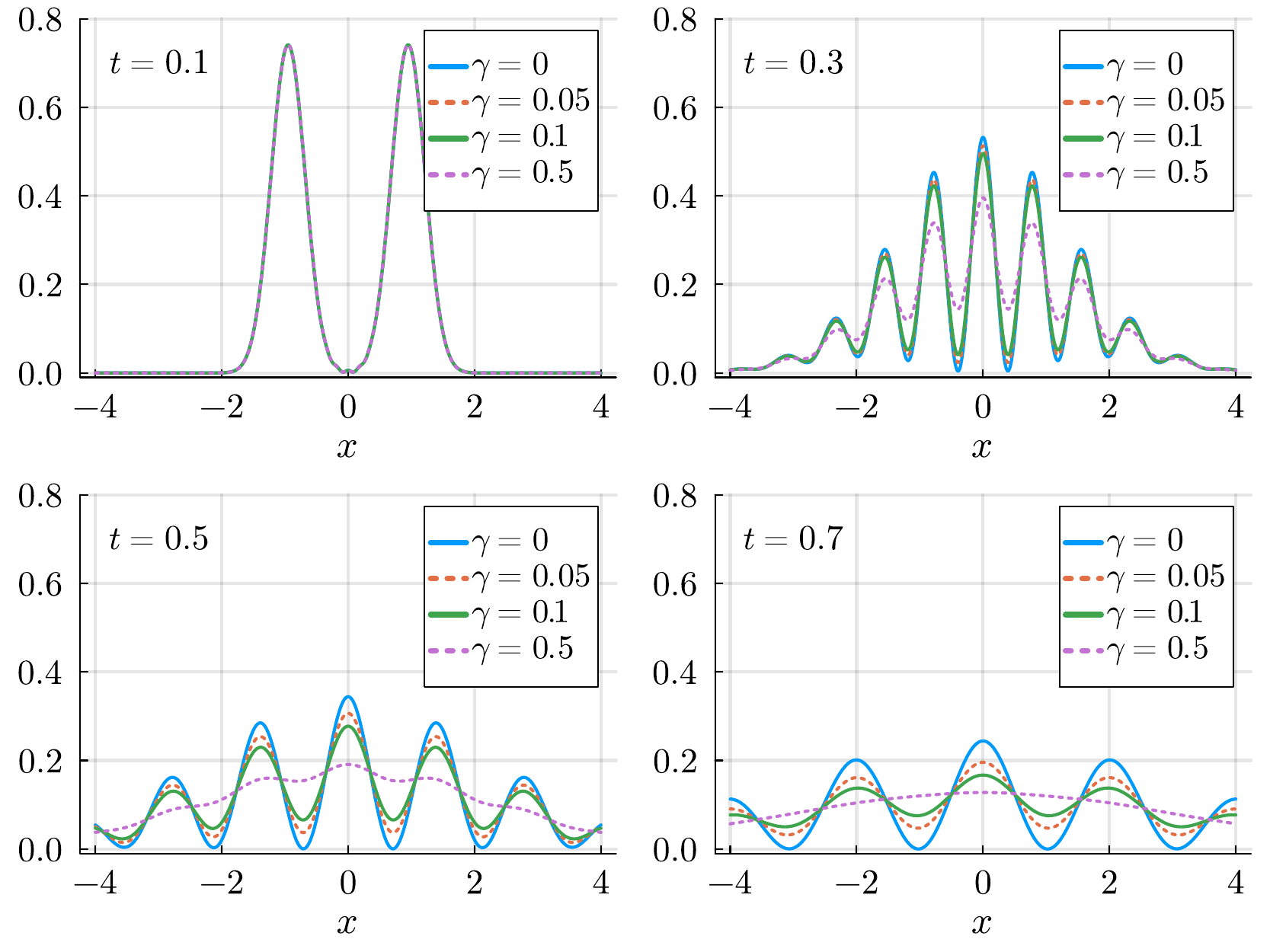}
  \subcaption{$\gamma$ dependence}
  \vspace{10pt}
  }
 	{
  \includegraphics[width=0.8\textwidth]{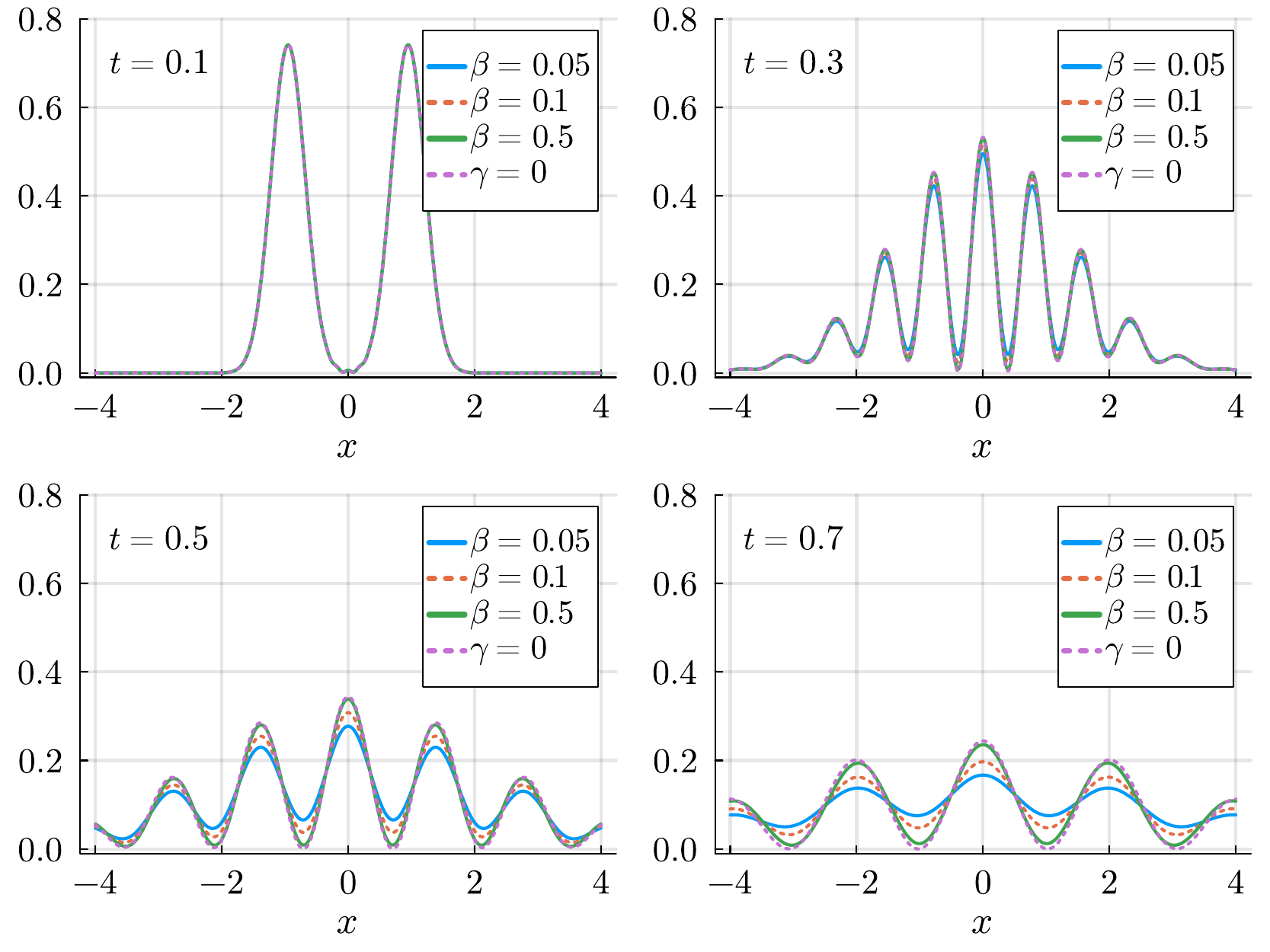}
  \subcaption{$\beta$ dependence}
  }
	\caption{The total reduced density matrix
          $\rho_\mathcal{S}(x,x;t)$ is plotted against $x$
          for various $\gamma$ with $\beta=0.05$ (a)
          and for various $\beta$ with $\gamma=0.1$ (b).
          In the latter, the result for $\gamma=0$ is also shown for comparison.}
        \label{fig:rho_total_gamma}
\end{figure}

Combining the direct terms and the interference term,
we obtain the total reduced density matrix $\rho_\mathcal{S}(x,x;t)$
defined by \eqref{init-rho-system-double-2-prev}.
In Fig.~\ref{fig:rho_total_gamma} we plot it
against $x$
at $t=0.1$, $0.3$, $0.5$ and $0.7$ 
for various $\gamma$ with $\beta=0.05$ (a)
        and for various $\beta$ with $\gamma=0.1$ (b).
As we have seen in Fig.~\ref{fig:plot_rho_00},
the two wave packets in the direct terms
start to overlap at $t=0.3$.
Correspondingly, we start to see a clear interference pattern at $t=0.3$.
At later times,
the interference pattern tends to disappear
for larger $\gamma$ and higher temperature $T=\beta^{-1}$
due to the effect of decoherence.


%
%

\section{Summary and discussions}
\label{sec:conclusion}

In this paper we have investigated quantum decoherence
numerically in the real-time path integral formalism.
While this has been thought to be extremely difficult due to the sign problem
that occurs in standard Monte Carlo methods,
recent developments of numerical methods such as the GTM
have made it possible practically.
In particular,
in the case of the Caldeira--Leggett model, which has been studied
extensively as a model of quantum decoherence,
we have pointed out that
the GTM simplifies drastically since there is only a single relevant saddle point,
which can be obtained numerically
by solving a linear equation with a sparse coefficient matrix,
and the integration over the Lefschetz thimble is nothing but the Gaussian
integral around the saddle point, which can be done analytically.
This enabled us to obtain the time-evolved reduced
density matrix reliably even for a long time and for a large
number of harmonic oscillators in the environment.

Unlike the previous works on the Caldeira--Leggett model,
we were able to obtain explicit results for completely general parameters
without any assumptions
or approximations.
In particular,
we have succeeded in reproducing the scaling behaviors of the decoherence
predicted from the
master equation at weak coupling and at high temperature.
We have also seen certain deviation from the predicted behavior
by lowering the temperature.
Furthermore, by increasing the degrees of freedom in the environment
and by investigating the long-time evolution of the system,
we were also able to see clear tendencies that the system of our interest
thermalizes through the interaction with the environment, which
plays the role of the heat bath.

Below we list some future directions.
%
%
First of all, it is interesting to explore the parameter regimes
that were not studied previously due to the limitation of the theoretical methods.
For instance, our calculations do not rely on 
the Markov approximation,
which implies that
there is no obstacle in investigating 
the case in which
the harmonic oscillators in the environment
has a spectral density $\rho(\omega)\, C^2(\omega) \propto \omega^p$
with $p<2$ (sub-Ohmic) or $p>2$ (super-Ohmic)\footnote{See
  Ref.~\cite{PhysRevD.45.2843} for previous work
  based on the master equation in this direction.}.
Furthermore,
in order to investigate
a more realistic model of quantum decoherence,
one can
investigate a system with an anharmonic potential and/or with a non-Gaussian
initial wave function
by using
the GTM \cite{Fukuma:2017fjq,Fukuma:2019uot,Fukuma:2020fez,Fukuma:2021aoo,Fujisawa:2021hxh,Fukuma:2023eru,Nishimura:2024bou}.
%
We hope that the present work provides
an ideal testing ground
for such calculations.

Last but not the least,
quantum decoherence discussed in this paper is expected to play
a crucial role also in quantum-to-classical transition \cite{Schlosshauer,Zurek:1991vd}.
For instance,
it would be interesting to investigate the mechanism of the transition
based on the environment-induced
superselection \cite{Zurek:1981xq,Zurek:1982ii,Zurek:2003zz}
(See also \cite{Konishi:2021nnt, Konishi:2022vzz} and references therein
for recent discussions.)
by performing explicit calculations
discussed in this paper.

\subsection*{Acknowledgements}
We would like to thank Yuhma Asano, Masafumi Fukuma, Kouichi Hagino,
Yoshimasa Hidaka, Katsuta Sakai, Hidehiko Shimada, Kengo Shimada, Hideo Suganuma
and Yuya Tanizaki for their helpful discussions and comments.
H.~W.~was supported by Japan Society for the Promotion of Science (JSPS) KAKENHI Grant numbers, 21J13014, 22H01218 and 23K22489.

\appendix

\section{Discretization effects due to finite lattice spacing $\epsilon$ and
finite $N_\mathcal{E}$}
\label{sec:limits_suppl}

\begin{figure}[th]
  \centering
  \includegraphics[width=\textwidth]{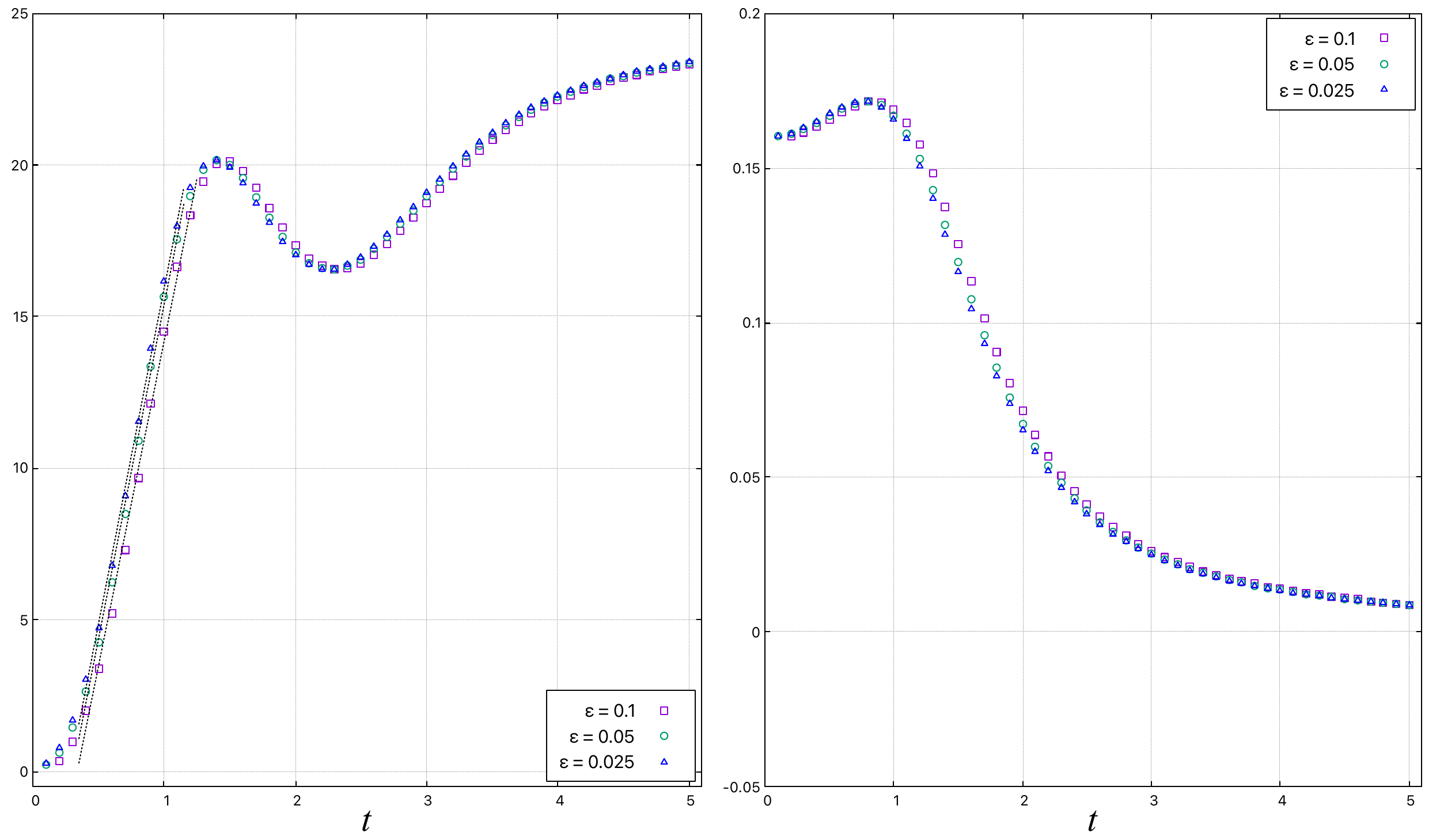}
  \caption{The quantities
    $\Gamma_\mathrm{off\mathchar`-diag}(t)$ (Left)
    and
    $\Gamma_\mathrm{diag}(t)$ (Right)
    are plotted against $t$
    for the lattice spacing $\epsilon = 0.025, 0.05, 0.1$ with
    $N_\mathcal{E}=64$, $\omega_\mathrm{cut}=2$, $\beta=0.05$ and $\gamma = 0.1$ fixed.
    The dotted lines in the left panel represent fits
    to the behavior $A \frac{16\gamma}{\beta} t + B$.
  }
    \label{fig:lattice-spacing_dependency}
\end{figure}


\begin{figure}[t]
	\centering
	\includegraphics[width=0.9\textwidth]{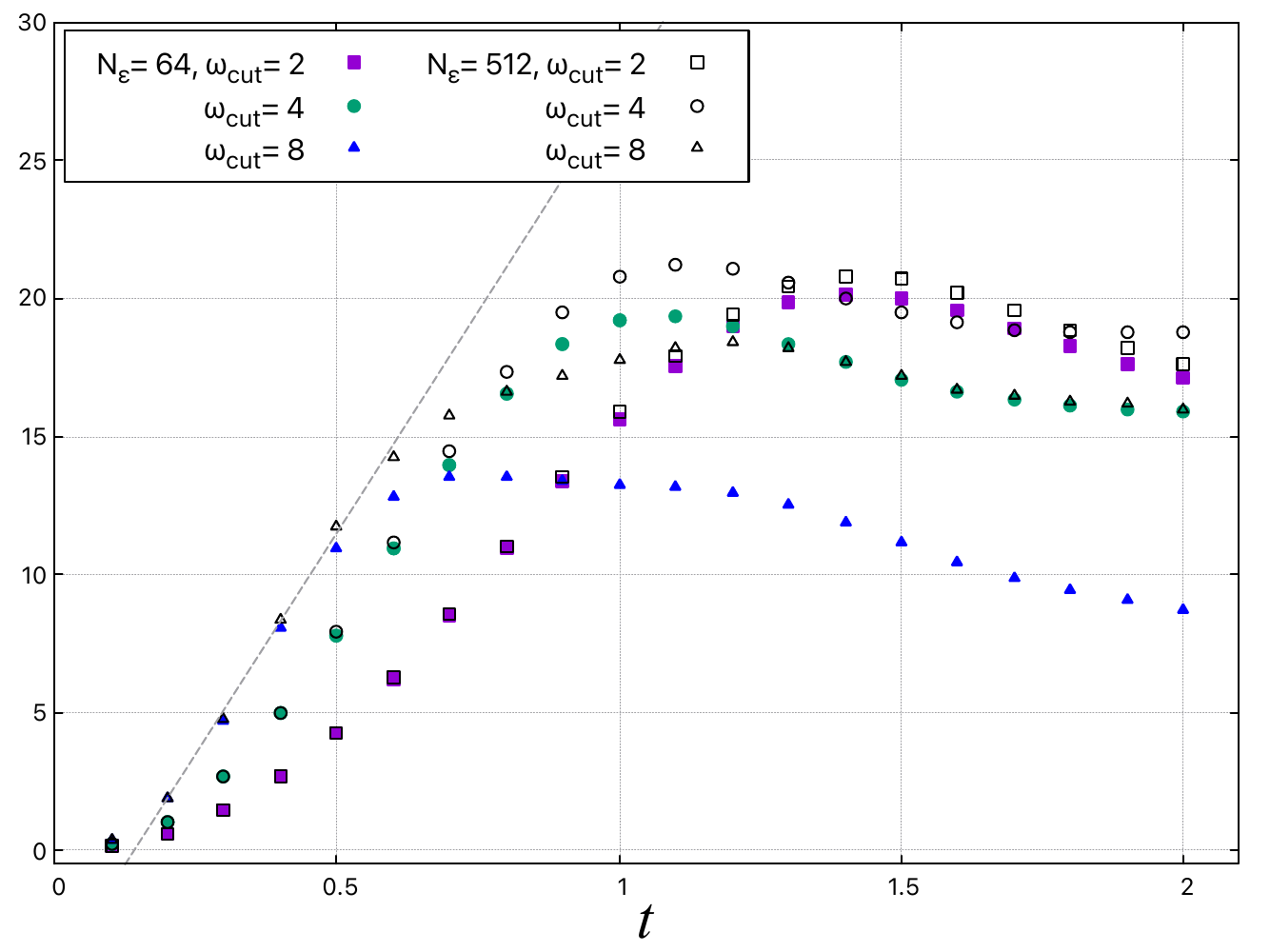}
	\caption{
  The quantity $\Gamma_\mathrm{off\mathchar`-diag}(t)$
  is plotted against $t \le 2$
  for various $\omega_\mathrm{cut}=2, 4, 8$ and $N_\mathcal{E} = 64, 512$
  with $\beta=0.05$ and $\gamma = 0.1$ fixed.
  The dashed line represents a guideline $16\frac{\gamma}{\beta} t +\mathrm{const}$
  for the prediction from the master equation.
  }
        \label{fig:cut_dependency_scaling}
\end{figure}

In this appendix,
we discuss the discretization effects due to
finite lattice spacing $\epsilon$ in time and finite $N_\mathcal{E}$ in the environment.

In Fig.~\ref{fig:lattice-spacing_dependency}, we plot
$\Gamma_\mathrm {off\mathchar`-diag}(t)$ (Left) and $\Gamma_\mathrm{diag}(t)$ (Right)
against $t$ for the lattice spacing $\epsilon = 0.025, 0.05, 0.1$
with $N_\mathcal{E}=64$, $\omega_\mathrm{cut}=2$, $\beta=0.05$ and $\gamma = 0.1$ fixed.
While we do see some $\epsilon$ dependence,
the value of the slope
and the convergence
to the thermal equilibrium are not drastically modified.



In Section \ref{sec:cutoff_dep}, we have seen that the condition
$\omega_\cut \gg \omega_\mathrm{r}$ is important in the agreement with the
prediction from the master equation.
Here we note that $N_\mathcal{E}$ has to be increased as we increase $\omega_\cut$
in order to regard the frequency spectrum \eqref{eq:omega_k_finiteNenv}
of the environment as continuous.
In Fig.~\ref{fig:cut_dependency_scaling}, we plot the quantity
$\Gamma_\mathrm{off\mathchar`-diag}(t)$ against $t$ for various choice of
$\omega_\cut$ and $N_\mathcal{E}$.
We find that the slope of the linear growth of $\Gamma_\mathrm{off\mathchar`-diag}(t)$
does not differ significantly between $N_\mathcal{E}=64$ and $N_\mathcal{E}=512$
for $2 \le \omega_\cut \le 8$.
Note, however, that the finite $N_\mathcal{E}$ effect
becomes more significant at late times
as we can see in Fig.~\ref{fig:Nenv_dependency} (Left)
due to the recurrence effects.

\section{More on decoherence for two wave packets}
\label{sec:more_double-slit}

In this appendix, we investigate
the decoherence between the two wave packets
discussed in Section \ref{sec:numerical_analysis}
from a slightly different point of view.
Here we pay attention to the off-diagonal elements of the
reduced density matrix, which was discussed in the case of a
single wave packet in Section \ref{sec:main_results_case1}.


For the off-diagonal elements of the reduced density matrix
corresponding to $x_\mathrm{F} = - \tilde{x}_\mathrm{F} = x$,
the exponent $\mathcal{A}_{ab}$ that appears in \eqref{eq:rho-detM-2}
reduces to
\begin{align}
  \mathcal{A}_{ab}
  &=  \frac{1}{2}
(c_\mu+\tilde{c}_\mu) (\mathcal{M}^{-1})_{\mu\nu} (c_{\nu}+\tilde{c}_\nu) x^2
-\ii E_\mu^{ab} (\mathcal{M}^{-1})_{\mu\nu}
( c_{\nu} + \tilde{c}_{\nu}) x
- \frac{1}{2} E_\mu^{ab} (\mathcal{M}^{-1})_{\mu\nu}  E_\nu^{ab} 
\ .
\end{align}
Thus each component of the density matrix is given by
\begin{align}
\rho_{00}(x,-x;t)
  &\simeq \exp \left\{ -  \frac{1}{2} K_6 \, x^2
  + \ii \left( k K_8 + p K_7\right)  x
+ \frac{1}{2} (
k^2  K_4
- p^2 K_5 )
\right\} \ , \\
\rho_{11}(x,-x;t)
&= \rho_{00}(-x,x;t) \ ,
\\
\rho_{01}(x,-x;t)
  &\simeq \exp \left\{ -  \frac{1}{2}
K_6 \, x^2
- \left( k  K_7  - p K_8 \right)  x
+ \frac{1}{2} (
k^2  K_5
- p^2 K_4 )
\right\} \ ,
\\
\rho_{10}(x,-x;t)
&= \rho_{01}(-x,x;t) \ , 
\end{align}
omitting the prefactors common to all components.
We have defined the real quantities
\begin{align}
K_6 &=
(c_\mu+\tilde{c}_\mu) \,  {\rm Re}
(\mathcal{M}^{-1})_{\mu\nu} (c_{\nu}+\tilde{c}_\nu) \ , \\
K_7 &=
  (e_{\mu}-\tilde{e}_\mu) \, {\rm Im} (\mathcal{M}^{-1})_{\mu\nu}
( c_{\nu} + \tilde{c}_{\nu}) \ ,  \\
K_8 &=
 (e_{\mu}+\tilde{e}_\mu) \,  {\rm Re} (\mathcal{M}^{-1})_{\mu\nu}
  ( c_{\nu} + \tilde{c}_{\nu}) \ . 
\end{align}

We would like to see how the off-diagonal elements
of the interference term $\rho_{01}(x,-x;t) $
decrease with time.
For that, we need to normalize the reduced density matrix
by the normalization factor \eqref{normalize-rho}.
The off-diagonal elements of the properly normalized reduced density matrix read
\begin{align}
  &~  \rho_{01}(x,-x;t) \nn \\
&=
\frac{1}{{\cal N}(t)} \exp \left\{ -  \frac{1}{2}
K_6 \, \left( x + \frac{k  K_7 - p K_8}{K_6} \right)^2
+  \frac{1}{2} \frac{(k  K_7 - p K_8)^2}{K_6} 
+ \frac{1}{2} (
k^2  K_5
- p^2 K_4 )
\right\} \ .
\end{align}

\begin{figure}[H]
	\centering
 	{
  \includegraphics[width=0.8\textwidth]{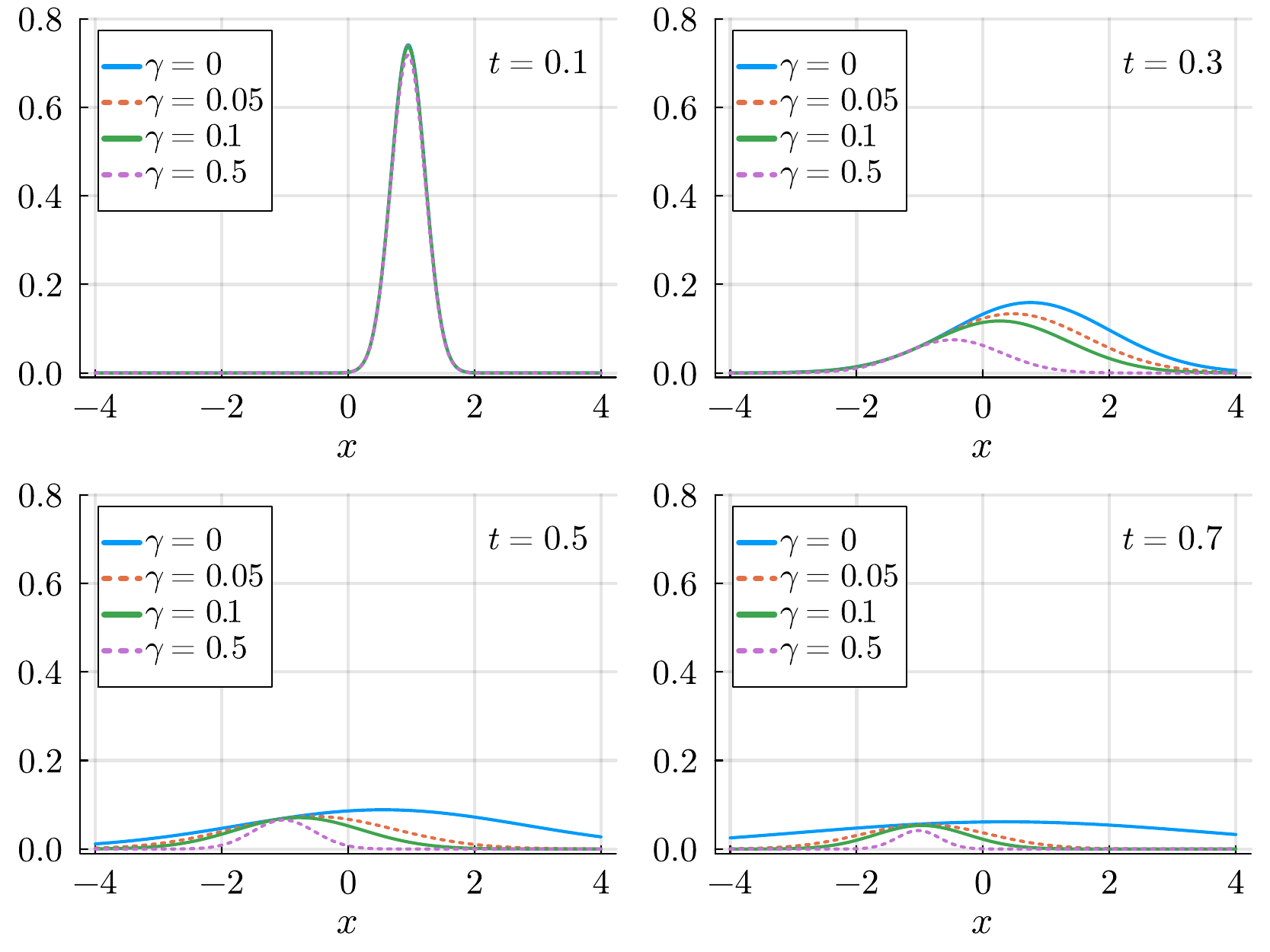}
  \subcaption{$\gamma$ dependence}
  \vspace{10pt}
  }
 	{
  \includegraphics[width=0.8\textwidth]{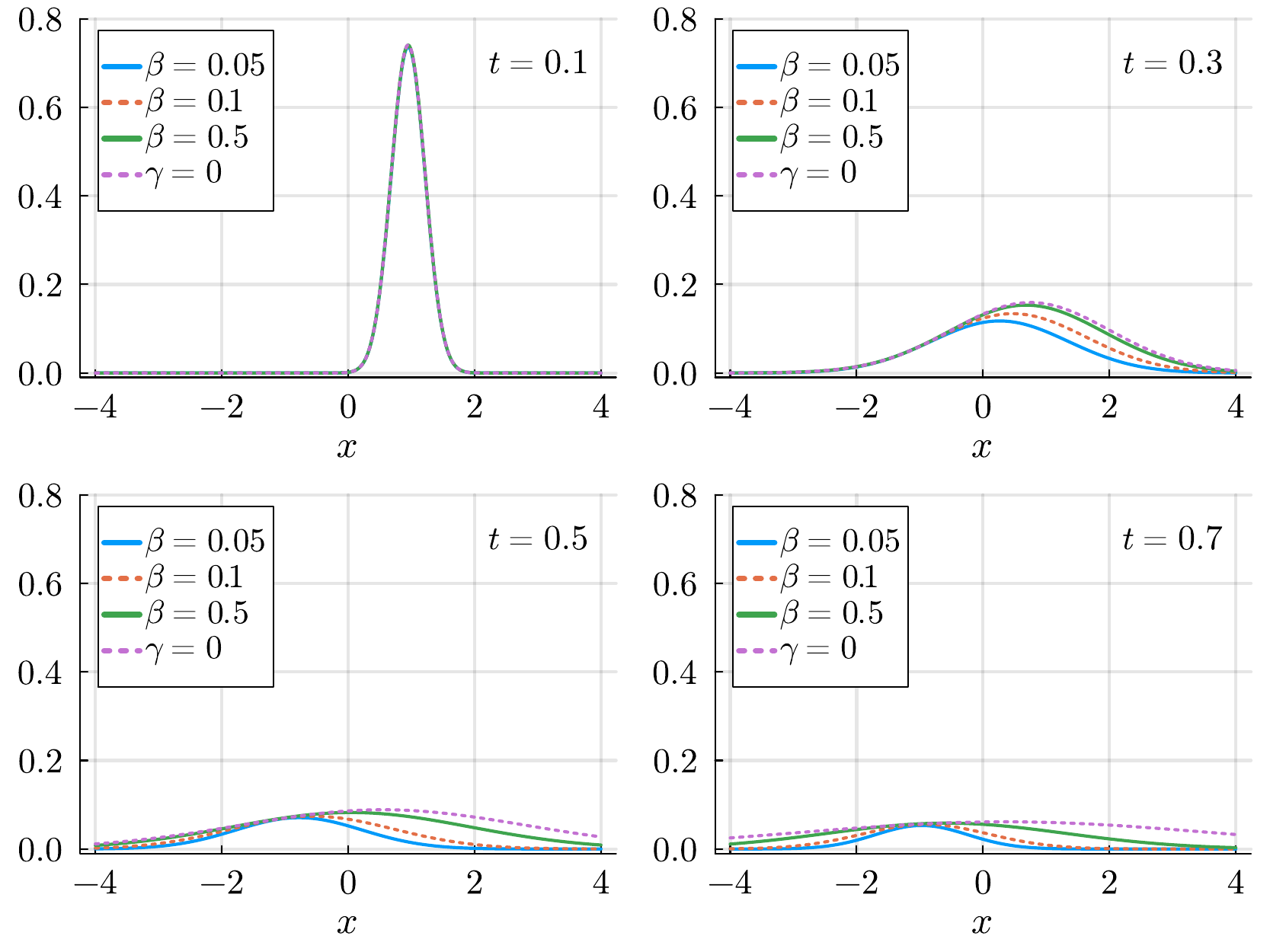}
  \subcaption{$\beta$ dependence}
  }
        \caption{The off-diagonal elements of the interference term
          $\rho_{01}(x,-x;t)$
          is plotted at $t=0.1,0.3,0.5,0.7$
          for various $\gamma$ with $N_\mathcal{E}=64$ and $\beta=0.05$ (Top)
          and
          for various $\beta$ with $N_\mathcal{E}=64$ and $\gamma=0.1$ (Bottom).
          In the latter, the result for $\gamma=0$ is also shown for comparison.
        }
        \label{fig:Re_rho_01_beta}
\end{figure}

In Fig.~\ref{fig:Re_rho_01_beta},
we plot
$\rho_{01}(x,-x;t)$ against $x$
for $t=0.1$, $0.3$, $0.5$, $0.7$.
Note that we do not see
the interference pattern unlike in Fig.~\ref{fig:re_rho_01}
since we are looking at the off-diagonal components of the density matrix.
We find that the amplitude is suppressed
compared to the $\gamma = 0$ case without the environment,
and this suppression is stronger for larger $\gamma/\beta$.

\begin{figure}[t]
	\centering
 	\scalebox{0.35}{\includegraphics{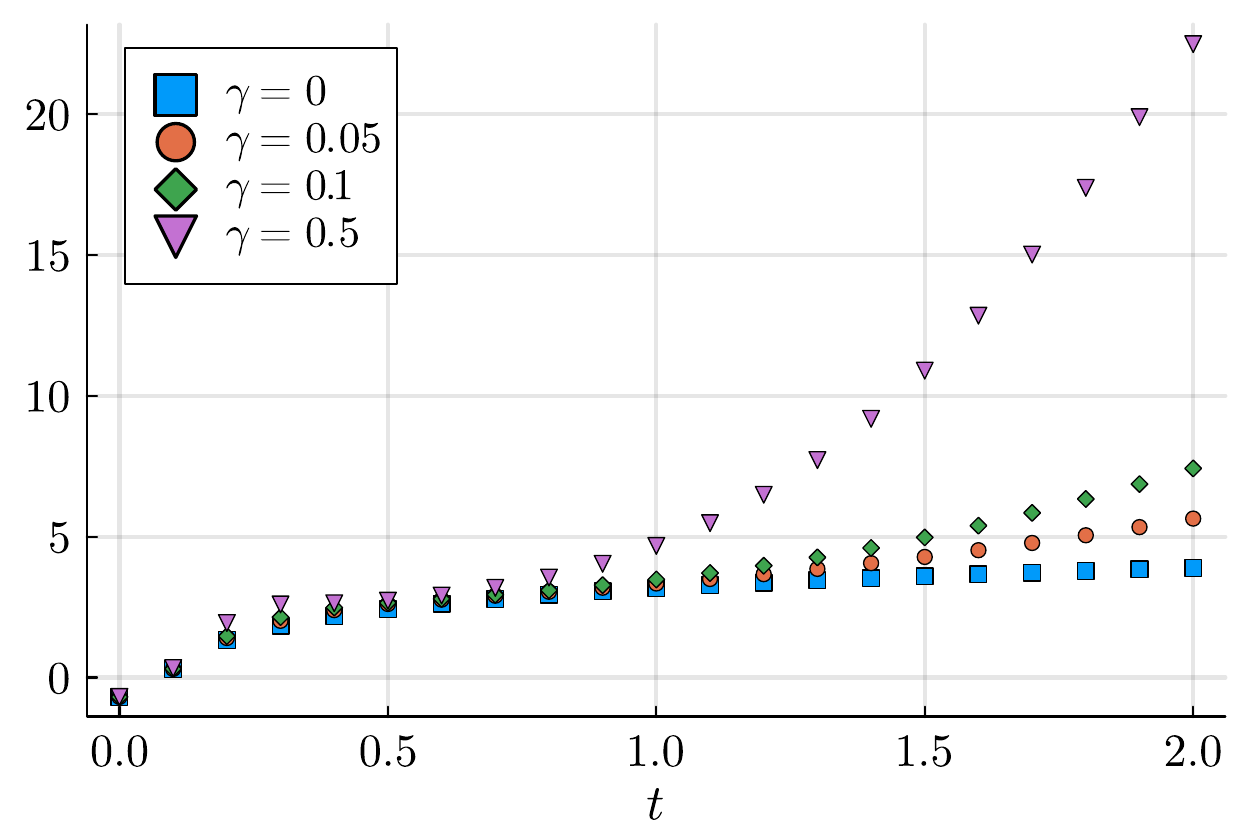}}
	\scalebox{0.35}{\includegraphics{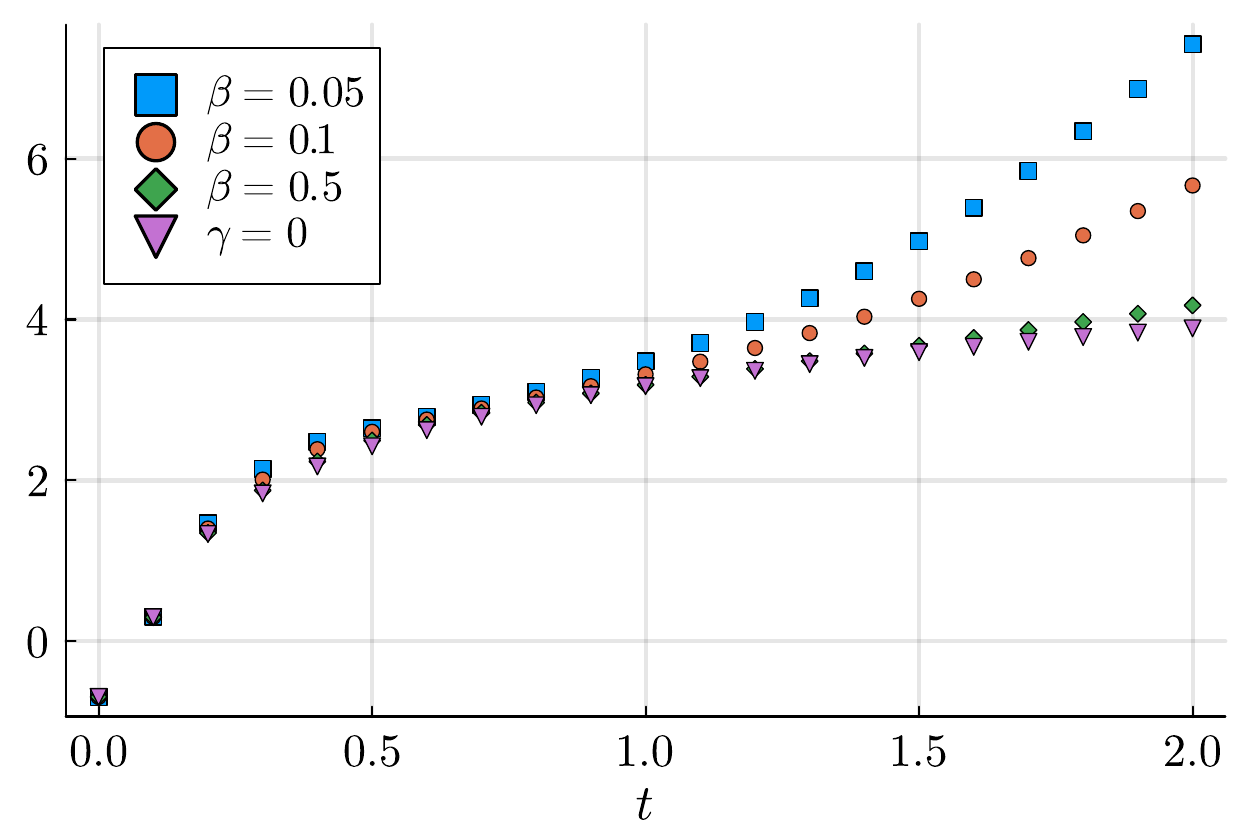}}
	\caption{The quantity $-\log h(t)$
          with $h(t)$ being the peak height of
          $\rho_{01}(x,-x;t)$
          is plotted against $t$
          for various $\gamma$ with $\beta=0.05$ (Left)
          and for various $\beta$ with $\gamma=0.1$ (Right).
          In the latter, the result for $\gamma=0$ is also shown for comparison.}
        \label{fig:height_log_param}
\end{figure}

\begin{figure}[t]
	\centering
	{
  \scalebox{0.35}{\includegraphics{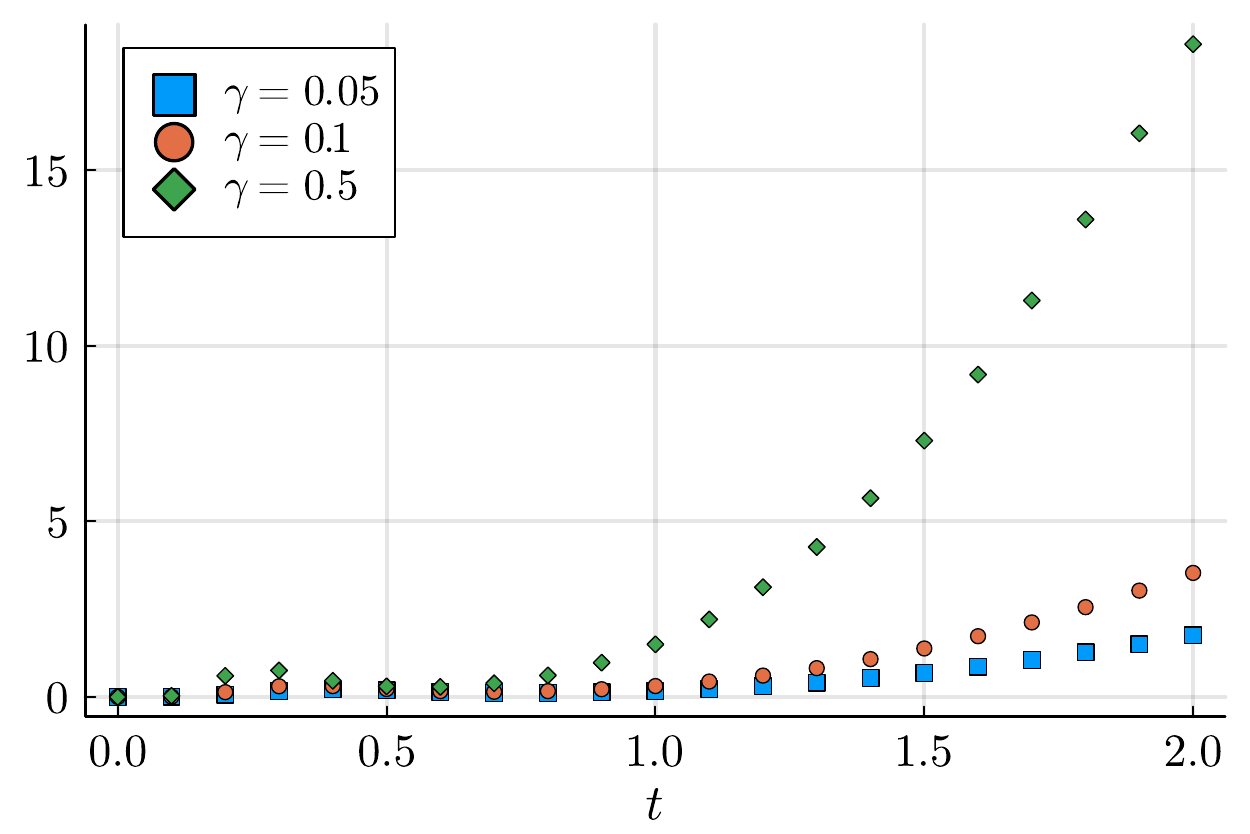}}
 	\scalebox{0.35}{\includegraphics{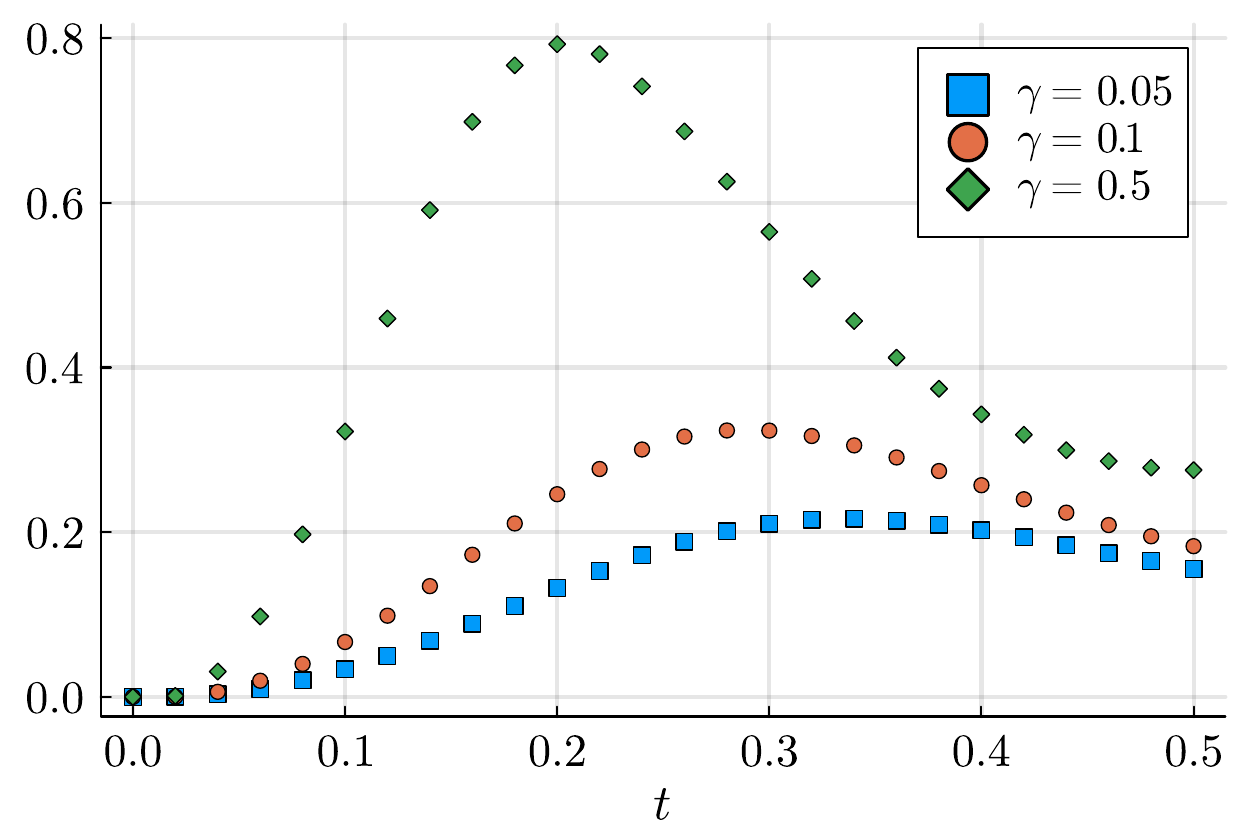}}
  \subcaption{$\gamma$ dependence}
  \vspace{10pt}
  }
  {
	\scalebox{0.33}{\includegraphics{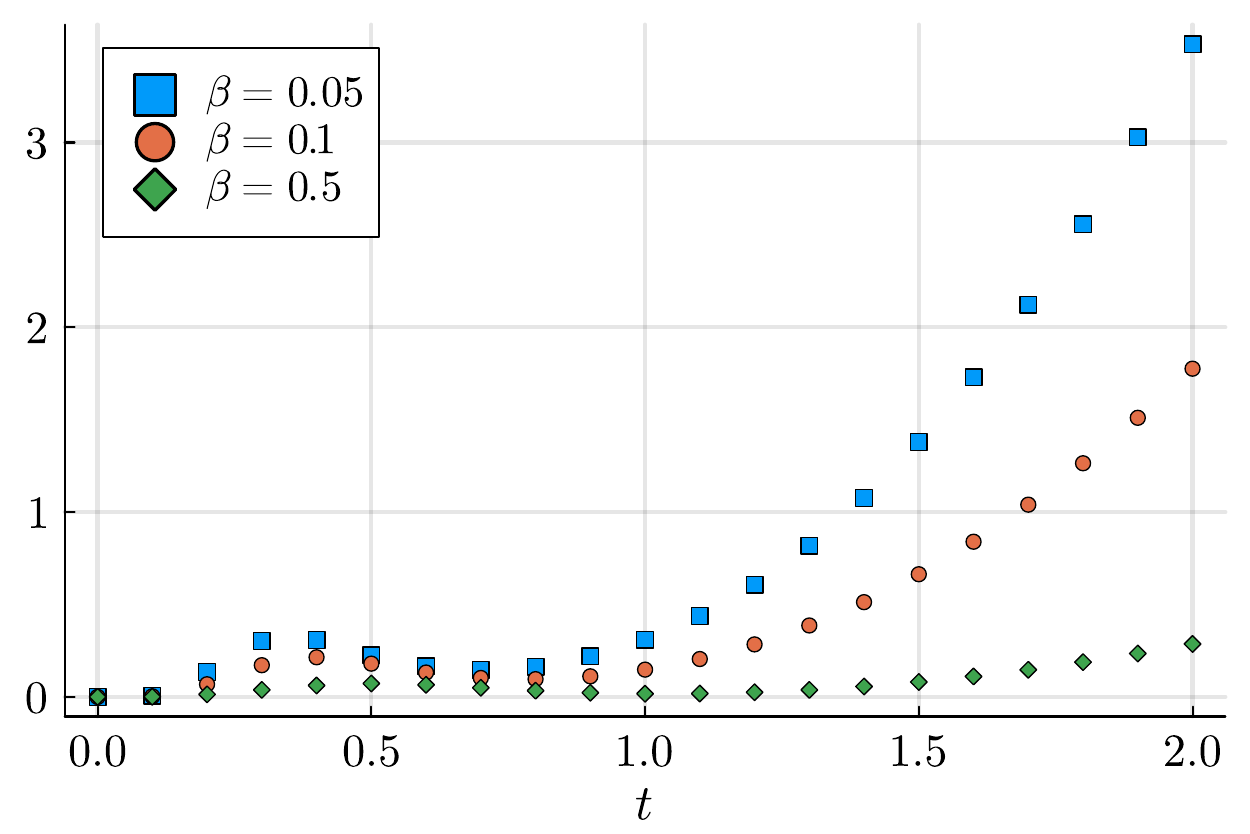}}
 	\scalebox{0.33}{\includegraphics{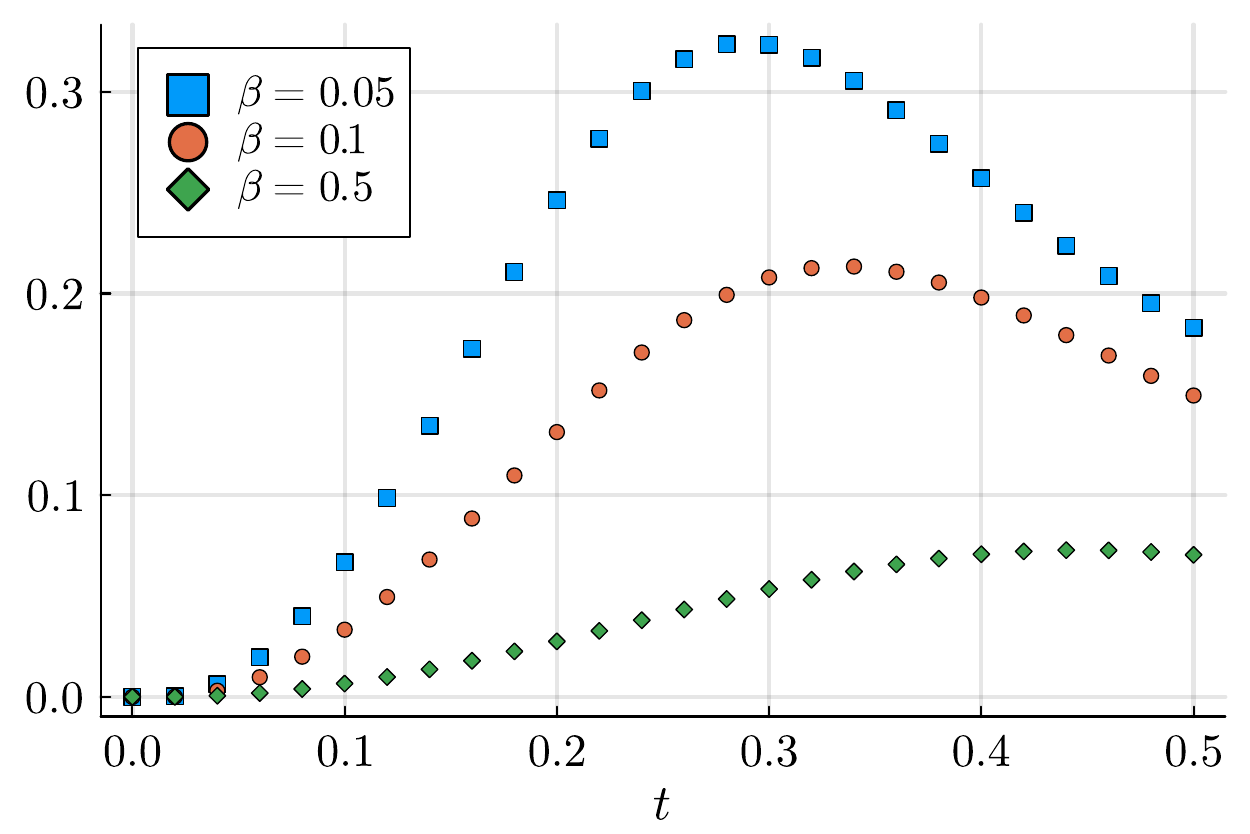}}
  \subcaption{$\beta$ dependence}
  }
	\caption{The quantity
          $-(\log h(t)-\left.\log h(t)\right|_{\gamma=0})$
          is plotted
          for various $\gamma$ with $\beta=0.05$ (Top)
          and for various $\beta$ with $\gamma=0.1$ (Bottom).
          On the left, we show a longer time region $0 \le t \le 2$
          obtained with the lattice spacing $\epsilon = 0.05$,
          whereas on the right, we show a shorter time region $0 \le t \le 0.5$
          obtained with the lattice spacing $\epsilon = 0.01$.}
        \label{fig:log_height_subtracted_beta}
\end{figure}

\begin{figure}[H]
	\centering
	\includegraphics[width=0.75\textwidth]{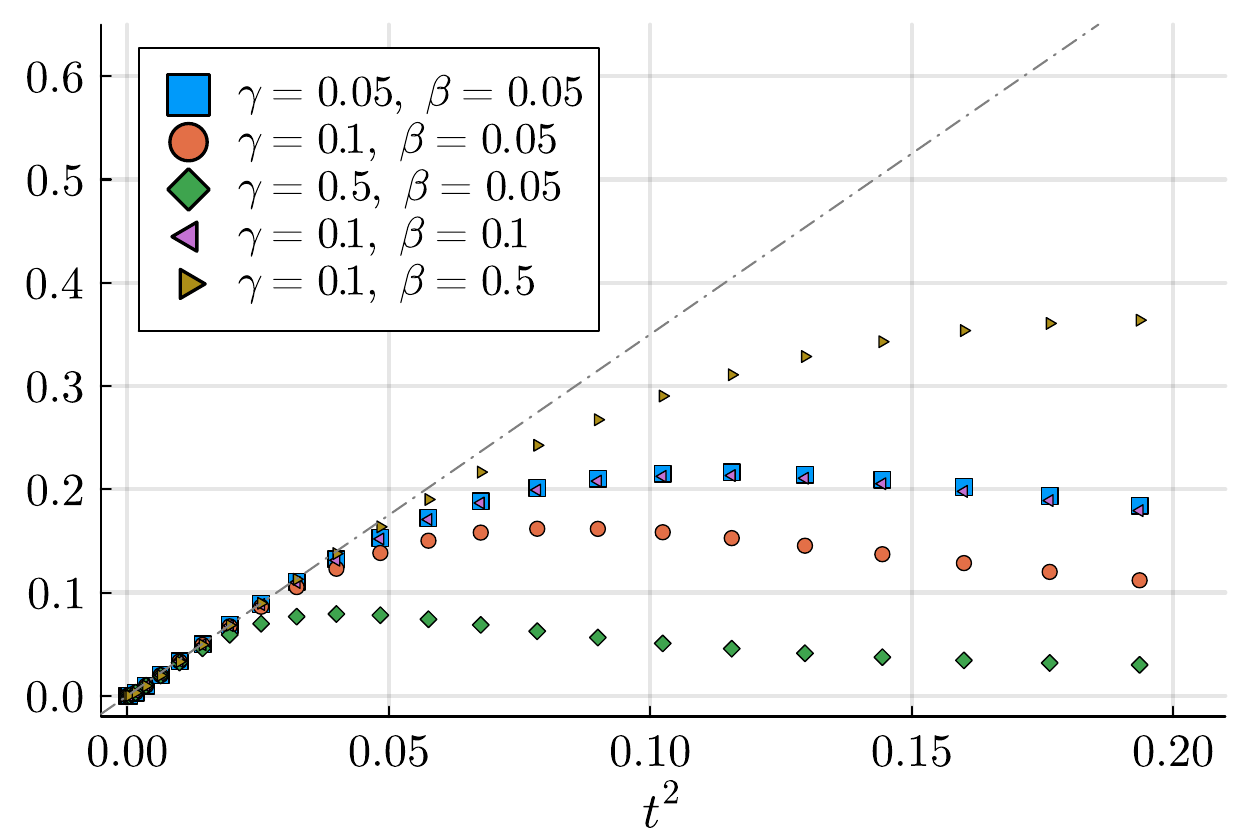}
	\caption{
          The rescaled quantity
          $-\frac{\beta}{\gamma}(\log h(t)-\left.\log h(t)\right|_{\gamma=0})$
          is plotted against $t^2$
          for various $\beta$ and $\gamma$
          with $N_\mathcal{E}=64$.
          The dash-dotted line represents a fit to a linear behavior
          $c \, t^2$
          in the small-$t$ region.
        }
        \label{fig:log_height_subtracted_comparison}
\end{figure}

In order to clarify
the effect of decoherence more quantitatively,
we calculate the peak height
$\rho_{01}(x,-x;t)$, which is obtained as
\begin{align}
  h(t) &= \max_x
   \rho_{01}(x,-x;t)
= \frac{1}{{\cal N}(t)} \exp \left\{
 \frac{1}{2} \frac{(k  K_7 - p K_8)^2}{K_6} 
+ \frac{1}{2} (
k^2  K_5
- p^2 K_4 )
\right\} \ .
\end{align}
In Fig.~\ref{fig:height_log_param},
we plot $-\log h(t)$ against $t$
              for various $\gamma$ with $\beta=0.05$ (Left)
          and for various $\beta$ with $\gamma=0.1$ (Right).
Let us extract the effect of decoherence
by considering the ratio $R(t) \equiv h(t)/\left.h(t)\right|_{\gamma = 0}$.
In Fig.~\ref{fig:log_height_subtracted_beta}
we plot the quantity
\begin{align}
  - \log R(t)
&=   - \Big(\log h(t)-\left.\log h(t)\right|_{\gamma=0} \Big)
    \label{def-R-ratio}
\end{align}
against $t$.
The growth of this quantity
at $t \lesssim 0.2$ 
is faster for larger $\gamma$ and for higher temperature $T=1/\beta$.
Fig.~\ref{fig:log_height_subtracted_comparison} shows that
$R(t) \sim \exp \left( - c \, (\gamma/\beta) \, t^2  \right)$,
which is analogous to the behavior of \eqref{def-A-ratio}.
Thus we find that the effect of decoherence is visible also
in the off-diagonal component of the density matrix
although the effect is weaker than what we have seen
in the interference pattern in Section \ref{sec:numerical_analysis}.

%
%

%


\bibliographystyle{JHEP}
\bibliography{ref}

\end{document}